\documentclass[aps,prb,reprint,twocolumn,showpacs,superscriptaddress]{revtex4-1}
\usepackage{amsmath, amssymb, braket, dsfont, times, units}
\usepackage[mathscr]{euscript}
\usepackage{graphicx}
\usepackage{subfigure}     

\newcommand{\texorpdfstring}[2]{#1}		

\usepackage[usenames,dvipsnames]{color}         
\definecolor{Zcolour}{rgb}{0.992, 0.588, 0.22}
\definecolor{dkgreen}{rgb}{0,0.5,0}
\definecolor{purple}{rgb}{0.5,0,0.5}

\newcommand{\norder}[1]{ {\mkern1mu\colon\mkern-4mu{#1}\colon\mkern-3mu} }
\newcommand{\RR}{\ensuremath{\mathrm{RR}_3}}
\newcommand{\aRR}{\ensuremath{\overline{\mathrm{RR}}_3}}
\newcommand{\dsone}{\mathds{1}}
\newcommand{\Z}[1]{\texorpdfstring{\ensuremath{\mathbb{Z}_{#1}}}{Z#1}}
\newcommand{\EC}{E_{\textrm{C}}}
\newcommand{\shift}{\mathscr{S}}
\newcommand{\HallVis}{{\eta_{_H}}}			

\begin{document}

\title{Fibonacci anyons and charge density order in the 12/5 and 13/5 plateaus}
\author{Roger S. K. Mong}
\affiliation{Department of Physics and Astronomy, University of Pittsburgh, Pittsburgh, Pennsylvania 15260, USA}
\author{Michael P. Zaletel}
\affiliation{Station Q, Microsoft Research, Santa Barbara, California, 93106-6105, USA}
\author{Frank Pollmann}
\affiliation{\mbox{Max-Planck-Institut f\"ur Physik komplexer Systeme, N\"othnitzer Str.\ 38, 01187 Dresden, Germany}}
\author{Zlatko Papi\'c}
\affiliation{School of Physics and Astronomy, University of Leeds, Leeds, LS2 9JT, United Kingdom}

\begin{abstract}

The $\nu=12/5$ fractional quantum Hall plateau observed in $\mathrm{GaAs}$ semiconductor wells is a suspect in the search for  non-Abelian Fibonacci anyons.
Using the infinite density matrix renormalization group, we find clear evidence that fillings $\nu = 12/5$ and $\nu=13/5$ are in the $k = 3$ Read-Rezayi phase in the absence of particle-hole symmetry-breaking effects.
The lowest energy charged excitation is identified as a non-Abelian Fibonacci anyon, distinguished from its Abelian counterpart by its local quadrupole moment.
However, several experiments at $\nu = 13/5$ observe a re-entrant integer quantum Hall effect, implying particle-hole symmetry is broken.
We rule out spin polarization as the origin of the asymmetry.
Further, we point out extremely close energetic competition between the Read-Rezayi phase and a re-entrant integer quantum Hall phase.
This competition suggests that even small particle-hole symmetry-breaking perturbations can explain the experimentally observed asymmetry between $\nu = 12/5$ and $13/5$.
We find that at $\nu=12/5$ Landau level mixing favors the Read-Rezayi phase over the re-entrant phase.

\end{abstract}

\maketitle

\section{Introduction}

The richness of the emergent excitations in many-body  systems can belie the simplicity of their underlying interactions.
This precept underlies the continued effort to realize  quantum materials that exhibit fractionalized non-Abelian quasiparticles.
When a finite number of such non-Abelian quasiparticles are introduced to a system, they give rise to a set of degenerate energy levels which cannot be distinguished by any local observable, and thus can encode quantum information resistant to decoherence.
Such  phases of matter, apart from their fundamental interest,  are the proposed building blocks of quantum computers resilient to decoherence.~\cite{Kitaev:QC:03, Nayak:TQCReview:2008}

Two types of non-Abelian quasiparticles, Majorana zero modes and Fibonacci anyons, stand out for their potential experimental relevance.
While significant progress has been made towards realizing emergent Majorana zero modes in a variety of experimental systems,
	\cite{Mourik:12, Das:12, Rokhinson:12, MDeng:12, Finck:13, Churchill:13}
	the number of candidate hosts for Fibonacci anyons---which are in some sense an interacting generalization of Majoranas---remains much more limited.
When two Fibonacci anyons `$\tau$' approach each other, either one or both of them is annihilated: $\tau \times \tau \rightarrow \dsone + \tau$.
These two possibilities can be used to encode a qubit of information, which is naturally protected from decoherence when the pair of $\tau$ are spatially far separated.\cite{Kitaev:QC:03}
The Hilbert space size grows with the number of Fibonacci anyons, following the Fibonacci sequence.
Unlike Majoranas, Fibonacci anyons have ``universal'' braiding statistics: braiding Fibonaccis alone is sufficient to approximate any quantum gate acting on their space of degenerate states. \cite{FreedmanLarsenWang:2002a}
Various lattice models have been proposed realizing Fibonacci anyons, but all such models require complex many-body interactions.\cite{Fidkowski:DFib:09, PhysRevB.88.205101, PhysRevB.91.125138, Barkeshli:GeneralizedKitaevModel:15, Stoudenmire:Fibonacci:15}


Remarkably, there has long been a suspicion that Fibonaccis already exist as the low-energy excitations of a fractional quantum Hall state in the $\nu=\frac{12}{5}$ plateau of $\mathrm{GaAs}$ quantum wells.\cite{Xia_12_5, Kumar:12fifths, ChoiPhysRevB.77.081301, PanPhysRevB.77.075307}
Experiments by Kumar \emph{et al.}~\cite{Kumar:12fifths} have observed an incompressible state at $\nu=\frac{12}{5}$ with a gap of about $\unit[80]{mK}$, larger than expected from the model of non-interacting composite fermions.~\cite{jainbook}
A theoretical proposal for a novel phase at $\nu=\frac{13}{5}$ and $\frac{12}{5}$ filling was first put forward by Read and Rezayi.\cite{ReadRezayi:99}
They described a class of incompressible phases---the \Z{k} ``Read-Rezayi'' sequence---which occur at filling fractions $\nu=N \pm \frac{k}{k+2}$.
The $k=3$ member (``$\Z3$ state") at filling $\nu=2+\frac35$ and $\nu = 3-\frac35$, which we call \RR\ and \aRR\ respectively, involve ``pairing" of triplets of particles and support Fibonacci excitations.

There are potentially less exotic explanations of the $\nu = \frac{12}{5}$ plateau,\cite{Halperin83, Haldane1983, BondersonSlingerland:08} and one worries nature may operate with a principle of parsimony.
While a number of interferometry experiments have provided suggestive evidence for non-Abelian braiding,\cite{Willett:5HalvesInterf:2009, Kang:5HalvesInterf:2011, WillettNayak:5HalvesInterf:2013}
	direct unambiguous detection of these anyons remains challenging.
Thus realistic numerical simulations play an important role.
Refs.~\onlinecite{ReadRezayi:99, RezayiRead:twelvefifths:09} presented some evidence that the ground state of the Coulomb interaction at $\nu=\frac{13}{5}, \frac{12}{5}$ is described by the \RR\ phase. 
These exact diagonalization studies were limited to small systems and required two assumptions: (\textbf{1}) the electron spin is fully polarized; and (\textbf{2}) the cyclotron energy is infinite (i.e., the absence of `Landau level mixing').
Taken together, these assumptions cannot account for a striking experimental observation.
In contrast to the $\nu = \frac{12}{5}$ plateau, the  $\nu=\frac{13}{5}$ plateau is replaced by a reentrant integer quantum Hall state (RIQH) with quantized Hall conductance $\sigma^{xy} = 3\frac{e^2}{h}$, which is believed to imply the formation of charge density order (CDO).\cite{Eisenstein2002, Kumar:12fifths, PanPhysRevB.77.075307, Deng2012_1, Deng2012_2}
Yet under assumptions (\textbf{1}) and (\textbf{2}), fillings $\frac{12}{5}$ and $\frac{13}{5}$ are related by a particle-hole symmetry.
Thus a compelling numerical case for Fibonacci anyons at $\frac{12}{5}$ must drop these assumptions in order to account for the RIQH phase observed at $\nu = \frac{13}{5}$.
Furthermore, it has thus far been impossible to measure the energy and nature of the anyonic excitations.
These are the main aims of this work.

In this work we use recent advances in our understanding of quantum entanglement  and the infinite density matrix renormalization group (iDMRG) to determine the nature of the $\nu=\frac{12}{5}$ and $\nu = \frac{13}{5}$ plateaus.
These developments allow us to measure the properties of individual anyonic excitations and to relax assumptions (\textbf{1}) and (\textbf{2}), and our findings are as follows. 
After verifying the existence of the $\aRR$ phase under assumptions (\textbf{1}) and (\textbf{2}) (Sec.~\ref{sec:vanillaRR}), we compute for the first time the energies and charge distributions of the anyon excitations (Sec.~\ref{sec:anyon}).
We find that the lowest energy charged excitation is a Fibonacci anyon, with a (disorder free) charge gap of approximately $\Delta \sim \unit[2]{K}$. 
Fortuitously, the two Fibonacci fusion outcomes have strikingly different charge distributions, implying the local quadrupole moment could be used for read-out of the fusion outcome.
We then relax the assumption of spin polarization  (\textbf{1})  by simulating both spin species, and verify that the system spontaneously polarizes at both $\nu = \frac{12}{5}, \frac{13}{5}$ in the absence of external Zeeman field.
Finally, we find that there is \emph{exceptionally} close competition between the \aRR\ phase and a RIQH phase; in fact, changing the width of the quantum well drives a phase transition between the two, which may explain the absence of a $\nu=\frac{12}{5}$ plateau in narrow wells.
Previous studies likely did not detect this transition because the CDO of the RIQH is strongly frustrated on the traditionally-used sphere geometry; exploring this competition is well-suited to the infinitely-long cylinder geometry used in iDMRG.
By relaxing assumption (\textbf{2}) and including the effect of Landau level mixing, we show that the $\aRR$ becomes preferred over RIQH at $\nu = \frac{12}{5}$, while within the limits of our numerical accuracy at $\nu = \frac{13}{5}$, it does not.
This provides a plausible explanation for the experimentally observed asymmetry, further  strengthening the interpretation of the $\nu = \frac{12}{5}$ as the $\aRR$ phase, and sheds light on the probable nature of the $\nu = \frac{13}{5}$ RIQH phase.


\section{\texorpdfstring{Identification of $\nu=\frac{12}{5}$ and $\frac{13}{5}$ quantum Hall states as the $k = 3$ Read-Rezayi phase}{Identify RR}}
\label{sec:vanillaRR}

We study electrons in a strong magnetic field $B \hat{z}$ interacting via the Coulomb potential:~\cite{prangegirvin}
\begin{align}
	V^\textrm{Coulomb} &= \int_{\mathbf{r},\mathbf{r}'} \frac{e^2}{4\pi\epsilon} \frac{\norder{\rho(\mathbf{r}) \rho(\mathbf{r}')}}{|\mathbf{r} - \mathbf{r'}|},
	\label{eq:ham}
\end{align}
where $\mathbf{r} = (x,y,z)$ is a position vector.
The electron gas is confined in the $z$-direction by an infinite square well potential of width $w$, and so we project into the lowest subband of the square well.~\cite{prangegirvin, PhysRevB.78.155308}
In a magnetic field the single particles states organize into Landau levels (LL) separated by cyclotron energy $\hbar \omega_c$.
In this Section, we assume the Coulomb interaction is projected into the spin-polarized $N=1$ Landau level (LL). We will later enlarge the Hilbert space by including both spin-species and additional LLs. Energies are expressed in units of $\EC = e^2/4\pi\epsilon \ell_B$, where $\ell_B$ is the magnetic length.
For reference, a $\unit[5]{T}$ magnetic field  corresponds to $\EC \sim \unit[120]{K}$ for $\mathrm{GaAs}$ samples.
Without internal degrees of freedom (spins or LL indices), the $\frac{12}{5}$ and $\frac{13}{5}$ states are related by an exact particle-hole symmetry, so to begin we only present results for $\nu=\tfrac{12}{5}$.

We use infinite DMRG \cite{McCulloch:2008, ZaletelMixing} to study the Coulomb interaction~\eqref{eq:ham} on an infinitely long cylinder of circumference $L$, where its ground state is expected to have topological degeneracy.
Rather than trying to identify the phase via the ground states' overlap with trial wavefunctions, we examine patterns in the quantum entanglement of the ground states which serve as the ``order parameter'' for topological order. For the \aRR\ phase, we expect ten degenerate ground states \cite{ReadRezayi:99} which split into two groups of five; within each quintuplet the states are related by translating the center of mass of the particles.
We label the representative ground states from each quintuplet as $\ket{\Omega_\dsone}$ and $\ket{\Omega_\tau}$.
\footnote{We choose $\ket{\Omega_\dsone}$ to have lower entanglement entropy than that of $\ket{\Omega_\tau}$.}

\begin{figure}[t]
	\includegraphics[width=85mm]{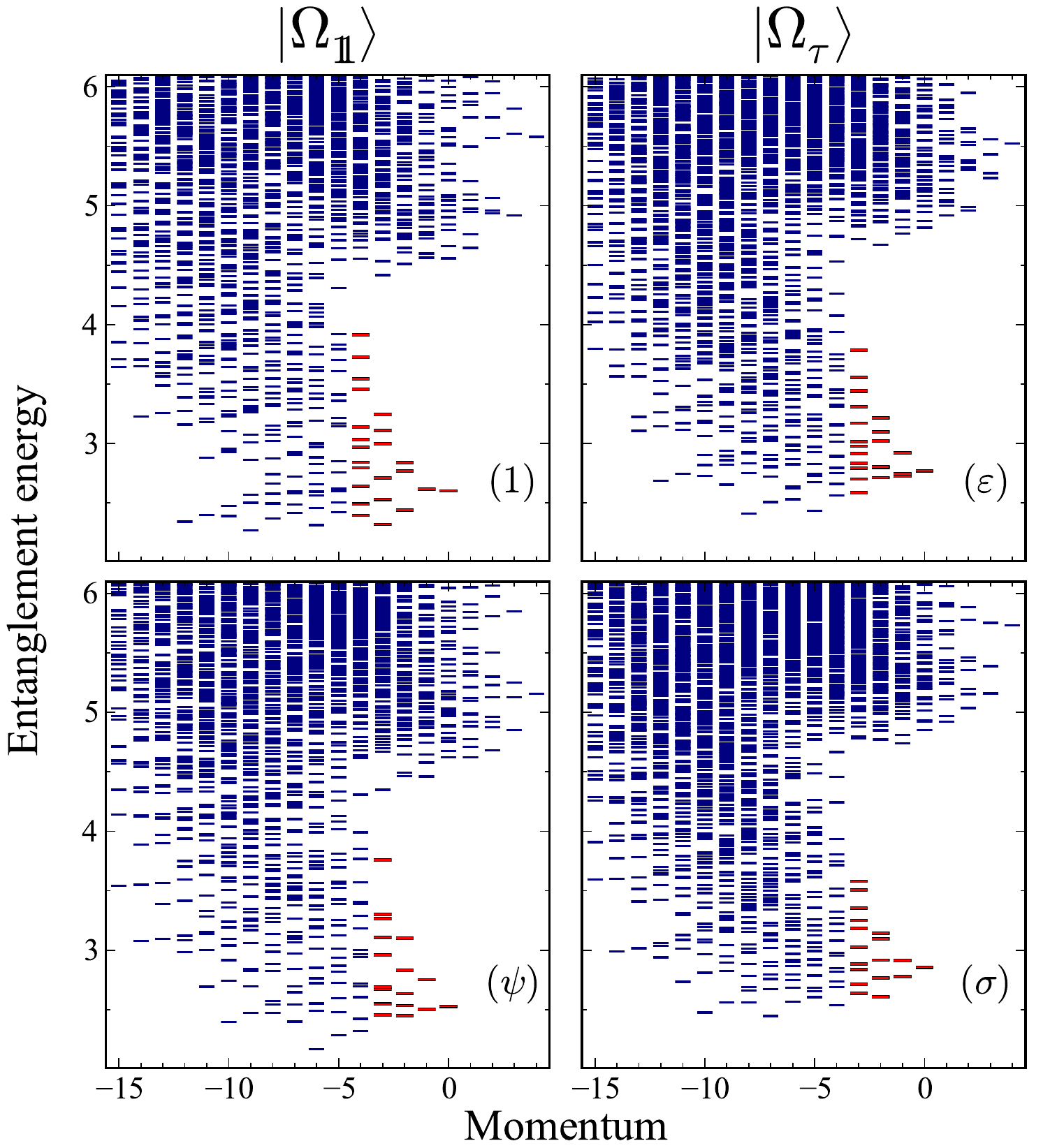}
	\caption{%
		Orbital entanglement spectra for the two degenerate ground states at circumference $L=32\ell_B$, well-width $w=3\ell_B$.
		For each ground state, we pick two orbital cuts and plot the entanglement energies vs.\ angular momentum.
		The low-lying spectra (highlighted in red) agree with the CFT prediction given in Eq.~\eqref{eq:RRcounting}, and thus provide an unambiguous identification of the \aRR\ phase.
	}
	\label{fig:RR_entspec}
\end{figure}
The first evidence for the \aRR\ phase comes from the entanglement spectra (Fig.~\ref{fig:RR_entspec}).
We partition the infinite cylinder into left (L) and right (R) halves, each semi-infinitely long.
Given a wavefunction $\ket{\Psi}$ on the cylinder, the entanglement spectrum~\cite{LiHaldane} $\{\epsilon_\alpha\}$ is the set of eigenvalues of $-\log \rho_L$, where $\rho_L = \operatorname{Tr}_R \ket{\Psi}\!\bra{\Psi}$ is the reduced density matrix for the left half of the system.
Each eigenvalue $\epsilon_\alpha$ is called an ``entanglement energy'' level, and carries quantum numbers of charge and angular momentum, just as a physical edge of the cylinder would.
Generically the low-lying levels of the entanglement spectra (along with their quantum numbers) mimic the physical edge theory of the phase,\cite{KitaevPreskill, LiHaldane, QiKatsuraLudwig} which can be used to identify the topological order.
For the \aRR\ phase, the corresponding edge structure is a product of the $\Z3$ parafermion conformal field theory (CFT)\cite{ZamolodchikovParafermion} with a $\mathrm{U(1)}$ boson.
Each entanglement spectra corresponds to one of six primary fields of the $\Z3$ parafermion CFT: $1$, $\psi$, $\psi^\dag$, $\varepsilon$, $\sigma$, $\sigma^\dag$.
The predicted level countings (i.e., the number of low-energy states for each momentum) are
\begin{align}\begin{aligned}\label{eq:RRcounting}
	1 &:\;		1, 1, 3, 6, 12, \dots ,
&	\varepsilon &:\;		1, 3, 6, 13, 24, \dots,
\\	\psi/\psi^\dag &:\;		1, 2, 5, 9, 18, \dots ,
&	\sigma/\sigma^\dag &:\;		1, 2, 5, 10, 20, \dots.
\end{aligned}\end{align}
Figure~\ref{fig:RR_entspec} shows the orbital entanglement spectra for the ground states $\ket{\Omega_{\dsone/\tau}}$ at $L=32\ell_B$.
The pattern of the low-lying levels (highlighted in the figure) is indicative of a chiral edge mode, with level counting consistent with the corresponding CFT of the Read-Rezayi phase.
In each spectra, the first four levels unambiguously match the theoretical prediction.

\begin{figure}[t]
	\begin{minipage}{85mm}
		\hspace{4.0mm}\includegraphics[width=76mm]{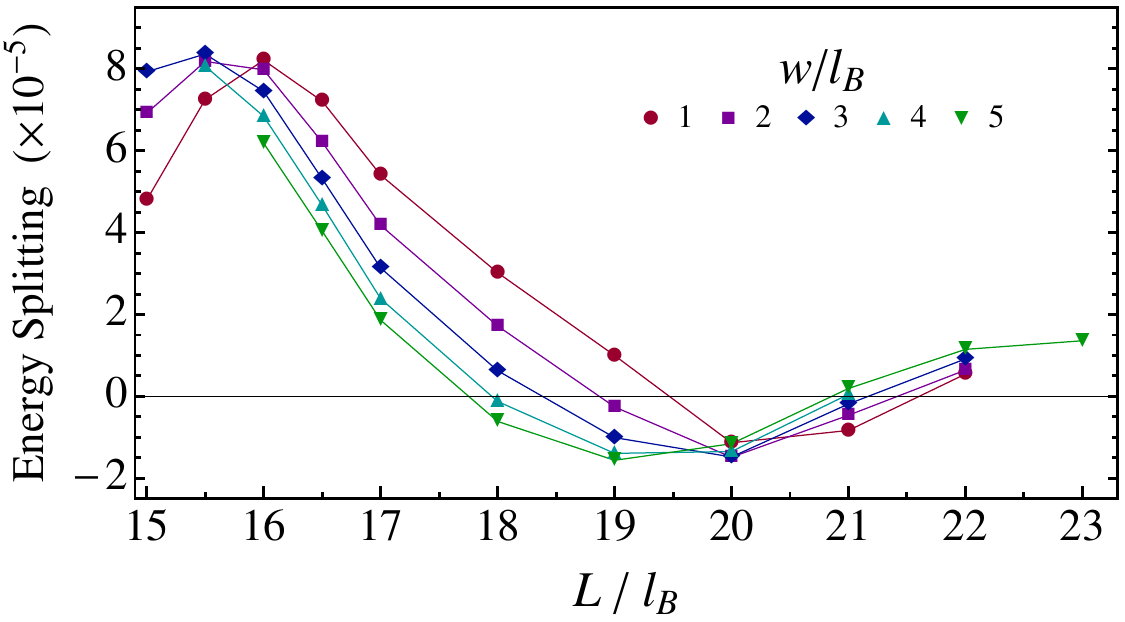}
		\\[-2ex]\hspace{-\textwidth}\begin{minipage}{0mm}\vspace{-75mm}\subfigure[]{\label{fig:RR_topo_splitting}}\hspace{-27mm}\end{minipage}
	\end{minipage}
	\begin{minipage}{85mm}
		\hspace{1.7mm}\includegraphics[width=78mm]{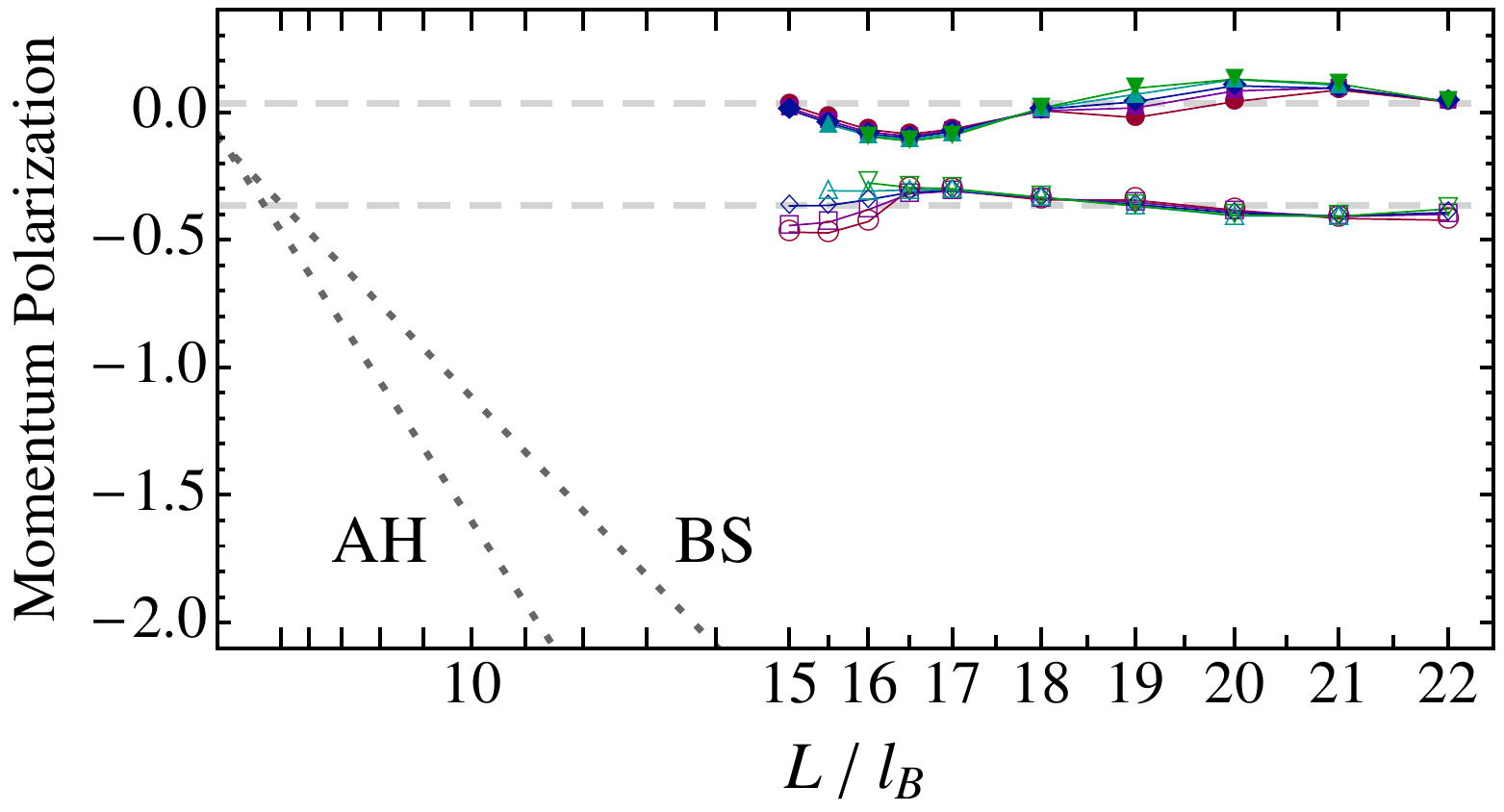}
		\\[-3ex]\hspace{-\textwidth}\begin{minipage}{0mm}\vspace{-75mm}\subfigure[]{\label{fig:RR_mompol_big}}\hspace{-27mm}\end{minipage}
	\end{minipage}
	\caption{%
		(a) The splitting of the topological degeneracy $E(\ket{\Omega_\tau}) - E(\ket{\Omega_\dsone})$ as a function of cylinder circumference $L$ is consistent with an exponential decrease. $w$ is the width of the well.
		(b) Scaling of the momentum polarization $M$ with circumference squared $L^2$, which reveals the shift $\shift$.
		Dashed lines indicate the theoretical predictions for the $\dsone$ and $\tau$ ground states of $\aRR$, and dotted lines the competing hierarchy (AH) and Bonderson-Slingerland (BS) states.
		}
	\label{fig:RR_char}
\end{figure}

Further evidence for the \aRR\ state comes from the vanishing splitting of topological degeneracy in the thermodynamic limit, as shown in Fig.~\ref{fig:RR_topo_splitting}. For circumferences $L \geq 17\ell_B$, we consistently find ten ground states, as expected for the \aRR\ phase. \footnote{For data in the \aRR\ phase, we impose translational invariance of the ground state wavefunction to bias against CDO states.} Moreover, the splitting of the degeneracy is consistent with an exponential decrease with system size.

Finally, a robust quantitative evidence for the \aRR\ phase is given by the ``momentum polarization,''\cite{ZaletelQHdmrg13, HHTuMomPol:13, YeJeHaldane2014} which effectively computes the modular $T$-transformation (Fig.~\ref{fig:RR_mompol_big}). The momentum polarization $M$ is defined to be $M = \operatorname{Tr}[\rho_L\hat{K}]$ with $\hat{K}$ the angular momentum operator on the cylinder; it measures the average amount of momentum carried in the left half of the system.
As explained in Ref.~\onlinecite{ZaletelMixing}, $M$ encodes three topological invariants: the ``shift''~\cite{wenzee} $\shift$, the chiral central charge $c$, and the anyon topological spin $h_a$:
\begin{align}
	M \big[ \ket{\Omega_a(L)} \big]
		&= -\frac{\nu \, \shift \, L^2}{(4\pi\ell_B)^2} + h_a - \frac{c}{24} + \mathcal{O}(e^{-\frac{L}{\xi}}) .
	\label{eq:mompol_shift}
\end{align}
Figure~\ref{fig:RR_mompol_big} shows $M(L)$ plot vs.\ $L^2$ for ground states $\ket{\Omega_{\dsone/\tau}}$.
From the slope of the data, we see that $\shift = 0$, consistent with the \aRR\ phase and definitively ruling out the $\nu=\frac{2}{5}$ hierarchy phase ($\shift=6$) and the Bonderson-Slingerland (BS) phase ($\shift=4$).\cite{BondersonSlingerland:08, BSTwelveFifths:12}
The intercepts $h_a - \frac{c}{24}$ are also consistent with $\aRR$ (cf.\ Appendix~\ref{app:RR_entanglement}).

\section{The nature and energetics of the anyonic excitations}
\label{sec:anyon}
	
\begin{figure}[t]
	\includegraphics[width=87mm]{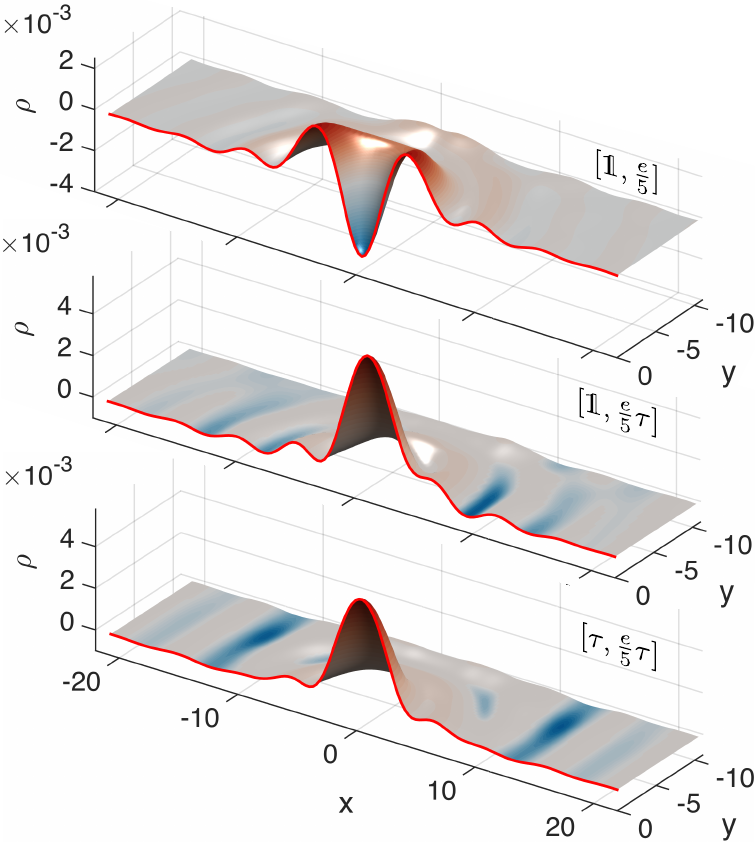}
	\caption{
		The density profile of the charge-$e/5$ anyons in the presence of a Gaussian pinning potential of width $\approx 4 \ell_B$, computed on a $L=21\ell_B$ cylinder. 
We enforce the presence of an anyon in the central region by using various `topological boundary conditions' $[l,r]$, and minimize the energy subject to this constraint.
		(\textbf{top}) The Abelian $\frac{e}{5}$ anyon.
		(\textbf{middle}) The Fibonnaci $\frac{e}{5} \tau$ anyon.
		(\textbf{bottom})
		When using topological boundary conditions $[\tau,\frac{e}{5}\tau]$, the fusion rule $\tau \times \tau = \dsone + \tau$ implies that there are two distinct anyon types that could arise: either an Abelian $\frac{e}{5}$ or the non-Abelian $\frac{e}{5}\tau$. We find the non-Abelian type has lower energy. Consquently, the local properties of the excitation (e.g.\ the charge density and energy) agree between the (\textbf{middle}) and (\textbf{bottom}).
			These quasiparticles are about $15\ell_B$ in diameter.
	}
	\label{fig:anyons}
\end{figure}

Having established the properties of the ground states, we turn to the anyonic excitations.
While certain properties  can be inferred from theory---e.g., the anyon types,  charges, and fusion rules---we are interested in non-topological aspects which depend on microscopic details.
We first determine which anyon type is the lowest energy charged excitation. The gap to this excitation controls the stability of the phase, and is measured in  activated transport measurements.
Second, we measure the charge density profile of the anyons; if distinct, the two fusion outcomes $\tau \times \tau = \dsone + \tau$ could potentially be detected by an electrostatic measurement.


There are ten anyon types in the $\aRR$ phase.
For each fractional charge $m\frac{e}{5}$ ($m\in\Z5$), there is one Abelian and one Fibonacci type, which we label $\frac{me}{5}$ and $\frac{me}{5}\tau$ respectively.
The Coulomb interaction will generally drive charges to fractionalize into the smallest possible units, with charge $\pm\frac{e}{5}$, but there remains two possibilities: $\pm \tfrac{e}{5}$  and $\pm \tfrac{e}{5} \tau$.
If the lowest energy excitation is $\pm \tfrac{e}{5} \tau$, then charged defects would naturally nucleate Fibonacci anyons; braiding and fusion of the $\tfrac{e}{5} \tau$ excitations would in principle be sufficient for universal topological quantum computing.

We use the recently developed ``defect-DMRG'' method \cite{ZaletelQHdmrg13} in order to trap a single anyonic excitation `$a$' on an infinite cylinder.
Referring to App.~\ref{app:defect-dmrg} for details, we take advantage of the DMRG variational ansatz to ensure that to the left and right of some large central region, the state relaxes to one of the degenerate ground states $l$ and $r$ respectively; we call the lowest energy state with this `topological boundary condition' $\ket{[l, r]}$.
The anyon $a$ trapped in the central region must appear in the fusion product: $r \times \bar{l} = a + \dots$, where $\bar{l}$ is the anti-particle of $l$.
In particular, when $l = \bar{l} = \dsone$ is the boundary condition corresponding to the trivial anyon, the trapped anyon $a$ is fixed by the right boundary condition, $a = r$.
By measuring the energy of this configuration relative to the vacuum, we determine the energy of the anyon $a$.
The experiment can be done both with and without an electrostatic trapping potential.

\begin{figure}[b]
	\includegraphics[width=85mm]{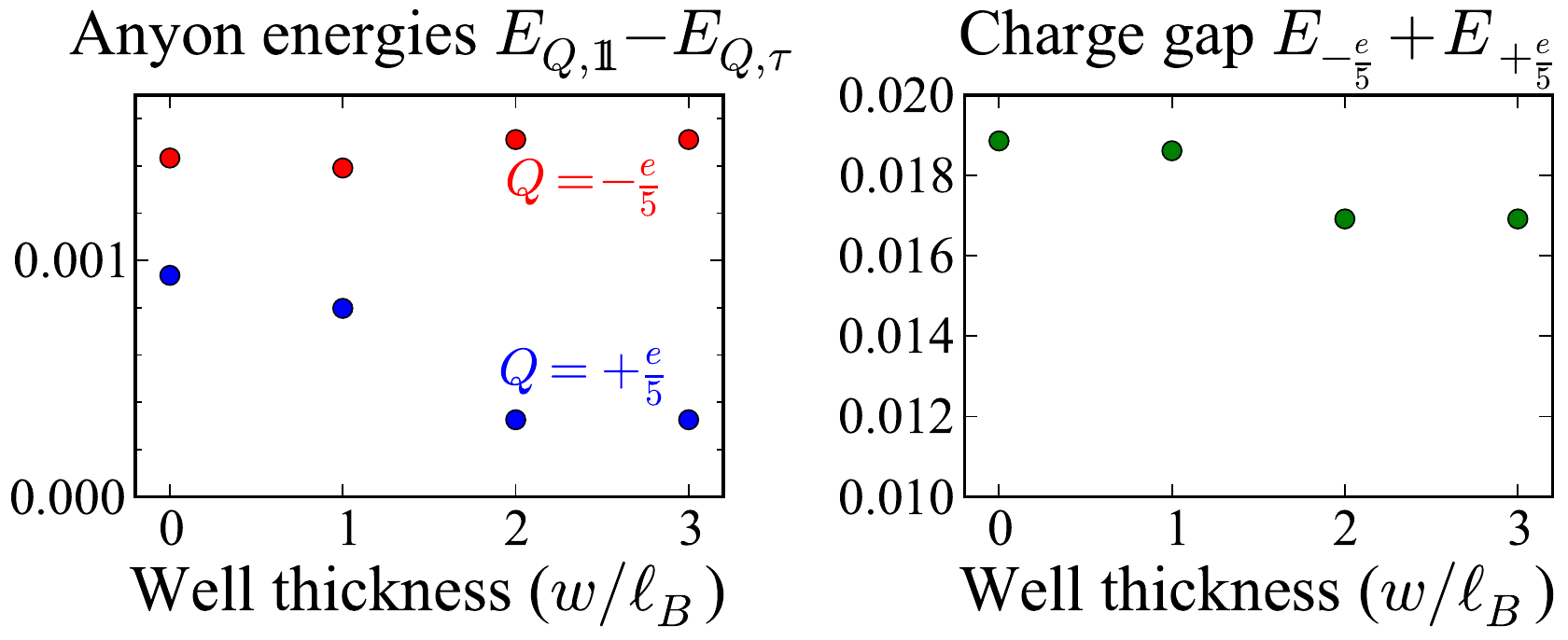}
	\caption{
		Anyon excitation energies $E_{Q,a}$ at various quantum well thicknesses, for the \aRR\ phase in absense of disorder.
		(Left) The difference between the Abelian and Fibonacci anyons for charges $Q = \pm\frac{e}{5}$.
		The data shows that for a fixed charge $Q$, it is energetically favorable to trap a Fibonacci rather than an Abelian anyon.
		(Right) The energy require to disassociate a pair of $\pm\frac{e}{5}$ quasiparticles.
		See the main text for caveats in regards to this data.
	}
	\label{fig:anyons_energies}
\end{figure}

Beginning without a trapping potential, we define the energy $E_a$ of anyon $a$ to be the energy required to add $a$ to the groundstate in the absence of a pinning potential plus the electrostatic interaction between a charge $Q_a$ point-charge and a neutralizing medium.
With this definition, we find $E_{\frac{e}{5}} = -0.0508$ and $E_{\frac{e}{5}\tau} = -0.0511$ at $L=21\ell_B, w=3\ell_B$.
The energy of the Fibonacci particle is lower for all well-widths $w = 0\mbox{--}3\ell_B$ and for both charges $\pm\frac{e}{5}$.
We show this energetic difference (at $L=21\ell_B$) in Fig.~\ref{fig:anyons_energies}.
Thus, in the absence of a pinning potential, it is energetically favorable to bind a Fibonacci $\tau$ to a $\pm\frac{e}{5}$ charge.

The ``charge gap'' for the \aRR\ phase is the energy required to create and separate a pair of charge $+\frac{e}{5}, -\frac{e}{5}$ particles from the ground state, $\Delta = E_{+\frac{e}{5} \tau} + E_{-\frac{e}{5} \tau}$, which we find numerically to be about $0.017$.
Note that while the chemical potential $\mu$ (which we have set to zero) factors into each $E_{\pm\frac{e}{5} \tau}$ individually, the dependence cancels in $\Delta$.
This corresponds to about \unit[2]{K} at $\unit[5]{T}$, much larger than the \unit[80]{mK} activated-transport gap observed in experiments.\cite{Kumar:12fifths}
A similar discrepancy was found for the Moore-Read state at $\nu = \tfrac{5}{2}$, where numerics find a charge gap of about 0.022 ($\approx$ \unit[2.5]{K} at \unit[5]{T})\cite{Morf2002} while the  experiments of Dean \emph{et al.}~\cite{Dean2008} report an activated-transport gap of about 0.0047 ($\approx$  \unit[540]{mK} at \unit[5]{T}).
At $\nu = \tfrac{5}{2}$ the discrepancy is reduced somewhat by including the effects of Landau-level mixing,\cite{Morf2002} but it is believed that disorder broadening plays a large role as well, for example an estimate of $\Gamma \approx \unit[1.75]{K}$ in the sample of Kumar \emph{et al}. \cite{Kumar:12fifths}
Furthermore, we will show below that the size of the charged excitations is comparable to the length scale of the disorder, which could complicate the relation between the our clean charge gap and the gap observed from activated transport.

To localize the anyons, we include a  weak Gaussian pinning potential of width $4 \ell_B$.
The charge density of the $\frac{e}{5}$ and $\frac{e}{5} \tau$ anyons is shown in Fig.~\ref{fig:anyons}.
Their charge distributions are qualitatively distinct, so in principle they could be distinguished by their quadrupole moment $Q_{zz}$, where $\hat{z}$ points normal to the plane.
Because of the concentrated charge distribution of the Fibonacci anyon, it receives a significantly lower potential energy from the pin than the Abelian charge.
This suggests an array of Fibonacci anyons can be pinned by trapping $\pm\frac{e}{5}$-charged particles with a weak one-body potential and slowly cooling the system.
Such a setup may be employed in a measurement-only quantum computing scheme.\cite{Bonderson:MeasurementOnlyQC:08}

\section{Evidence for spontaneous spin-polarization}
\label{sec:spinpol}
If the system does not spin-polarize at either $\nu = \frac{12}{5}$ or $\tfrac{13}{5}$, the two plateaus are not related by any symmetry, since the spinful particle-hole symmetry exchanges $2 + \tilde{\nu} \leftrightarrow 4 - \tilde{\nu}$.
There have been few experimental studies on the spin-polarization of the $\tfrac{12}{5}$ plateau.~\cite{PhysRevB.85.241302}
The experiments of Ref.~\onlinecite{PhysRevB.85.241302} found that applying an in-plane $B$-field, which increases the Zeeman splitting, drove the $\tfrac{12}{5}$ plateau through a transition; at a critical in-plane $B$-field the gap measured from activated transport closed, then reopened at larger field.
One explanation, believed to explain similar physics at $\nu = \frac23, \frac25$,\cite{Eisenstein90, Hirayama2004} is that the Coulomb point is spin-unpolarized until a critical effective Zeeman field spin-polarizes the phase.
Ref.~\onlinecite{PhysRevB.85.241302} pointed out that the spin transition observed at $\nu=\frac{12}{5}$ resembles to some extent the transition at $\nu=\frac25$, which likely occurs between the 332-Halperin state~\cite{Halperin83} and the Abelian hierarchy (composite fermion) state. 
However, the transport is anisotropic after the transition, and tilt-field experiments are complicated to interpret because the in-plane $B$ field  combines with the finite well width to change the interaction in  important ways, for example by mixing in higher sub-bands with a $N=0$ LL character and reducing the effective well-width.\cite{PhysRevB.87.245315}

Rather than attempting to model the experiment in Ref.~\onlinecite{PhysRevB.85.241302}, we determine whether the ground state of the Coulomb Hamiltonian \eqref{eq:ham} spontaneously spin polarizes  when the Zeeman splitting vanishes.
Indeed, the typical Zeeman splitting is small in comparison to the Coulomb energy, $g \mu_B B / \EC \approx \frac{1}{70}$, hence the spin-singlet vs.\ spin-polarized character of a plateau is largely determined by interactions.

We check the spontaneous spin-polarization of the Coulomb state by keeping the full Hilbert space of both spin species with valence density $\tilde \nu = \frac{1}{5} + \frac{1}{5}$ and enforcing number conservation of each spin separately (we ignore fully filled $\nu=2$ and LL-mixing).
At $w = 3\ell_B$ and $L = 21\ell_B$ we observe long-range ordering of the fermion spin in the XY plane, signaling spontaneous breaking of $\mathrm{SO(3)}$ (see Appendix~\ref{app:spinpol}).
Despite the larger Hilbert space, the energy obtained agrees with the spin-polarized filling to excellent precision.
For small system sizes that can be studied by exact diagonalization we also find that the ground state as well as the low-lying energy spectrum are fully spin polarized.
In contrast, an identical numerical experiment at $\nu = \frac{2}{5}$ is clearly spin-unpolarized, with an energy difference of $0.002$ per flux below to the polarized phase, in agreement with experiment.

It would be strongly desirable to revisit the experimental issue with a probe  other than tilt-field, such as Knight shift, as proved effective at $\nu = \frac{5}{2}$.\cite{KnightShift5halves,Tiemann5halvesSpin}

\section{The RIQH plateaus and the \texorpdfstring{$\frac{12}{5}$, $\frac{13}{5}$}{12/5, 13/5} asymmetry}
\label{sec:cdo}
\begin{figure*}[t]
	\begin{minipage}[t][53mm][t]{72mm}
		\includegraphics[width=72mm]{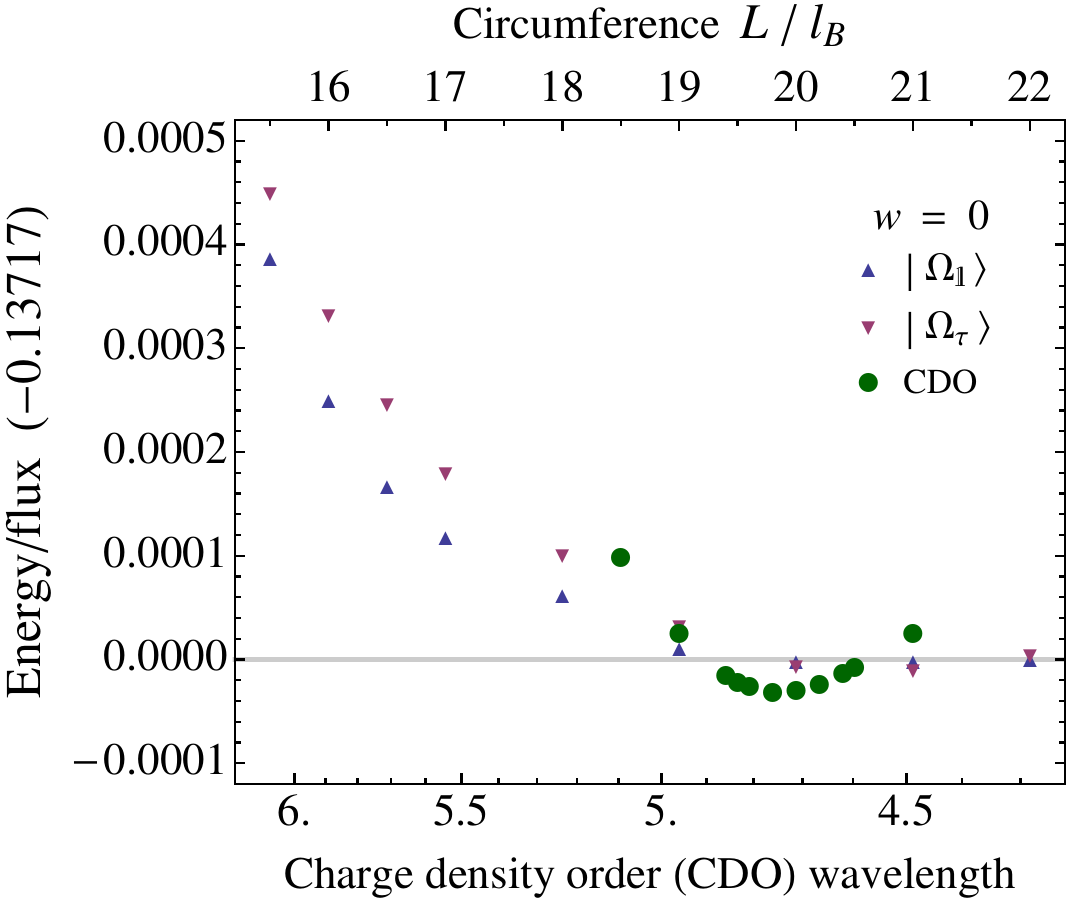}
		\\[-3ex]\hspace{-\textwidth}\begin{minipage}{0mm}\vspace{-110mm}\subfigure[]{\label{fig:cdo_lambda}}\hspace{-27mm}\end{minipage}
		\\[-51mm]\hspace{7mm}\fbox{\includegraphics[width=30mm]{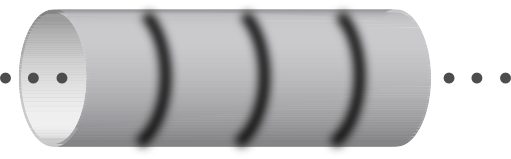}}
	\end{minipage}
	\;\;
	\begin{minipage}[t][53mm][t]{56mm}
		\raisebox{-1.7mm}{\includegraphics[width=56mm]{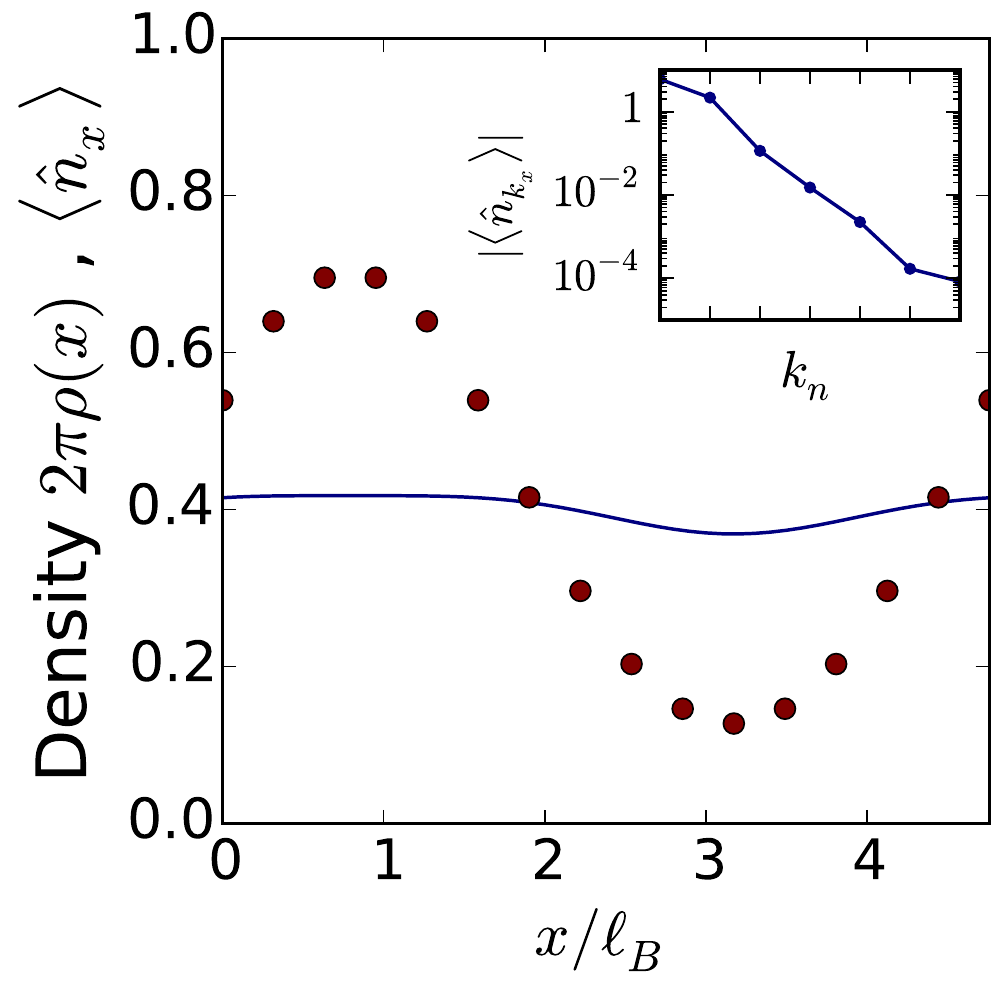}}
		\\[-3ex]\hspace{-\textwidth}\begin{minipage}{0mm}\vspace{-110mm}\subfigure[]{\label{fig:cdo_density}}\hspace{-27mm}\end{minipage}
	\end{minipage}
	\;\;
	\begin{minipage}[t][53mm][t]{42mm}
		\includegraphics[width=42mm]{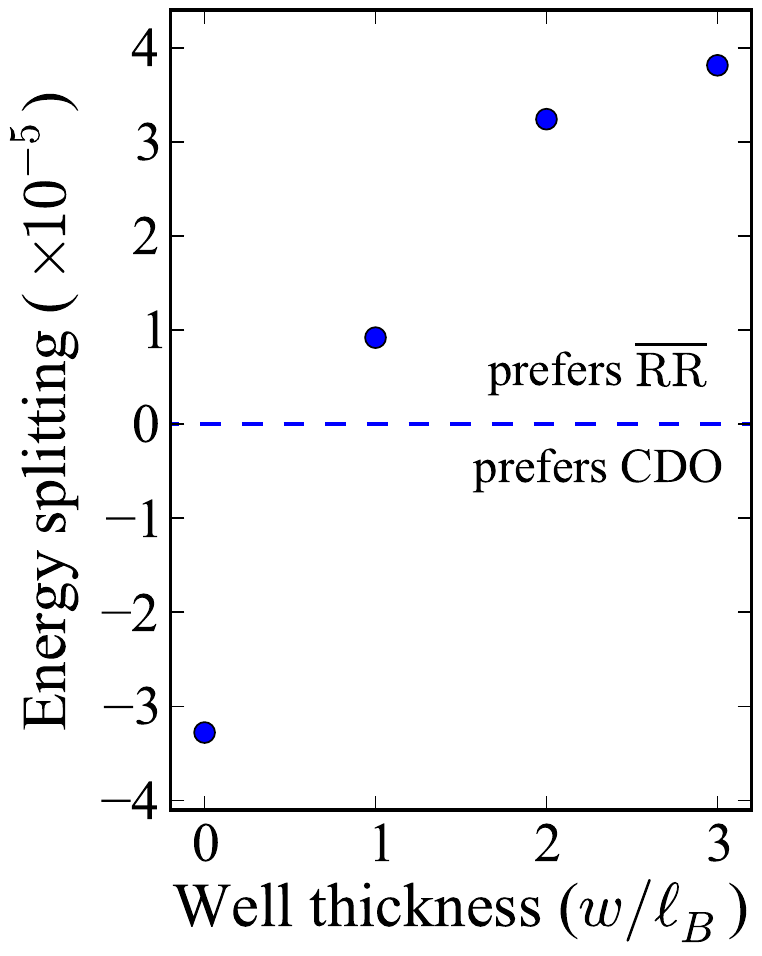}
		\\[-3ex]\hspace{-\textwidth}\begin{minipage}{0mm}\vspace{-110mm}\subfigure[]{\label{fig:cdo_w}}\hspace{-27mm}\end{minipage}
	\end{minipage}
	\caption{%
		(a) Energies for the \aRR\ states (red and blue triangles) as a function of $L$ and CDO (green circles) as a function of wavelength $\lambda$, at with zero well-width.
		For data points denoted by triangles, the energies are computed while enforcing translational invariance of the wavefunctions.
		The two ground states $\ket{\Omega_{\dsone/\tau}}$ are metastable, and their energy approaches $E_\textrm{RR} \approx -0.13717$ in the thermodynamic limit.
		For data points denoted by circles, a charge density wave of wavelength $\lambda$ is imposed along the length of the cylinder (illustrated in the inset).
		Here the CDO minimum $E_\textrm{CDO}$ dips below $E_\textrm{RR}$, which suggests that at $w=0$, the quantum Hall phase is unstable to formation of CDO.
		(b) Orbital (dots) and real-space (line) density profile of the CDO minima at $\lambda\approx4.8\ell_B$.
		Inset shows the Fourier transform of the orbital density, with an exponential decay.
		(c) Energy splitting (per flux) between the \aRR\ phase and the CDO phase $(E_\textrm{CDO}-E_\textrm{RR})$, as a function of $w$.
		The data shows that increased well-width $w$ tends to stabilize the Read-Rezayi phase.
	}
	\label{fig:RR_CDO}
\end{figure*}
In several experiments, the filling $\nu = \tfrac{13}{5}$ lies within a re-entrant integer quantum Hall plateau (called ``R2c'') with Hall conductance $\sigma^{xy} = 3 \frac{e^2}{h}$. \cite{Eisenstein2002, Deng2012_1}  Since Galiean invariance implies $\sigma^{xy} = \frac{e^2}{h} \nu$,  translation invariance must be strongly broken.
It is thought that charge-density order spontaneously develops and is pinned by disorder, rendering inert the fractional filling of electrons/holes in the valence LL.
A similar RIQH plateau (``R2b'') with $\sigma^{xy} = 2 \frac{e^2}{h}$ lies directly proximate to the $\nu = \tfrac{12}{5}$ plateau.
Ref.~\onlinecite{Deng2012_1} found that the partial filling at the center of the RIQH plateaus, $\nu_{2b}, \nu_{2c}$, very nearly satisfies  particle-hole symmetry: $1 - \nu_{2c} = \nu_{2 b} - 0.006$.
However, the width of the R2b plateau is a bit thinner than R2c, so $\nu = \tfrac{12}{5}$ filling just escapes the RIQH region and develops as a separate plateau.\cite{Kumar:12fifths}
Thus even small particle-hole breaking effects could account for the $\tfrac{12}{5} / \tfrac{13}{5}$ asymmetry by shifting the RIQH--\RR\ phase boundary.

The nature of the charge density order in the RIQH R2b and R2c phases is not known.\cite{HaldaneRezayiYang2000, Eisenstein2002, Deng2012_1, Deng2012_2, Rossokhaty2014}
A Hartree-Fock (HF) analysis predicts that among mean-field states either a two-particle ``bubble'' (a Wigner crystal of two-electron droplets) or a stripe (smectic) phase may be competitive at $2/5$ and $3/5$ partial filling.\cite{Fogler1996, Moessner1996, Goerbig2003}
For a pure $N=1$ Coulomb interaction the stripe is predicted to occur at a wavelength $\lambda_{\textrm{HF}} \approx 4.91 \ell_B$. \cite{Goerbig2003}
It is expected that effects beyond HF will spontaneously modulate the density along the stripes, effectively forming an anisotropic bubble phase without the symmetries of the triangular-lattice.\cite{MacDonald2000, Barci2002}

\subsection{Charge density order}

	To explain the asymmetry between $\nu = \tfrac{12}{5}$ and $\tfrac{13}{5}$, it is necessary to find the competing charge density order of the RIQH.
CDO is subtle to study in finite-size numerics:  the sphere geometry will strongly frustrate CDO, while the torus has to be tuned to the correct aspect ratio (this is further discussed in Appendix~\ref{app:torus_ed}).
On the infinite cylinder we must consider the finite circumference $L$ and the unit cell used in the infinite DMRG. At filling $p/q$ the unit cell must contain $m \cdot q$ flux ($m \in \mathbb{Z}$), corresponding to a period $\lambda_{m} = m \cdot q \frac{2 \pi \ell_B^2}{L}$  along the length of the cylinder.
The $\aRR$ phase is commensurate with $m = 1$ for all $L$.

In the numerics presented so far, CDO was implicitly forbidden because we used the minimal DMRG unit cell $m=1$. 
To remedy this, we run simulations (still neglecting LL-mixing) for a variety of  circumferences $L$ and unit cells $m \cdot q$ chosen to accommodate the bubble and stripe phases, even including pinning potentials to preference various orders in the initial stages of the DMRG.
For Coulomb interaction and finite well width, the only CDO found is a stripe phase with a wavelength $\lambda_{\textrm{CDO}}$ quite close to $\lambda_{\textrm{HF}}$.
Note that even though our geometry is quasi-1D, spontaneously breaking translation is not forbidden by Mermin-Wagner because the magnetic algebra renders the symmetry discrete along the length of the cylinder.

	When using a DMRG unit cell of $5 m$ fluc, the wavelength of the CDO is forced to be $\lambda_{m} = 5 m \frac{2 \pi \ell_B^2}{L}$.
As the circumference $L$ is changed, $\lambda_m$ deviates from the intrinsic wavelength  $\lambda_{\textrm{CDO}}$ of the CDO, and the stripes are forced to compress or stretch.
Consequently we expect the energy density of the CDO depends parabolically on $L$, via $E \sim (\lambda_m - \lambda_{\textrm{CDO}})^2$, as found in Fig.~\ref{fig:cdo_lambda}.
For $w=0$ we verified that this feature is found both for $m=3$, $L \sim 20\ell_B$ and the larger circumference $m = 4$, $L \sim 27\ell_B$.
The optimal $\lambda_{\textrm{CDO}}$ can be determined from the minimum of the parabola, and depends on $w$ at the level of 10\%, with larger $w$ preferring smaller $\lambda_{\textrm{CDO}}$.

As shown in Fig.~\ref{fig:cdo_w}, the CDO has remarkably close energy ($\Delta E/\textrm{flux} \sim 10^{-5}$) to the $\aRR$ phase.
In fact, for $L \sim 20\ell_B$ and $w=0$, the CDO is the true ground state.
Increasing $w$ favors the $\aRR$ state; our finite size numerics points to a transition around $w \sim 1\ell_B$. 
Our numerics on the infinite cylinder shows a sharp first order transition between the \aRR\ and CDO phases.

\begin{figure}[t]
	\includegraphics[width=0.99\columnwidth]{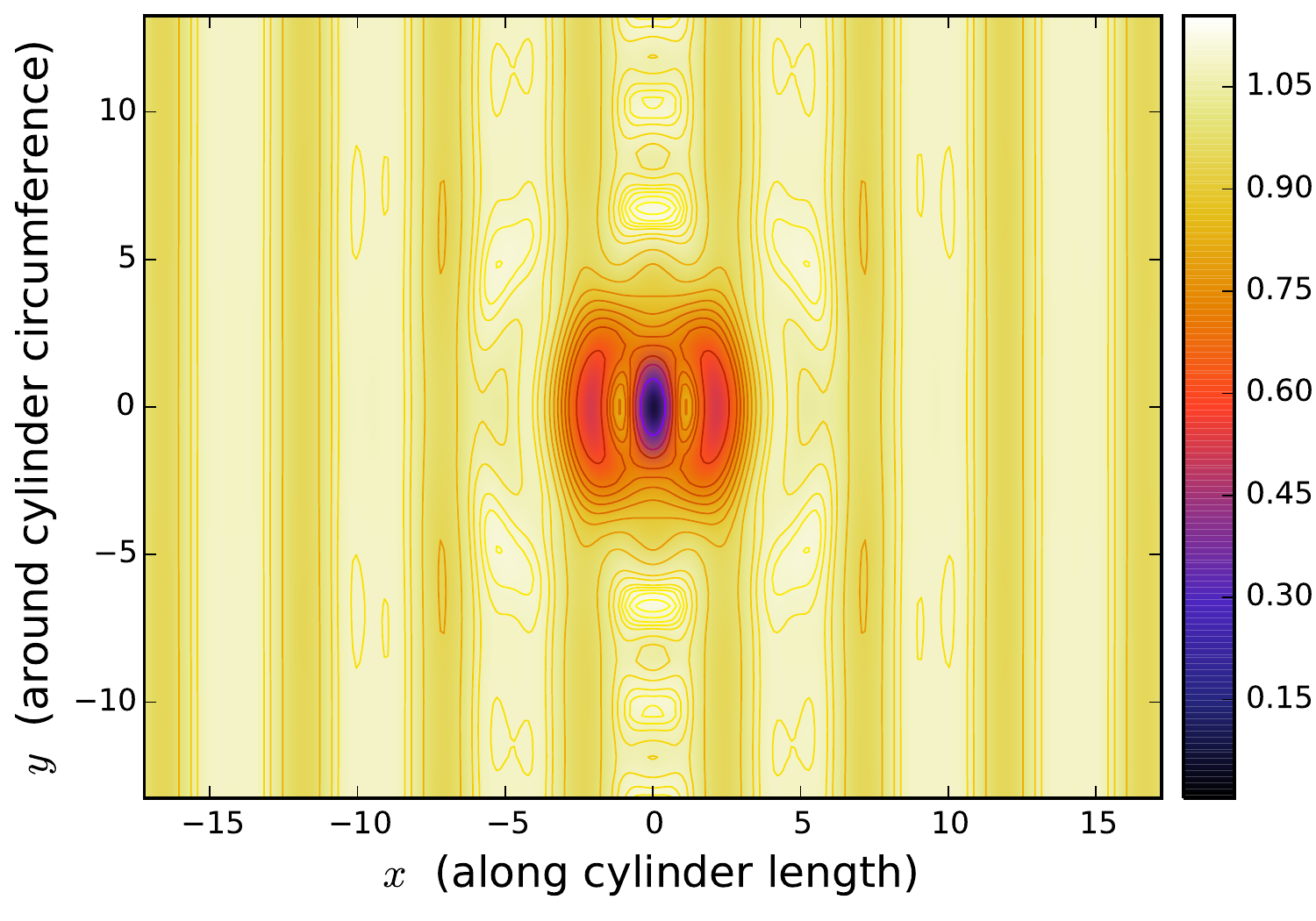}
	\caption{%
		The real space density-density correlation function $S(\mathbf{r}) = \braket{ \rho(\mathbf{r}) \rho(\mathbf{0}) }$ of the CDO phase, at $\nu = \frac{12}{5}$, for $w = 1\ell_B$, and $L = 26.5\ell_B$.
		The data is scaled by the mean density-squared $\big(\frac{\nu}{2\pi\ell_B^2}\big)^2$, and the point $\mathbf{r} = \mathbf{0}$ is taken to lie at the center of the high-density stripe. The horizontal axis is along the length of the cylinder; the vertical axis is around the circumference of the cylinder. At this circumference the unit cell along the cylinder contains $4\cdot5$ flux,  corresponding to eight electrons per stripe. The intra-stripe correlations have seven local maxima plus a correlation hole.
The nearest-stripe correlations have weaker maxima which are out of phase with the intra-stripe maxima, which suggests there is a tendency towards forming a body-centered rectangular lattice of single electrons.}
	\label{fig:cdw_sf}
\end{figure}

While $\lambda_{\textrm{CDO}} \sim 4.75 \ell_B$ is rather close to $\lambda_{\textrm{HF}}$,  the CDO we find is a highly entangled state rather different from the naive HF ansatz.
In the HF ansatz for a stripe, the orbital occupation alternates between $\nu = 1$ and $\nu =0$, effectively forming decoupled stripes of IQH. 
Because the IQH has chiral edge modes, the correlations are algebraic along the length of the stripe.
As shown in Fig.~\ref{fig:cdo_density}, we find the CDO orbital occupation instead forms a nearly perfect sine-wave, with higher Fourier components falling off exponentially in $k$. 
This implies that the Greens function decays exponentially along the length of the stripes, suggesting interaction effects have gapped out the chiral edge modes. 
This is consistent with theoretical predictions that the HF smectic order is unstable towards bubble (e.g.\ Wigner crystal) formation. \cite{MacDonald2000, Barci2002}
We can observe the onset of a Wigner-crystal phase in the density-density correlations $S(\mathbf{r}) = \braket{ \rho(\mathbf{r}) \rho(\mathbf{0}) }$ shown in Fig.~\ref{fig:cdw_sf}.
Within the same stripe, the correlations are consistent with a crystal with one-electron per unit cell. Between neighboring stripes the correlations are much weaker, but suggest a body-centered rectangular structure.
Thus while we cannot determine if true long-range order develops between the stripes in the thermodynamic limit,  the correlations are consistent with a body-centered rectangular lattice of electrons.

\subsection{Effect of Landau level mixing on the competition between the Read-Rezayi and re-entrant integer quantum Hall effect}
\begin{figure}[t]
	\includegraphics[width=0.9\columnwidth]{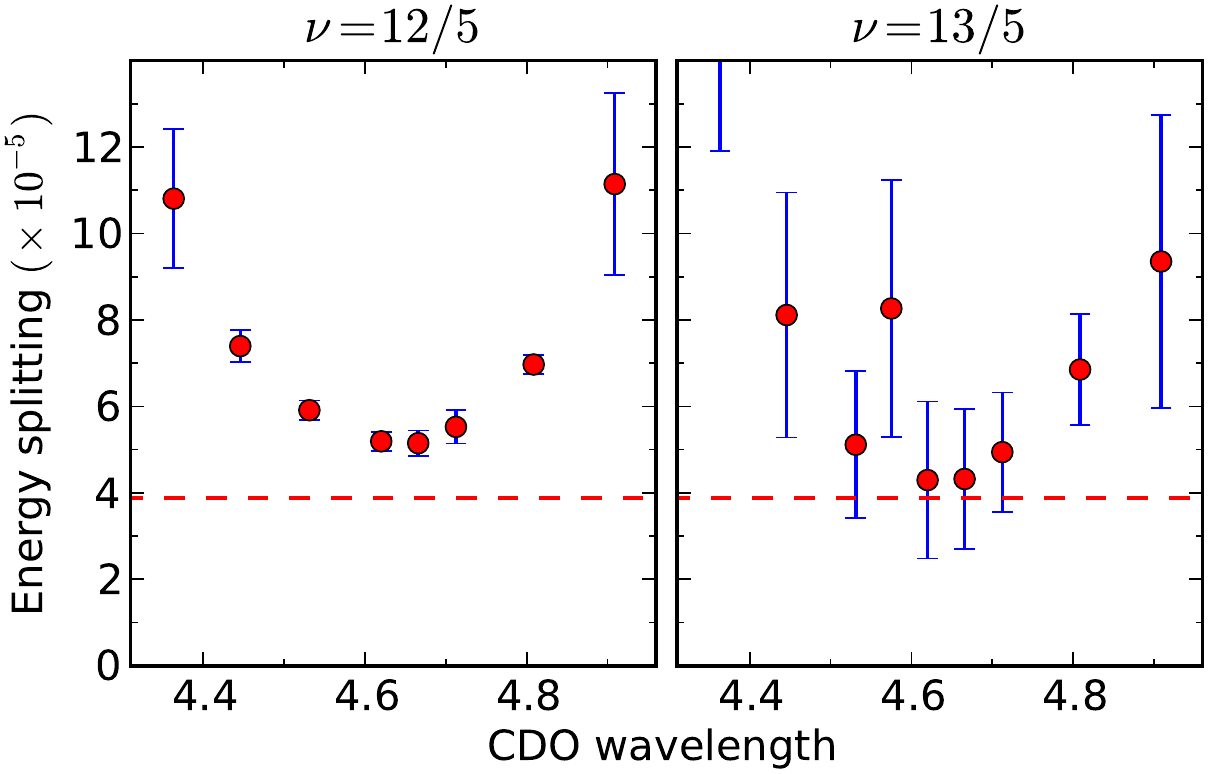}
	\caption{%
		The effect of Landau level mixing on energy splitting (per flux) between the CDO and the Read-Rezayi states.
		The data is given for $w=3\ell_B$ and $\kappa = 0.4$, where $\kappa$ is the ratio of Coulomb to cyclotron energy.
		For reference, the dashed line denotes the energy splitting in the absence of LL-mixing.
		For $\nu=\frac{12}{5}$, the splitting increases with the inclusion of LL-mixing, further favoring the \aRR\ phase.
		For $\nu=\frac{13}{5}$ the data is more ambiguous due to our numerical uncertainty in the energy of the \RR\ phase, which becomes significantly less stable with the addition of LL-mixing.
		We conclude that LL-mixing stabilizes the \aRR\ phase at $\nu = \frac{12}{5}$.
	}
	\label{fig:mixing}
\end{figure}
Since the ground states at $\nu = \frac{12}{5}$ and $\frac{13}{5}$ are fully spin-polarized, in the absence of LL-mixing the two are related by particle-hole symmetry.
To account for the discrepancy seen in experiment we now include LL-mixing, whose strength is parameterized by the ratio of Coulomb to cyclotron energies, $\kappa = \frac{e^2}{4\pi\epsilon} / \hbar\omega_c $.
We include the effect of LL-mixing by performing DMRG in the full space of spin-polarized $N = 0, 1, 2$ LLs.\cite{ZaletelMixing}
Adding LL-mixing is computationally expensive; each data point takes several weeks.

We first outline the logic of our approach.
Since it appears that there is a first-order transition between the \RR\ states and CDO, the relevant task is \emph{not} to compare the properties of \aRR\ vs.\ \RR\ as LL-mixing is introduced, but rather to compare the energy of the \RR\ state to the competing CDO at the \emph{same} filling.
In particular, near a first order transition the quasiparticle gap and correlation functions needn't be related to the \emph{extensive} energy splitting between the competing phases which actually drives the transition.
Thus, we must compute the ground state energy density of the \RR\ states and CDO states, at both filling $\tfrac{12}{5}$ and $\tfrac{13}{5}$, as LL-mixing is added.

As LL-mixing is increased the CDO wavelength $\lambda_{\textrm{CDO}}$ changes slightly, so it is important to repeat the analysis of varying the circumference $L$.
In Fig.~\ref{fig:mixing} we compute the energy difference between the CDO and \RR\ states at $\kappa = 0.4$, $w = 3\ell_B$ for various circumferences.
We find an asymmetry between the two fractions.
At $\nu = \frac{12}{5}$, LL-mixing increases the energy of the CDO state relative to the \aRR\ state, thus stabilizing the \aRR\ quantum Hall plateau.
For our data at $\nu = \frac{13}{5}$, however, the $\RR$-numerics become much more unstable in the presence of LL-mixing and it is difficult to converge our numerics with the accuracy we require (with LL-mixing we have data up to $\chi = 14000$).
Our uncertainty in the DMRG ground state energy is indicated by error bars, which we determined as follows.
DMRG produces a variational energy $E(\chi)$ which depends on the DMRG bond-dimension $\chi$ as $E(\chi) = E_0 + \varepsilon \chi^{-b}$, where $E_0$ is the true ground state energy.\cite{Arad:AreaLaw}
A fit to this form is done with and without the largest available bond-dimension, and we consider the change in the fitting parameter $E_0$ to be the uncertainty.
While arguably a flawed measure, by this definition the change in the CDO--\RR\ energy splitting is inconclusive at $\frac{13}{5}$.
However, our best data point is consistent with a scenario which the CDO--\RR\ splitting changes significantly less than the CDO--\aRR\ splitting at $\frac{12}{5}$, if at all.

The data in Fig.~\ref{fig:mixing} is taken at LL-mixing $\kappa = 0.4$, smaller than that of the experiments in Ref.~\onlinecite{Kumar:12fifths} ($\kappa \approx 1$). At least perturbatively the change in splitting should scale linearly with $\kappa$, implying the true energy difference is even greater than  reported here, though we suspect $\kappa = 1$ is beyond the perturbative regime.
We only present data at $\kappa = 0.4$ because we are unable to converge the $\nu=\frac{13}{5}$ \RR\ state at larger $\kappa$, apparently because much larger bond dimension becomes required (the \aRR\ state at $\nu=\frac{12}{5}$  does not have as severe a dependence).
Note that while the computational difficulty of \RR\ increases relative to \aRR\, there is not much relative difference in the \emph{observable} properties like the correlation length, and no indication of a diverging correlation length.

\section{Summary and Discussion}

We numerically simulated the fractional quantum Hall effect at a filling factor $\nu = \frac{12}{5}, \frac{13}{5}$ using the infinite density matrix renormalization group and exact diagonalization methods.
Our simulations include realistic Coulomb interaction appropriate for $\mathrm{GaAs}$ quantum wells of finite width.
In the absence of Landau level mixing, the topological properties of the $\nu=\frac{12}{5}$ ground state are consistent with the $k=3$ Read-Rezayi phase.
The ground state remains spin-polarized for a range of well-widths, even as the Zeeman splitting vanishes.  
The lowest energy (spin-polarized) charged excitation was identified with a non-Abelian Fibonacci anyon, which supports universal braid statistics.

	Our findings reveal several  properties which could help make the 12/5-plateau suitable for a bulk implementation of universal measurement-only topological quantum computing.\cite{Bonderson:MeasurementOnlyQC:08} First, our numerically calculated gap of $\unit[2]{K}$ suggests that the 12/5 Fibonacci state is not intrinsically much more delicate than the  5/2 state. The smaller $\unit[80]{mK}$ observed in experiment may be related to slowly varying disorder potentials, and hence there is some hope this gap could be engineered.
Second, of the two types of charge $\pm e/5$ quasiparticles, the Fibonacci $\pm e/5 \tau$  particle has the lower energy. Furthermore, the $\pm e/5 \tau$ particles and $\pm e/5$ particles have a qualitatively different charge profile; the Fibonaccis are much more tightly localized at their center (see Fig.~\ref{fig:anyons}).
This implies that an electrostatic pinning potential will stably trap a Fibonacci anyon; in contrast, if it was the Abelian $\pm e/5$ which had a much lower energy when pinned, a trapped $\pm e/5 \tau$ would spontaneously  eject the neutral $\tau$, which would fly off in an uncontrolled manner.
This also provides a convenient way to initialize a lattice of Fibonaccis with a known fusion tree: adiabatically turning on two nearby pinning potentials $+V, -V$ would usually generate a $e/5 \tau, -e/5 \tau$ pair that fuse to the identity, which can then be separated. 
In addition, rather than using an edge-interferometer to detect fusion outcomes, one can try a more naive approach: bring two charged Fibonaccis together and  ``look'' at them. Since we find the charge distribution depends qualitatively on the fusion outcome, measuring, for instance, the local quadrupole moment or the magnetization density detects the fusion outcome.
Thus, given sufficient control over local pinning potentials, we have the basic required ingredients: 1) charged Fibonaccis can be stably trapped by electrostatic pinning potentials 2) an array of Fibonaccis can be pairwise created with a known fusion outcome 3) the charged Fibonacci and Abelian  particles have qualitatively different charge profiles, so fusion outcomes can be detected by any probe sensitive to the local charge or magnetization distribution.

The full spin polarization of the ground state and the large charge gap we obtain ($0.017 \approx \unit[2]{K}$) are encouraging but not in complete agreement with experiments.\cite{Kumar:12fifths, PhysRevB.85.241302}
The estimated gap in Ref.~\onlinecite{Kumar:12fifths} is $\approx \unit[80]{mK}$, and Ref.~\onlinecite{PhysRevB.85.241302} detected a spin transition upon tilting the field.
While it may be possible that the lowest charge excitation is actually a skyrmion and/or strongly renormalized by the LL-mixing, a better theoretical understanding of activated transport and spin polarization in the exotic $N=1$ LL plateaus is also desired. 
As discussed in Ref.~\onlinecite{Nuebler2010} in the context of $\nu = 5/2$ plateau, the size  of the quasiparticles ($d \sim 15 \ell_B$) is comparable to the length scale of disorder arising from the remote ionized donors. 
In this regime there is a larger tunneling amplitude across saddle points of the disorder potential; it would be interesting if these amplitudes could be numerically estimated using the single-anyon DMRG.

Perhaps most intriguingly, we find an exceptionally close competition between CDO and the $\aRR$ phase, the former being the likely origin of the RIQH phase experimentally observed at $\nu = \frac{13}{5}$.
Our numerics show that with increased width $w$ the \aRR\ phase is preferred over the CDO; it is advantageous to fabricate the quantum well as wide as possible to stabilize the $\aRR$ phase.
At the same time, with increased $w$ the 1\textsuperscript{st} excited subband LL also comes down in energy, crossing with the $N=1$ LL at $w \sim 3.8\ell_B$, which puts an upper limit on $w$.
Finally, Landau level mixing increases the energy of the CDO state relative to the $\aRR$ at $\nu = \frac{12}{5}$, which would explain the observed asymmetry between the $\nu = \frac{12}{5}$ and $\frac{13}{5}$ plateaus, giving further confidence to the numerics.

\emph{Note:} During the final preparation of this work, we learned of overlapping results by W.\ Zhu et al.,\cite{WZhuDSheng:ReadRezayi} and a study of Landau level mixing on the sphere by Pakrouski et al.\cite{Pakrouski12fifths}

\acknowledgments
We would like to thank X.\ L.\ Qi, C.\ Nayak, K. Pakrouski, and P.\ Bonderson for helpful conversations, and Ed Rezayi for advice and collaboration on past studies.
We are grateful to Dave Wecker for his time, debugging skills, and the  Microsoft Research QuArC cluster.
RM acknowledges the Walter Burke Institute for Theoretical Physics and Institute for Quantum Information and Matter at California Institute of Technology, as well as funding from the Sherman Fairchild Foundation.
MZ and RM acknowledge support from the visitors program of MPI-PKS Dresden.
ZP acknowledges support by DOE grant DE-SC0002140.

\clearpage
\appendix

\section{Further characterization of the Read-Rezayi phase via entanglement}
\label{app:RR_entanglement}

We discuss in details the various entanglement measures to identify the \aRR\ phase at filling $\nu=\frac{12}{5}$.
The data presented here are are computed assuming full spin-polarization and no Landau level mixing, i.e., a single $N = 1$ Landau level at $\frac{2}{5}$ filling.
(The data at $w=0$ are taken while enforcing a uniform density to stabilize the \aRR\ phase.  It appears rather ``noisy'', possibly due to a CDO instability.)

We first return to the momentum polarization for the ground states $\ket{\Omega_\dsone}$ and $\ket{\Omega_\tau}$.
As alluded to in the Sec.~\ref{sec:vanillaRR}, the momentum polarization reveals the shift, chiral central charge $c$, and topological spin $h_a$.
We numerically calculate the Berry phase $U_{T;a}$ of performing a $2\pi$ twist on the left half of the cylinder for ground state $\ket{\Omega_a}$: comparing to the theoretical prediction\cite{ZaletelQHdmrg13}
\begin{align}
	U_{T;a}
		&= \exp\left[ 2 \pi i \left( h_a - \frac{c}{24} - \frac{\HallVis}{2\pi\hbar} L^2 \right) + \dots \right].
	\label{eq:U_exact}
\end{align}
The ellipsis denotes term exponentially suppressed with circumference $L$, and $\HallVis$ is the ``Hall viscosity'',\cite{AvronSeilerZograf:HallViscosity:1995} related to the shift via $\HallVis = \frac{\hbar}{4}\frac{\nu}{2\pi\ell_B^2}\shift$.
(For the \aRR\ phase, the shift is computed at $\nu = \frac{2}{5}$ in the 1\textsuperscript{st} LL.)
The formula given in the main text is the ``logarithm'' of Eq.~\eqref{eq:U_exact}, and in principle ambiguous modulo 1.
In practice, because $U_{T;a}$ is a continuous function of the circumference, one can easily extract the $\shift$ by taking the slope and compute the combination $h-\frac{c}{24}$ modulo 1.

\begin{figure}[t]
	\begin{minipage}{82mm}
		\includegraphics[width=76mm]{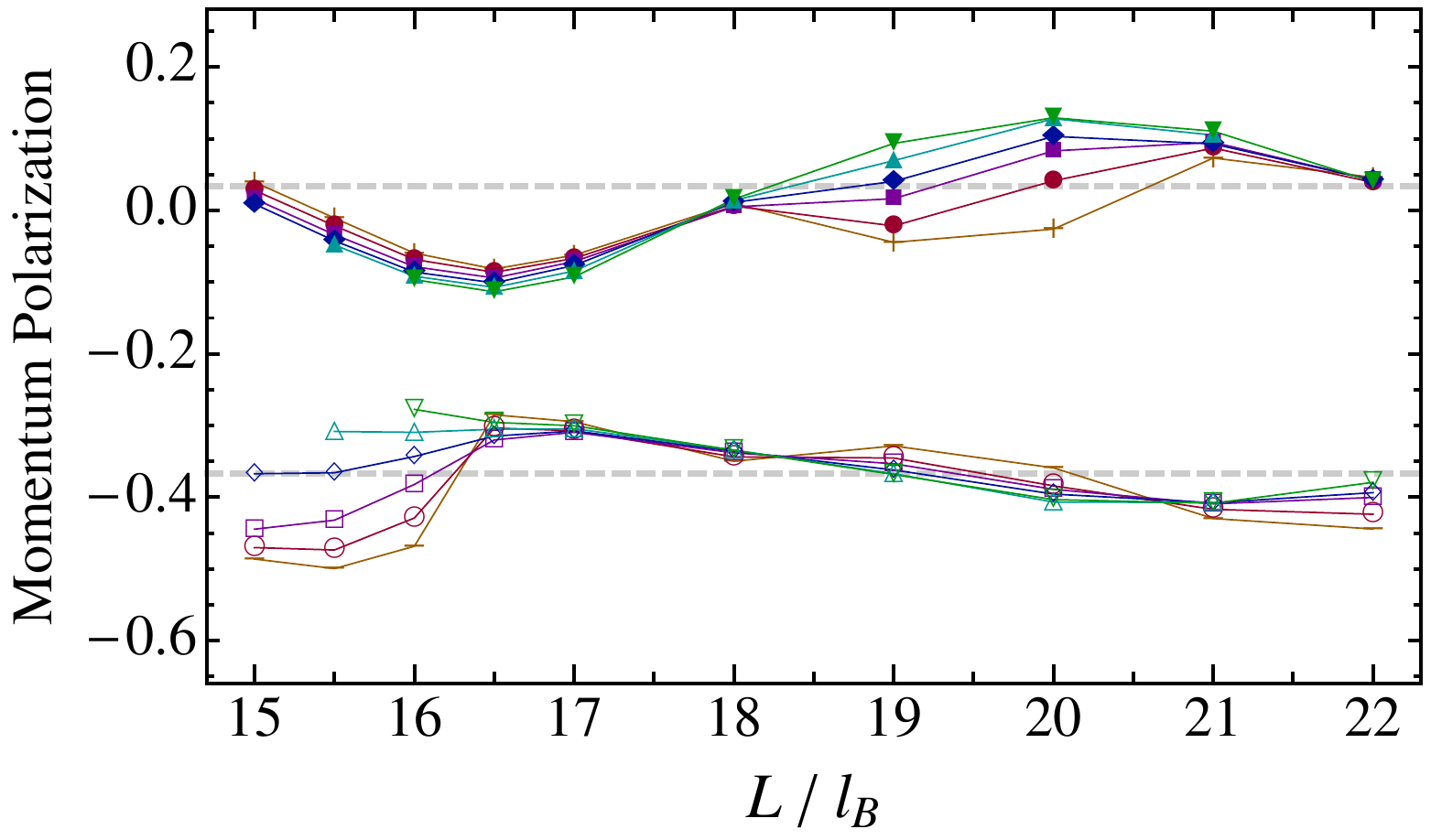}
		\\[-1ex]\hspace{-\textwidth}\begin{minipage}{0mm}\vspace{-90mm}\subfigure[]{\label{fig:RR_mompol}}\hspace{-27mm}\end{minipage}
	\end{minipage}
	\\
	\begin{minipage}{82mm}
		\hspace{2mm}\begin{minipage}[t]{78mm}
			\includegraphics[width=74mm]{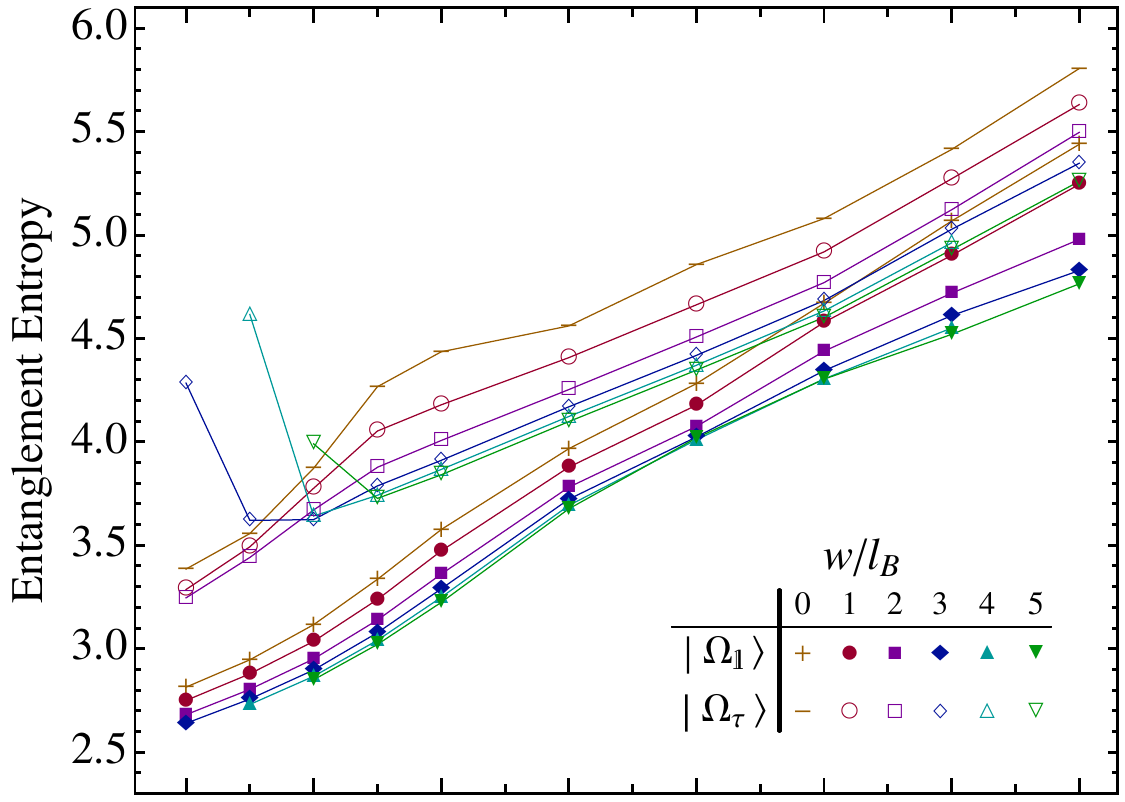}\\
			\includegraphics[width=74mm]{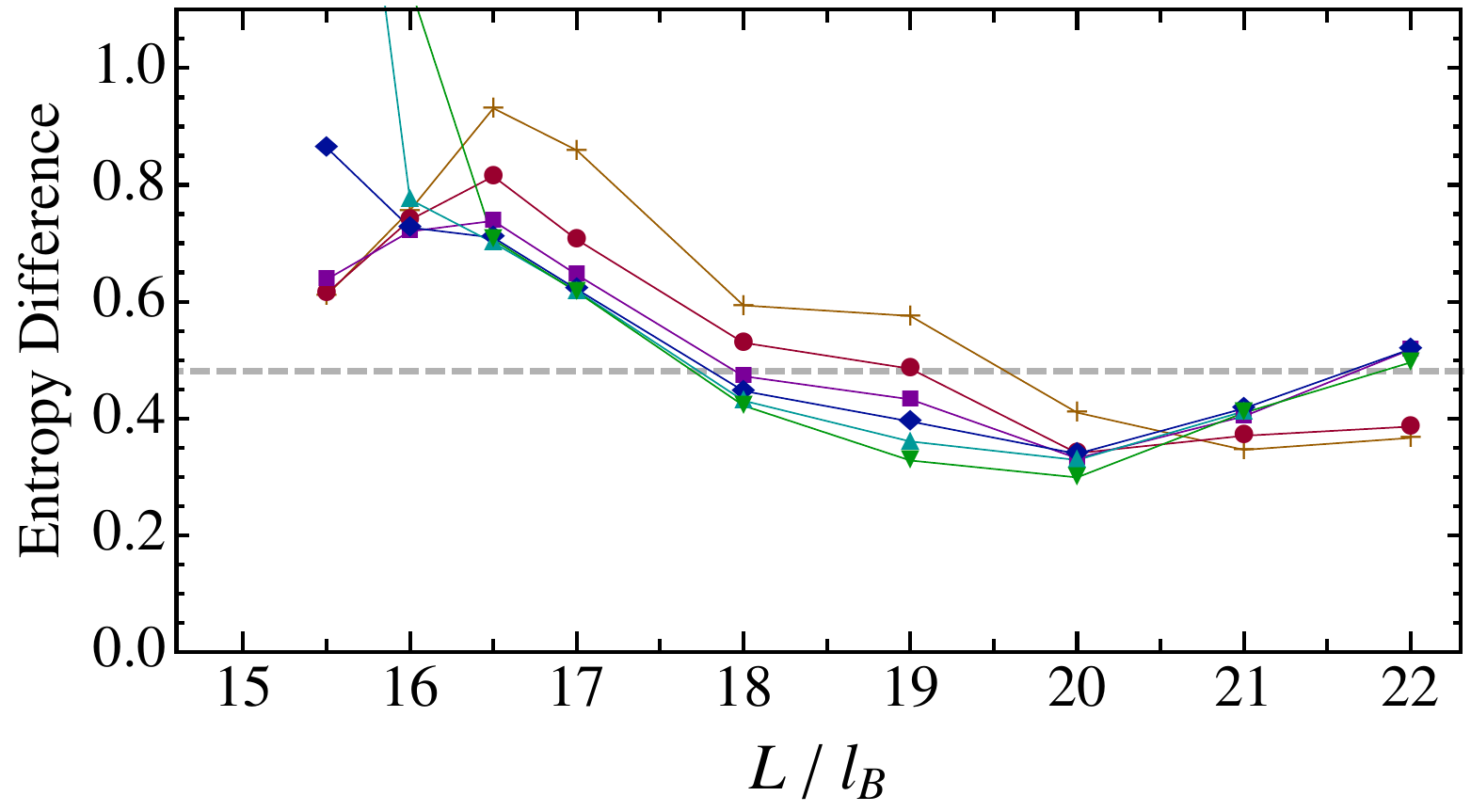}
		\end{minipage}
		\\[-1ex]\hspace{-\textwidth}\begin{minipage}{0mm}\vspace{-181mm}\subfigure[]{\label{fig:RR_S}}\hspace{-27mm}\end{minipage}
	\end{minipage}
	\caption{%
		(a) Momentum polarization for the ground states $\ket{\Omega_{\dsone/\tau}}$.
		The dashed lines are the predicted values of $h_a-\frac{c}{24}$ for the two ground states.
		(b) Entanglement entropies $S_\dsone$ and $S_\tau$, along with their difference $S_\tau-S_\dsone$.
		The dashed line at $\log(\varphi) \approx 0.48$ is predicted for the \aRR\ phase.
	}
	\label{fig:RR_entanglement}
	\vspace{6mm}
	\includegraphics[width=82mm]{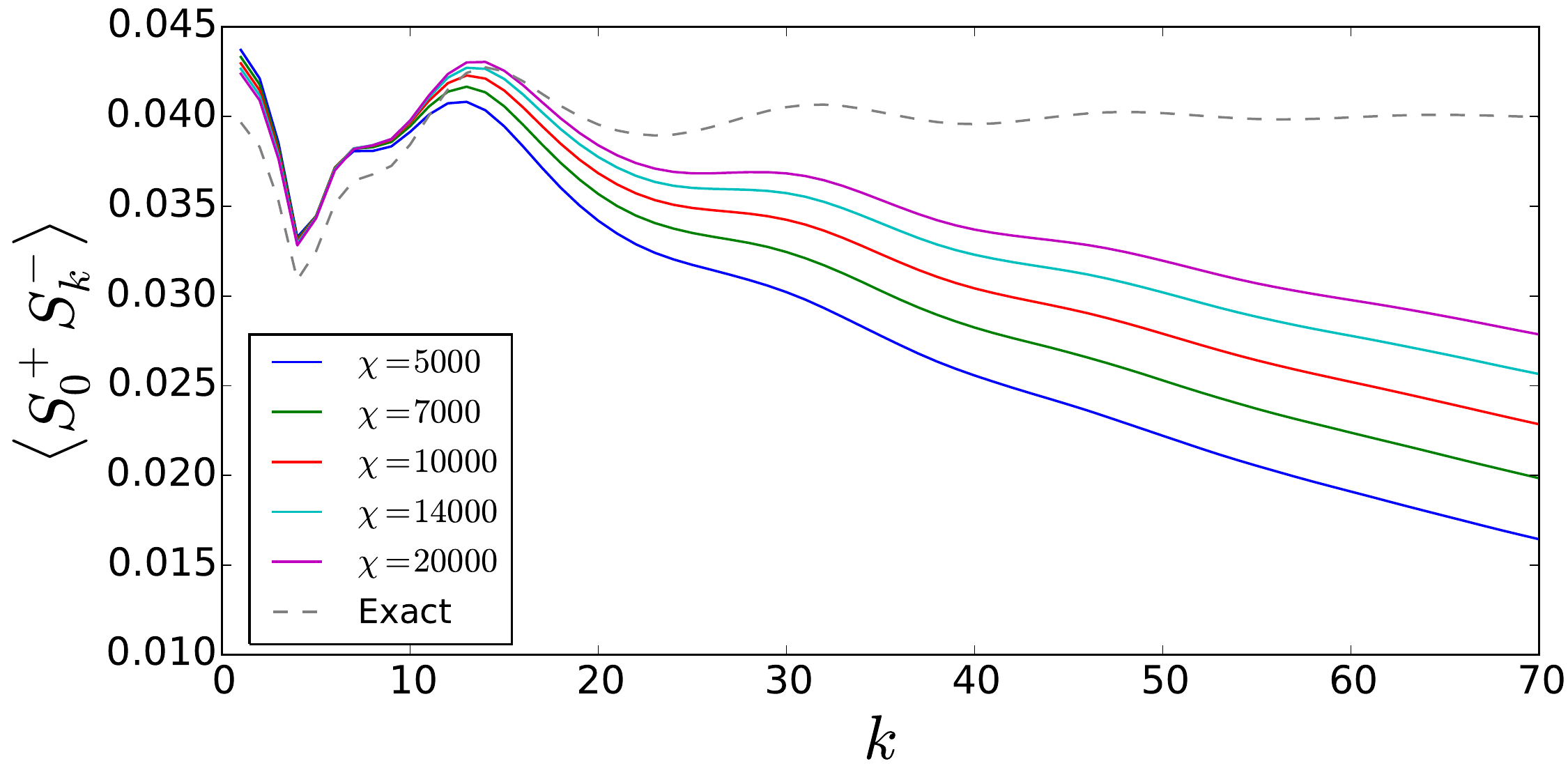}
	\\[-2ex]
	\caption{%
		Spin correlator $\braket{ S^+_0 S^-_k }$ for $L=21\ell_B$, $w=3\ell_B$ at filling $\tilde\nu^{\uparrow/\downarrow} = \frac{1}{5}$.
		Solid curves shows the data at various bond dimensions $\chi$, converging to the expected correlator given as the dashed curve.
		This is a signature of spontaneously broken spin-rotational symmetry, indicative of a fully spin-polarized ground state at $\nu=\frac{12}{5}$.
	}
	\label{fig:spin_correlation}
\end{figure}

Figure~\ref{fig:RR_mompol} shows the momentum polarization for $L = 15\mbox{--}22\ell_B$, using various range of thicknesses $w$.
From the slope of the data, we extract $\shift = 0 \pm 0.1$, in excellent agreement with the \aRR\ phase at zero shift.
Assuming the shift is exactly zero, the data shows the residue $h_a - \frac{c}{24}$.
These values corroborate well with those of the \aRR\ phase with $h_\dsone = 0$, $h_\tau = -\frac{2}{5}$ and chiral central charge $c = 1-\frac{9}{5} = -\frac{4}{5}$, shown as the pair of dashed lines.

Finally, we also study the entanglement entropy of the ground states, given by $S = \operatorname{Tr} [ -\rho_L \log \rho_L ]$.
The entanglement entropy is expected to scale with the circumference as $S_a \approx sL - \gamma_a$, where $\gamma_a$ is a constant called the ``topological entanglement entropy'' (TEE) associated with anyon type $a$.\cite{KitaevPreskill, LevinWen:06}
The TEE is given by $\gamma_a = \log \mathcal{D} - \log d_a$, where $d_{a}$ are the quantum dimensions of the anyon type $a$, and $\mathcal{D}$ is the total quantum dimension of the system given by $\mathcal{D}^2 = \sum_a d_a^2$.
Figure~\ref{fig:RR_entanglement} shows the entanglement entropy as a function of $L$.
Unfortunately the data $S_{\dsone/\tau}$ suffer from very strong finite size and finite entanglement effects, it is not possible to extract $\gamma_{\dsone/\tau}$ with any meaningful degree of certainty (the slope $s$ is sensitive to microscopic details and non-universal).
However, the difference $S_\tau - S_\dsone = \gamma_\dsone - \gamma_\tau$ is universal, predicted to be $\log(\varphi)$ where $\varphi = \frac{1+\sqrt5}{2}$ is the golden ratio.
While the entropy data alone cannot confirm the existence of a Fibonacci anyon, it is nevertheless consistent with the \aRR\ phase.

\clearpage

\section{Stability of the Read-Rezayi phase against short-range perturbations}
\label{sec:v1}

The ground state at $\nu = 2/5$  in the $N = 0$ LL is the spin-polarized Abelian hierarchy (AH) or composite fermion state, and an obvious competitor at $\nu = 12/5$.
Here we examine the stability of the $\aRR$ phase at $\nu=12/5$ as we perturb the $N=1$ LL projected Coulomb interaction by the short-range $V_1$ pseudopotential. 

\begin{figure}[b]
	\includegraphics[width=0.8\columnwidth]{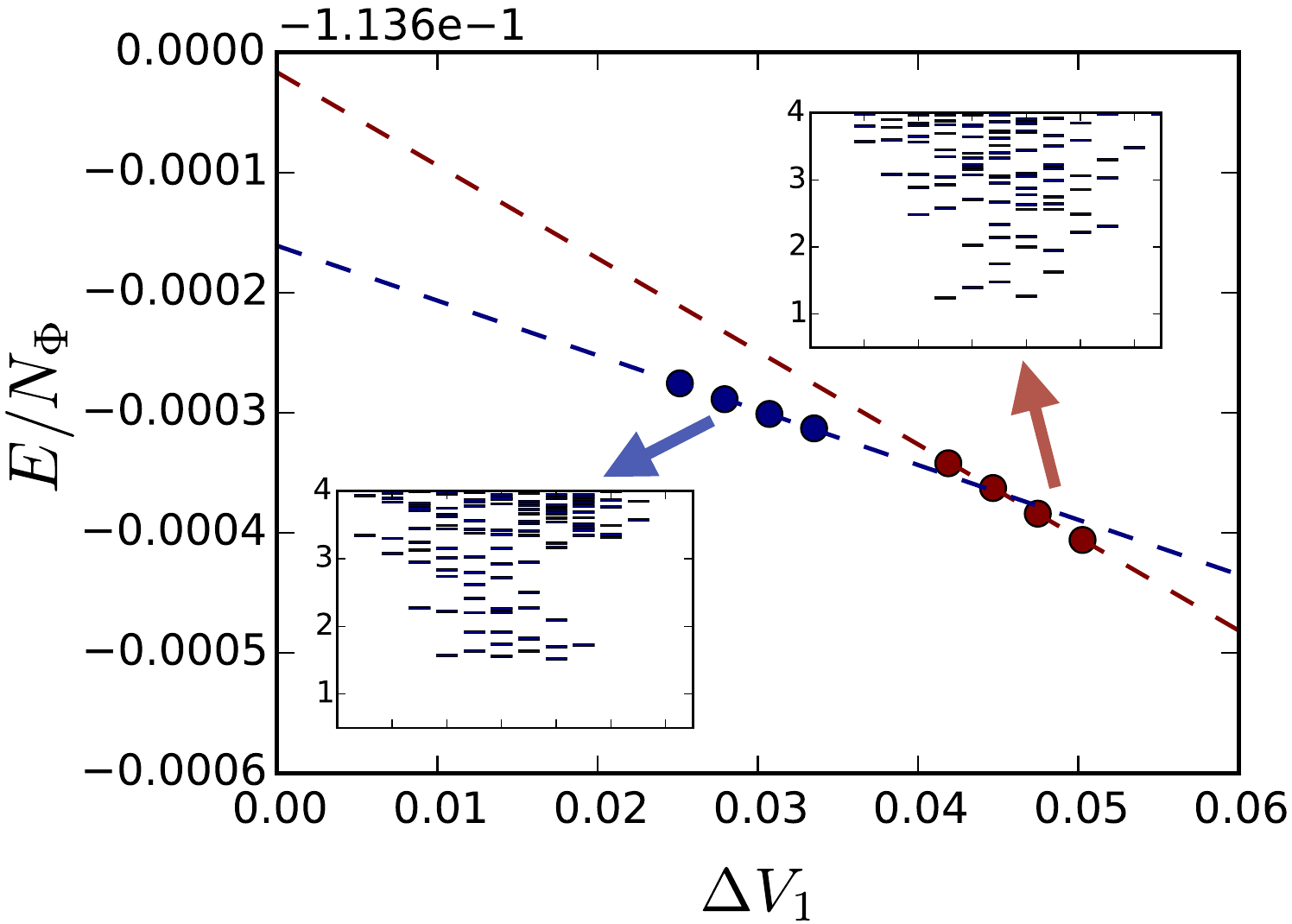}
	\caption{%
		Transition between the $\aRR$ and Hierarchy states as the Haldane pseudopotential $\Delta V_1$ is added to the Coulomb interaction.
		In the cylinder geometry the transition is first order.
		For perspective, in the $N=0$ LL, $V_1 / V_3 = 1.6$; in the $N = 1$ LL, $V_1 / V_3 = 1.3$; at the observed transition $V_1 / V_3 = 1.43$.
		Linearly extrapolating the energy of Hierarchy state to the Coulomb point, we obtain a splitting of $E_\textrm{AH} - E_\textrm{RR} \approx 1.5 \times 10^{-4} / \textrm{flux}$.
		Data is taken from DMRG at $L = 20\ell_B$, $w = 2\ell_B$.
	}
	\label{fig:RRvJain}
\end{figure}

As the node of $N=1$ LL wavefunctions softens the interaction, we expect that adding $\Delta V_1>0$ will drive the $\aRR$ phase back into the Hierarchy phase.
This is indeed what happens, Fig.~\ref{fig:RRvJain}.
In terms of the ratio $V_1 / V_3$, the transition between the RR and hierarchy phase is about 1/3 of the way between the Coulomb $N = 1$ and $N = 0$ LLs.
$\Delta V_1$ is a `best-case' perturbation for the Hierarchy state, as a typical real-space potential will be distributed across all $V_m$.
In fact, we have verified that projecting a $V(r) = \nabla^2 \delta(r)$ interaction into the $N=1$ LL favors a CDO phase, not the Hierarchy state.
Linearly extrapolating the energy of the Hierarchy state to the Coulomb point, we obtain a splitting between the $\aRR$ and Hierarchy state of about $\Delta E \approx 1.5 \times 10^{-4}$ per flux at well-width $w = 2\ell_B$. 

\begin{figure}[t]
	\includegraphics[width=\linewidth]{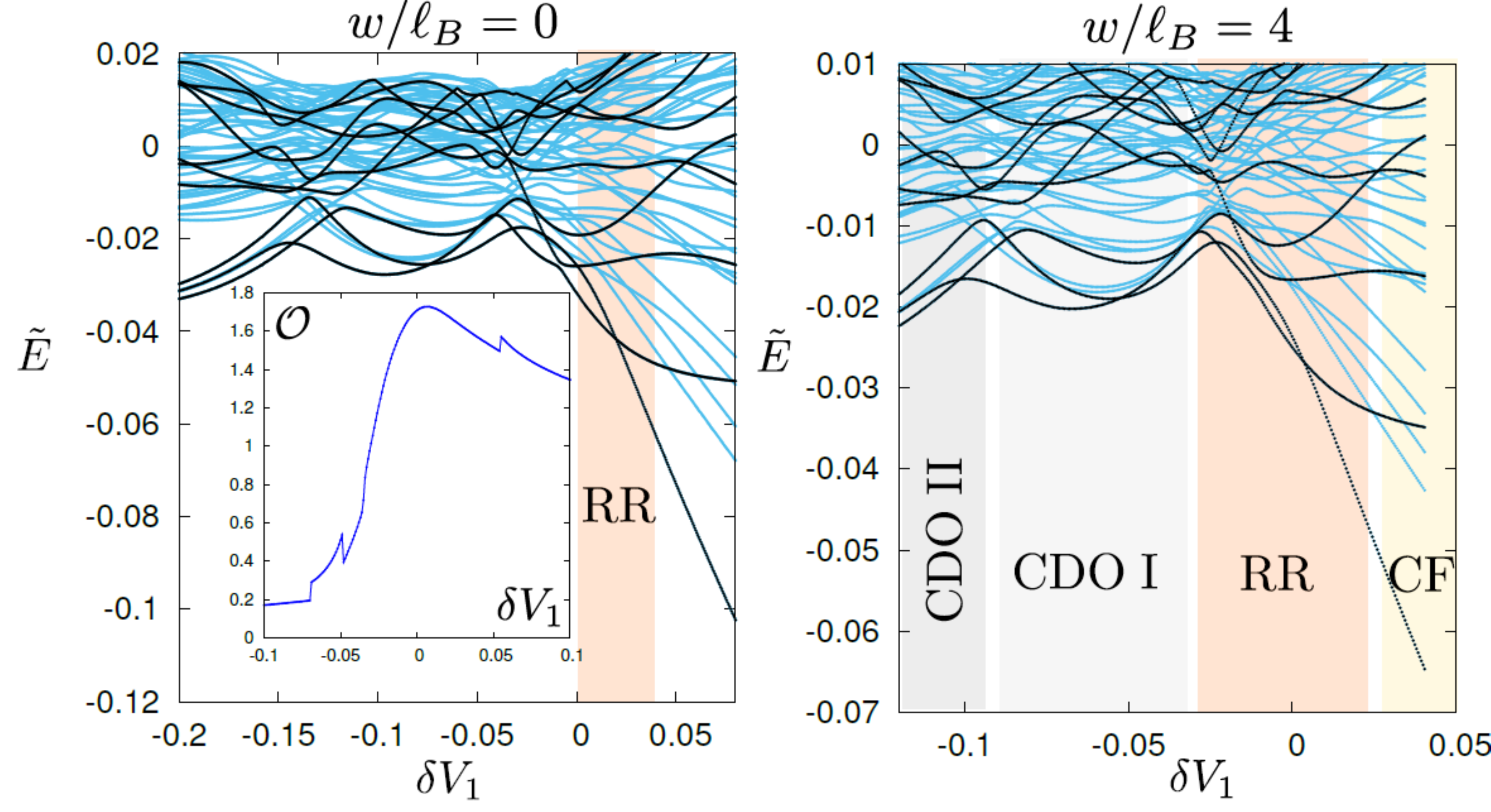}
	\caption{%
		Phase diagram as a function of modified $V_1$ pseudopotential, for zero width (left) and $w/\ell_B=4$ (right).
		Data is obtained by exact diagonalization of the system with 25 flux quanta through the hexagonal unit cell.
		We compute the lowest 10 energies per momentum sector and plot them relative to the average energy of the system $\tilde E = E - E_\textrm{avg}$.
		Black symbols denote levels in $\mathbf{K}=0$ sector.
		For zero width (left), we identify the $\Z3$ ground state degeneracy in a narrow shaded region around the Coulomb point $\delta V_1=0$.
		In this region the ground state has large overlap $\cal{O}$ with the model $\Z3$ wavefunction (inset).
		At width $w/\ell_B=4$ (right), the $\Z3$ phase widens and becomes more robust.
		It is surrounded by the hierarchy/composite fermion state for larger positive $\delta V_1$, and several charge density ordered phases for negative $\delta V_1$.
	}
	\label{fig:ED_V1}
\end{figure}

We directly probe the stability of the entire low-energy spectrum of the $\aRR$ phase upon varying $V_1$ using exact diagonalization (ED) [Fig.~\ref{fig:ED_V1}].
We assume complete spin polarization and no Landau level mixing.
The energy spectrum of the system is resolved as a function of pseudomomentum $\mathbf{K}$,\cite{Haldane-PhysRevLett.55.2095} and the $\aRR$ phase is characterized by a two-fold degenerate ground state in $\mathbf{K}=0$ sector.
Additionally, as discussed in Sec.~\ref{sec:vanillaRR}, there are five copies of those that are related by the center of mass translation and can be factored out.

In Fig.~\ref{fig:ED_V1} we show the phase diagram of the system as $V_1$ is modified, for zero width (left) and $w/\ell_B=4$ (right).
We compute 10 lowest energies per momentum sector of the system with 25 flux quanta through the hexagonal unit cell.
The energies are given in units of $\EC$, and for clarity we plot them relative to the average energy of the system $\tilde E = E - E_\textrm{avg}$.
Black symbols denote the levels belonging to $\mathbf{K}=0$ sector.
For zero width (left), we identify the $\aRR$ ground state degeneracy in a narrow shaded region around the Coulomb point $\delta V_1=0$.
In this region the ground state also has large overlap $\cal{O}$ with the model $\aRR$ wavefunction (inset).
Because of the strong mixing of four lowest energy levels with $\mathbf{K}=0$ around the Coulomb point, we define the overlap $\cal{O}$ as a sum of singular values of the $4\times 2$ overlap matrix $O_{ij}\equiv \braket{ \psi_{\rm exact}^i | \psi_{\Z3}^j }$, $i=1,\ldots,4$, $j=1,2$.
At larger width $w/\ell_B=4$, the $\aRR$ phase widens and becomes more robust as the two quasidegenerate $\mathbf{K}=0$ levels become better separated from the rest of the spectrum.
By inspecting the level degeneracy and computing the overlaps, we also deduce that the $\aRR$ phase is surrounded by the hierarchy/composite fermion state for larger positive $\delta V_1$, and several charge density ordered phases for negative $\delta V_1$.
The phase CDO I was identified with a stripe in Ref.~\onlinecite{RezayiRead:twelvefifths:09}.
The estimate of critical $V_1$ for the transition into the Hierarchy state is in agreement with DMRG estimate in Fig.~\ref{fig:RRvJain}.


\begin{widetext}
\section{Spin polarization}
\label{app:spinpol}
\begin{figure*}[t]
	\includegraphics[width=130mm]{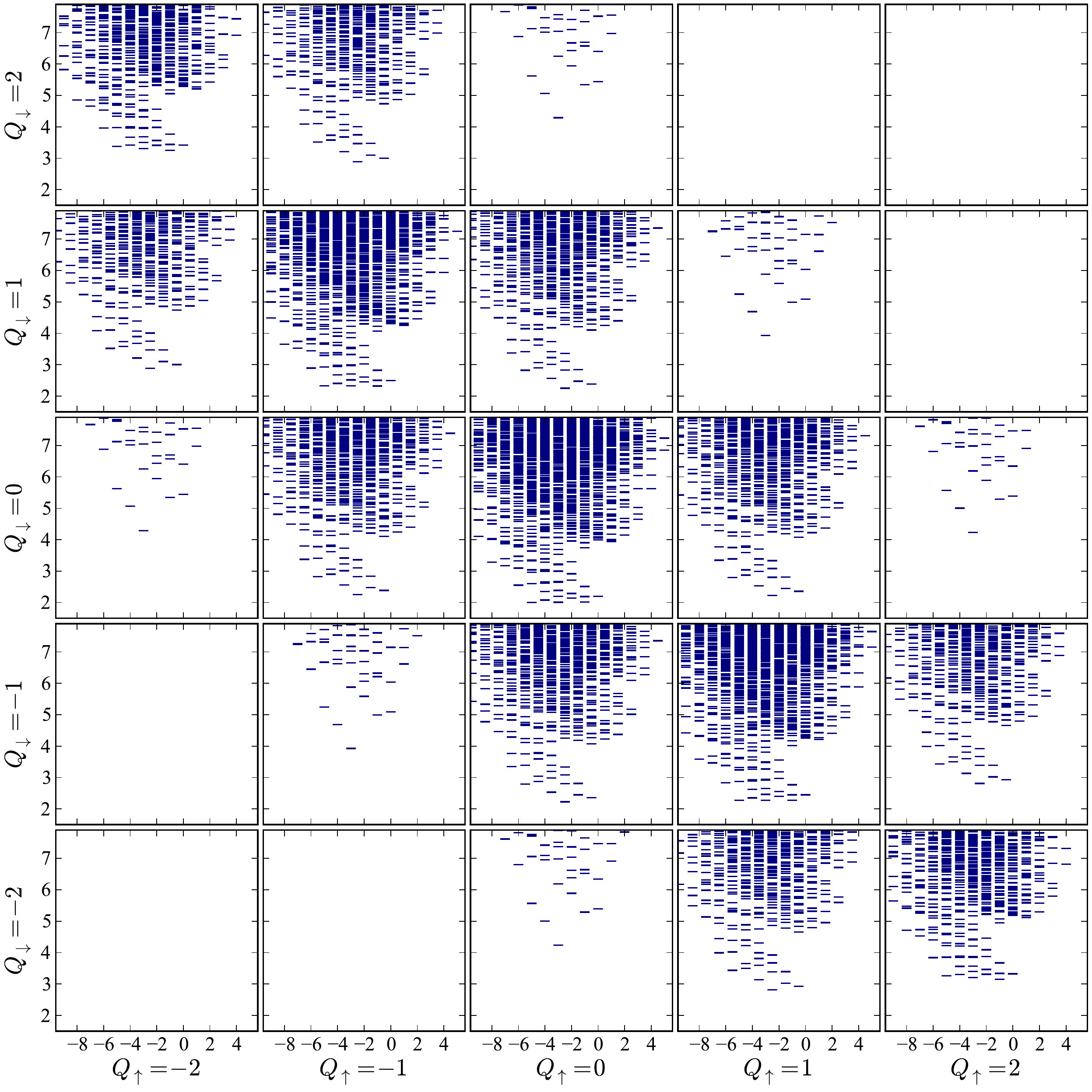}
	\hspace{9mm}
	\raisebox{-0.5mm}{\includegraphics[width=30.9mm]{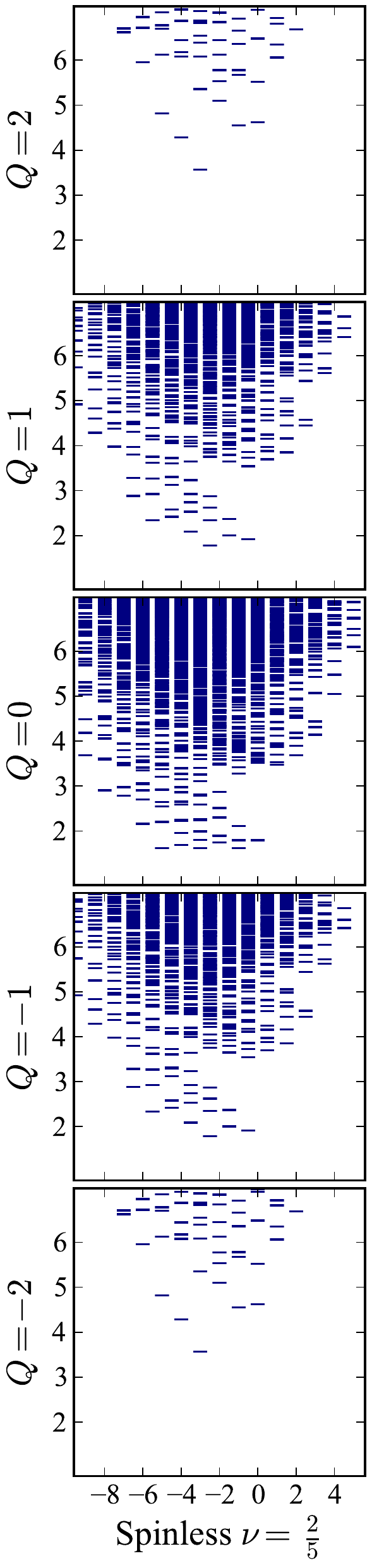}}
	\caption{%
		Signature of spontaneous spin rotation symmetry breaking.
		(Left) The entanglement spectrum of the spin-full system at $\tilde\nu^{\uparrow/\downarrow} = \frac{1}{5}$, sorted by charge sector $(Q^\uparrow,Q^\downarrow)$.
		(Right) The entanglement spectrum for a spinless $\tilde\nu = \frac{2}{5}$ system for comparison.
		Both spectra are taken at $L=21\ell_B$, $w=3\ell_B$.
	}
	\label{fig:spin_entspec}
\end{figure*}

	We first clarify the expected signatures of spin-polarization when explicitly preserving $S^z$ invariance at filling $\tilde\nu^{\uparrow} = \tilde\nu^{\downarrow} = \frac{1}{5}$.
If $\ket{\theta}_{\aRR}$ represents a $\aRR$ state spontaneously polarized in the XY plane at angle $\theta$, we  force the state into the superposition
\begin{align}
	\ket{\Psi} = \int_0^{2\pi} \!\! d \theta  \ket{\theta} _{\aRR}.
\end{align}
This state has infinite bipartite entanglement, so cannot be represented \emph{exactly} by a matrix product state (MPS).
However, we expect that the $\braket{ S^+(r) S^-(r') }$ correlations will be large and nearly constant out to a distance which will depend on the MPS bond-dimension $\chi$, after which it will decay exponentially. 
(We define $S^+ = c_{\uparrow}^\dag c_{\downarrow}$, and $S^-$ to be its Hermitian conjugate.)
This effect (in orbital space) is shown in Fig.~\ref{fig:spin_correlation} for $L=21\ell_B$, $w=3\ell_B$.
In the $\chi \rightarrow \infty$ limit, we expect the correlation function to approach the number-number correlator $\frac14\!\braket{n_0 n_k}$ for the fully spin-polarized \aRR\ phase (shown in dashed line).

The entanglement spectrum provides further evidence.
Recall that the entanglement spectrum can be sorted by the $\mathrm{U(1)}$ charges $(Q^\uparrow, Q^\downarrow)$.
In Fig.~\ref{fig:spin_entspec} we show the entanglement spectrum for a $5\times5$ grid of $(Q^\uparrow, Q^\downarrow)$ sectors.
For constant $Q$, the line $Q^\uparrow + Q^\downarrow = Q$ contains many copies of the charge $Q$ $\aRR$ spectrum, up to some cutoff $|Q^\uparrow - Q^\downarrow| < S_z^{\textrm{max}}(\chi)$ that depends on $\chi$.
This is because in order to support long-range XY correlations, there must be very large fluctuations in $S^z$ across any bipartition of the system.
The entanglement spectrum of a symmetry broken phase was discussed in Ref.~\onlinecite{Metlitski2011}, where it was shown that the spectrum  contain a ``tower of states'' associated with the broken symmetry; the many identical copies of $\aRR$ we find is this tower of states.

We also note that similar computation was perform for $\nu=\frac{13}{5}$ and we also found evidence for spontaneous spin-polarization.

%
%
%
%
%

\section{Structure factor of the Read-Rezayi phase}
\label{app:RRS}
\begin{figure*}[t]
	\includegraphics[width=\linewidth]{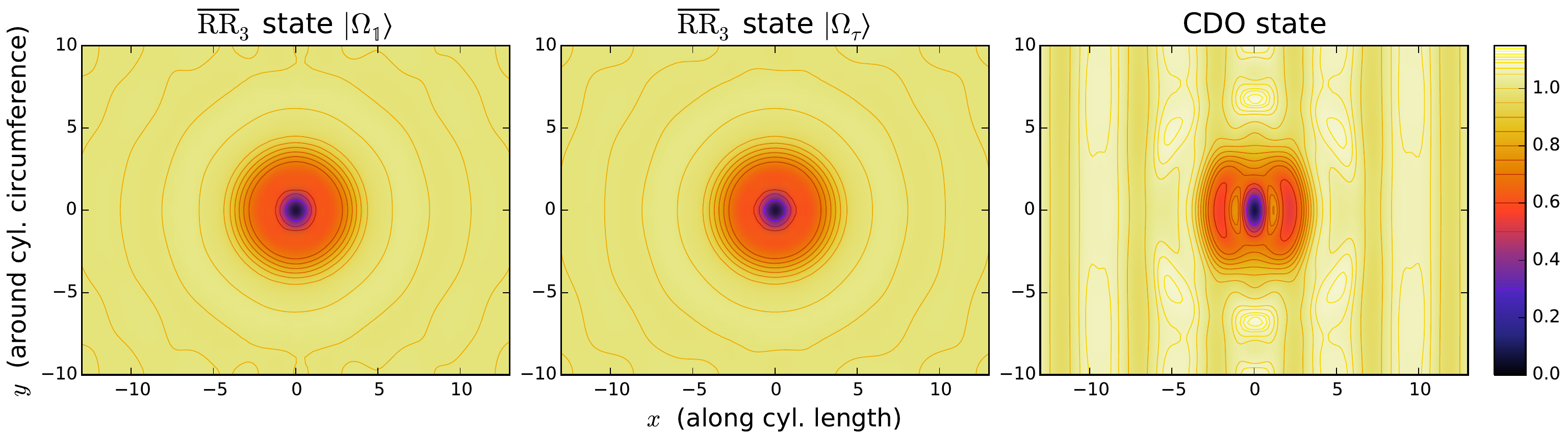}
	\caption{%
		Real-space structure factor $S(\mathbf{r}) = \braket{ \rho(\mathbf{r}) \rho(0) } / \braket{\rho}^2$ of the $\nu=\tfrac{12}{5}$ Read-Rezayi and CDO states at $L=20\ell_B$, $w=1\ell_B$.
		The degenerate \aRR\ states are mostly identical which is a necessary condition for topological order.
		The \aRR\ states are also nearly rotationally symmetric in contrast to the striped structure of the competing CDO phase;
			the residual anisotropy is presumably due to the finite circumference of the cylinder.
	}
	\label{fig:pair_corr_RR}
\end{figure*}
In Fig.~\ref{fig:pair_corr_RR}, we show the structure factor $S(\mathbf{r})$ for the \aRR\ states and the CDO state.
The ground states $\ket{\Omega_\dsone}$ and $\ket{\Omega_\tau}$ have nearly identical structure factor, a requisite for topological order in the \aRR\ phase.
At $L = 20\ell_B$ and $w = 1\ell_B$, the energies of the \aRR\ and CDO states are within $2\times10^{-5}$ of one another.
However, despite having such close energies, the \aRR\ states and CDO state have much different structure factors.
Furthermore, while the CDO state has higher energy than the \aRR\ states, it remains metastable in our DMRG simulations indicative of a 1\textsuperscript{st} order transition.

\end{widetext}
\clearpage
\newpage

\section{Exact diagonalization: effects of finite well-width and torus aspect ratio}
\label{app:torus_ed}

\begin{figure}[t]
	\includegraphics[width=\linewidth]{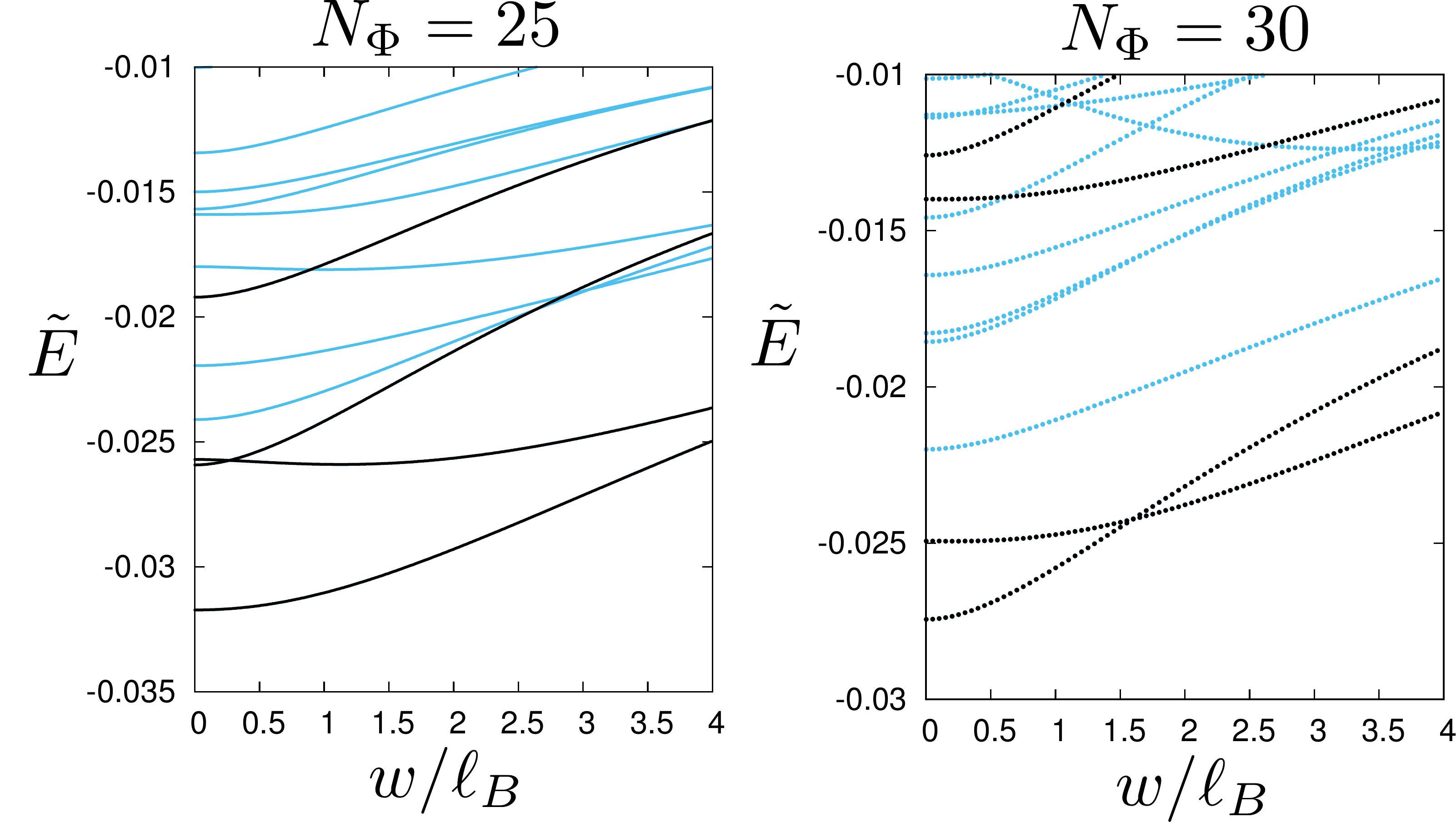}
	\caption{%
		The effect of finite width $w/\ell_B$ on the hexagonal torus threaded by $N_\phi=25$ flux quanta (left) and $N_\Phi=30$ (right).
		Energy spectrum is obtained by exact diagonalization and plotted relative to the average energy of the system.
		Black symbols denote levels belonging to $\mathbf{K}=0$ sector.
	}
	\label{fig:ED_w}
\end{figure}
We showed in Sec.~\ref{sec:vanillaRR} and \ref{sec:cdo} that finite width stabilizes the homogeneous fluid phase.
In Fig.~\ref{fig:ED_w}(left) we systematically study the effect of well width on the energy spectrum (in particular, the topological ground state degeneracy) via exact diagonalization.
We consider a hexagonal torus threaded by $N_\Phi=25$ and $N_\Phi=30$ flux quanta. 

For a smaller system ($N_\Phi=25$), the topological degeneracy is not well-resolved for zero width due to the mixing with a higher level belonging to $\mathbf{K}=0$ sector (black symbols).
In this case, the main effect of non-zero $w/\ell_B$ is to lift the spurious level in energy, leaving a robust two-fold degenerate manifold of ground states.
The gap separating these two quasidegenerate states from the rest of the spectrum further widens as $w/\ell_B$ is increased.
At the same time, the overlap with the model Read-Rezayi state slightly increases as a function of width (not shown).
For a larger system ($N_\Phi=30$), the two-fold degeneracy appears to be present already for zero width, but it gets better resolved for moderate widths $w/\ell_B \approx 1.7$.  

\begin{figure}[t]
	\includegraphics[width=\linewidth]{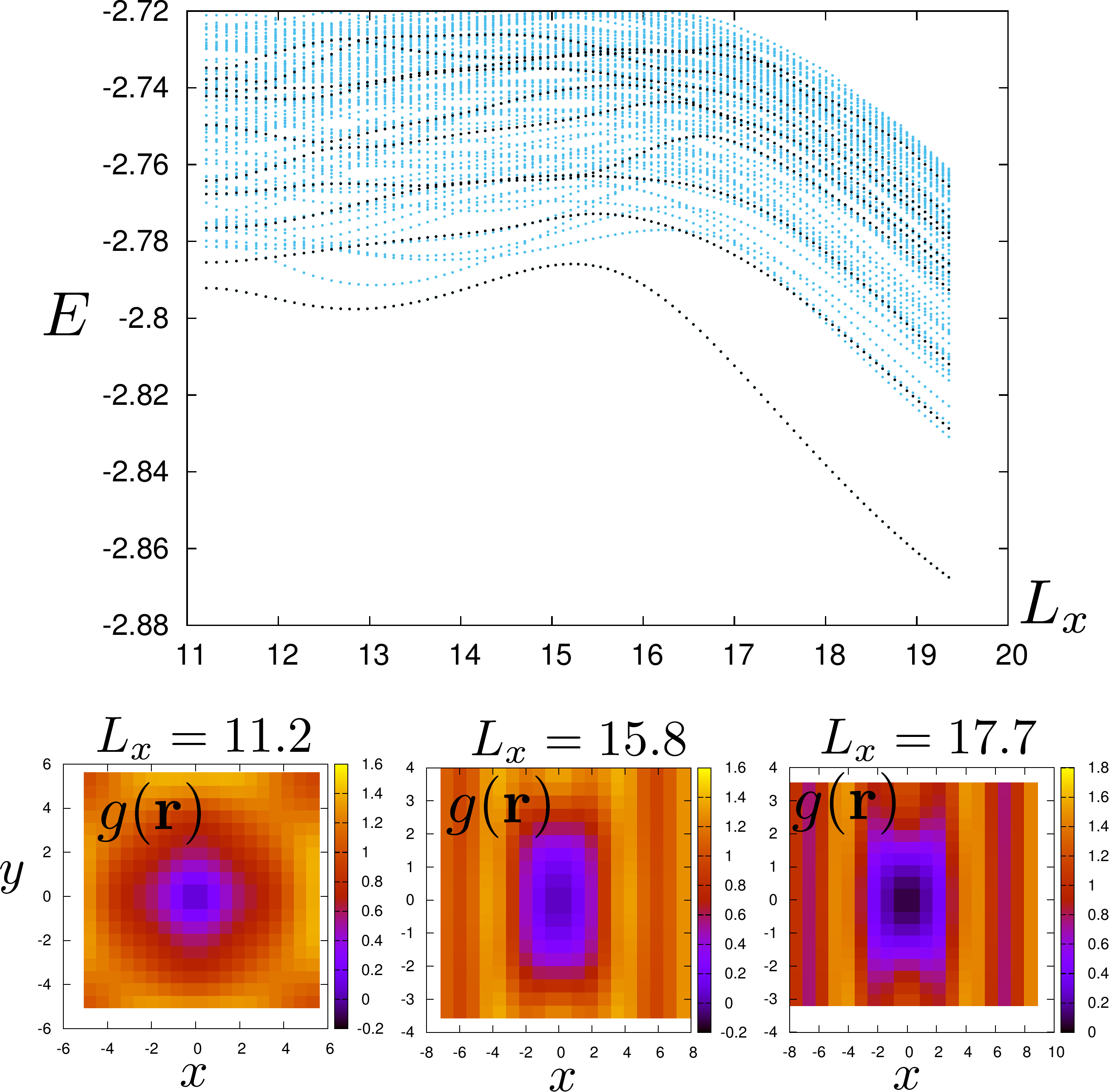}
	\caption{%
		The effect of changing the geometry (one linear dimension $L_x$ of the torus) on the energy spectrum (top) and the ground state (bottom row) of the system.
		Energy spectrum is obtained by exact diagonalization of the rectangular torus with $N_\Phi=20$ flux quanta, and black symbols denote levels belonging to $\mathbf{K}=0$ sector.
		Varying $L_x$ induces charge density order in the ground state, as seen in the pair correlation function $g(\mathbf{r})$ for several values of $L_x$ (bottom).
	}
	\label{fig:ED_r}
\end{figure}

In Fig.~\ref{fig:ED_r}(top) we show the effect of changing the geometry of the torus unit cell.
We consider a rectangular $L_x \times L_y $ torus in this case, whose area is fixed due by the condition $L_x L_y=2\pi \ell_B^2 N_\Phi$.
By changing one of the linear dimensions of the torus ($L_x$), we can drive a transition between the homogenous phase and the CDO.
In Fig.~\ref{fig:ED_r}, $L_x \approx 11\ell_B$ corresponds to an isotropic torus ($L_x = L_y$) where the ground state is approximately two-fold degenerate and belongs to the Read-Rezayi phase.
Beyond $L_x\approx 19\ell_B$, the system evolves towards another, much deeper, energy minimum, which was identified in Sec.~\ref{sec:cdo} with the CDO phase.
The difference in ordering between the two phases is captured by the pair correlation function
	$g(\mathbf{r}) = \frac{L_xL_y}{N_e(N_e-1)} \braket{\delta(\mathbf{r}-\mathbf{R}_i+\mathbf{R}_j)}$ shown in Fig.~\ref{fig:ED_r}(bottom).
Unfortunately, because the area of the torus must be preserved as we change $L_x$, this implies that correlations along the $y$-direction in the ground state of the system at $L_x\approx 19\ell_B$ are artificially truncated because of small $L_y$.
Therefore, the CDO ground state in this case is likely not faithfully reproduced due to finite-size effects and has significantly less entanglement than what we found by DMRG in Sec.~\ref{sec:cdo}.  

\section{Defect-DMRG}
\label{app:defect-dmrg}
	Here we briefly discuss the method used to generate pinned anyons.
The iMPS ansatz for a single anyon excitation was discussed in Ref.~\onlinecite{ZaletelQHdmrg13}.
Because the anyons are charged, in the absence of a pinning potential the single-anyon states form a Landau level.
It is convenient to study the anyons in the Landau gauge  so that anyon $a$  forms a plane wave with momentum $k$ around the cylinder and is localized  near $x \propto k\frac{\ell_B^2}{L}$ along the length.
This choice allows us to conserve momentum in the DMRG simulations (interestingly, $k$ is actually fractional because of  the topological spin of the anyon).
After DMRG we obtain single-anyon states $\ket{k}_a$; just like a Landau level, their energy is independent of $k$, and different $\ket{k}$ are related by a magnetic translation $T_x$ along the length of the cylinder.
The energy $E_a$ reported in the main text is the energy of $\ket{k}_a$ relative to the ground state, combined with the interaction of a point-charge with a neutralizing background.

We caution that we have not performed finite size scaling of the gaps.
If one is familiar with gap calculations on the sphere, this might seem problematic: on a sphere with $N$ electrons, the corrections are of order $1/N$ and it is imperative to extrapolate gaps in $1/N$.
However, the absence of curvature on the cylinder leads to  a much more favorable scaling.
One can show that if the quasiparticle density is bounded by an exponential tail, our estimate of the gap will converge exponentially quickly in the cylinder circumference to its Coulomb value.
(Strictly speaking, for a Coulomb interaction the quasiparticles will have $1/r^5$ tails, but this will lead to weak corrections.)
As a benchmark,  if we apply our procedure to calculate the energy of a quasihole in the integer quantum Hall effect at the same circumference ($L=21\ell_B$) and precision as the results quoted in the text, we obtain a quasihole energy of $E \approx 1.2531$; the exact result is $\sqrt{ \pi / 2 } \approx 1.2533$.

To pin the anyons in 2D we need to introduce a pinning potential $\phi(x, y)$, but it is too expensive to work in the full Hilbert space without conserving momentum.
In the limit of a weak pinning potential we can use first-order degenerate perturbation theory and project $\phi$ into the variational space spanned by the anyonic Landau level $\{ \ket{k}_a \}$.
We compute the many-body matrix elements $H^{\textrm{eff}}_{k k'} = \bra{k'} \hat{\phi} \ket{k}_a$ using standard MPS techniques,  diagonalize $H^{\textrm{eff}}$, and compute the real-space density of the lowest energy state $\sum_k \Psi_k \ket{k}_a$.

The total energy of the pinned excitation is the energy $E_a$ reported in the main text plus the potential energy $H^{\textrm{eff}}$ from the pin.
For a width 2$\ell_B$ Gaussian, the pin lowers the energy of the non-Abelian $\frac{e}{5}\tau$ excitation by 85\% more than the Abelian $\frac{e}{5}$; for a width 6$\ell_B$ Gaussian, the difference is 25\%.


\bibliography{bibtwelve}

\begin{thebibliography}{68}%
\makeatletter
\providecommand \@ifxundefined [1]{%
 \@ifx{#1\undefined}
}%
\providecommand \@ifnum [1]{%
 \ifnum #1\expandafter \@firstoftwo
 \else \expandafter \@secondoftwo
 \fi
}%
\providecommand \@ifx [1]{%
 \ifx #1\expandafter \@firstoftwo
 \else \expandafter \@secondoftwo
 \fi
}%
\providecommand \natexlab [1]{#1}%
\providecommand \enquote  [1]{``#1''}%
\providecommand \bibnamefont  [1]{#1}%
\providecommand \bibfnamefont [1]{#1}%
\providecommand \citenamefont [1]{#1}%
\providecommand \href@noop [0]{\@secondoftwo}%
\providecommand \href [0]{\begingroup \@sanitize@url \@href}%
\providecommand \@href[1]{\@@startlink{#1}\@@href}%
\providecommand \@@href[1]{\endgroup#1\@@endlink}%
\providecommand \@sanitize@url [0]{\catcode `\\12\catcode `\$12\catcode
  `\&12\catcode `\#12\catcode `\^12\catcode `\_12\catcode `\%12\relax}%
\providecommand \@@startlink[1]{}%
\providecommand \@@endlink[0]{}%
\providecommand \url  [0]{\begingroup\@sanitize@url \@url }%
\providecommand \@url [1]{\endgroup\@href {#1}{\urlprefix }}%
\providecommand \urlprefix  [0]{URL }%
\providecommand \Eprint [0]{\href }%
\providecommand \doibase [0]{http://dx.doi.org/}%
\providecommand \selectlanguage [0]{\@gobble}%
\providecommand \bibinfo  [0]{\@secondoftwo}%
\providecommand \bibfield  [0]{\@secondoftwo}%
\providecommand \translation [1]{[#1]}%
\providecommand \BibitemOpen [0]{}%
\providecommand \bibitemStop [0]{}%
\providecommand \bibitemNoStop [0]{.\EOS\space}%
\providecommand \EOS [0]{\spacefactor3000\relax}%
\providecommand \BibitemShut  [1]{\csname bibitem#1\endcsname}%
\let\auto@bib@innerbib\@empty
\bibitem [{\citenamefont {Kitaev}(2003)}]{Kitaev:QC:03}%
  \BibitemOpen
  \bibfield  {author} {\bibinfo {author} {\bibfnamefont {A.~Y.}\ \bibnamefont
  {Kitaev}},\ }\href {\doibase 10.1016/S0003-4916(02)00018-0} {\bibfield
  {journal} {\bibinfo  {journal} {Ann. Phys. (NY)}\ }\textbf {\bibinfo {volume}
  {303}},\ \bibinfo {pages} {2} (\bibinfo {year} {2003})}\BibitemShut {NoStop}%
\bibitem [{\citenamefont {Nayak}\ \emph {et~al.}(2008)\citenamefont {Nayak},
  \citenamefont {Simon}, \citenamefont {Stern}, \citenamefont {Freedman},\ and\
  \citenamefont {Das~Sarma}}]{Nayak:TQCReview:2008}%
  \BibitemOpen
  \bibfield  {author} {\bibinfo {author} {\bibfnamefont {C.}~\bibnamefont
  {Nayak}}, \bibinfo {author} {\bibfnamefont {S.~H.}\ \bibnamefont {Simon}},
  \bibinfo {author} {\bibfnamefont {A.}~\bibnamefont {Stern}}, \bibinfo
  {author} {\bibfnamefont {M.}~\bibnamefont {Freedman}}, \ and\ \bibinfo
  {author} {\bibfnamefont {S.}~\bibnamefont {Das~Sarma}},\ }\href {\doibase
  10.1103/RevModPhys.80.1083} {\bibfield  {journal} {\bibinfo  {journal} {Rev.
  Mod. Phys.}\ }\textbf {\bibinfo {volume} {80}},\ \bibinfo {pages} {1083}
  (\bibinfo {year} {2008})}\BibitemShut {NoStop}%
\bibitem [{\citenamefont {Mourik}\ \emph {et~al.}(2012)\citenamefont {Mourik},
  \citenamefont {Zuo}, \citenamefont {Frolov}, \citenamefont {Plissard},
  \citenamefont {Bakkers},\ and\ \citenamefont {Kouwenhoven}}]{Mourik:12}%
  \BibitemOpen
  \bibfield  {author} {\bibinfo {author} {\bibfnamefont {V.}~\bibnamefont
  {Mourik}}, \bibinfo {author} {\bibfnamefont {K.}~\bibnamefont {Zuo}},
  \bibinfo {author} {\bibfnamefont {S.~M.}\ \bibnamefont {Frolov}}, \bibinfo
  {author} {\bibfnamefont {S.~R.}\ \bibnamefont {Plissard}}, \bibinfo {author}
  {\bibfnamefont {E.~P. A.~M.}\ \bibnamefont {Bakkers}}, \ and\ \bibinfo
  {author} {\bibfnamefont {L.~P.}\ \bibnamefont {Kouwenhoven}},\ }\href
  {\doibase 10.1126/science.1222360} {\bibfield  {journal} {\bibinfo  {journal}
  {Science}\ }\textbf {\bibinfo {volume} {336}},\ \bibinfo {pages} {1003}
  (\bibinfo {year} {2012})}\BibitemShut {NoStop}%
\bibitem [{\citenamefont {Das}\ \emph {et~al.}(2012)\citenamefont {Das},
  \citenamefont {Ronen}, \citenamefont {Most}, \citenamefont {Oreg},
  \citenamefont {Heiblum},\ and\ \citenamefont {Shtrikman}}]{Das:12}%
  \BibitemOpen
  \bibfield  {author} {\bibinfo {author} {\bibfnamefont {A.}~\bibnamefont
  {Das}}, \bibinfo {author} {\bibfnamefont {Y.}~\bibnamefont {Ronen}}, \bibinfo
  {author} {\bibfnamefont {Y.}~\bibnamefont {Most}}, \bibinfo {author}
  {\bibfnamefont {Y.}~\bibnamefont {Oreg}}, \bibinfo {author} {\bibfnamefont
  {M.}~\bibnamefont {Heiblum}}, \ and\ \bibinfo {author} {\bibfnamefont
  {H.}~\bibnamefont {Shtrikman}},\ }\href {\doibase 10.1038/nphys2479}
  {\bibfield  {journal} {\bibinfo  {journal} {Nat. Phys.}\ }\textbf {\bibinfo
  {volume} {8}},\ \bibinfo {pages} {887} (\bibinfo {year} {2012})}\BibitemShut
  {NoStop}%
\bibitem [{\citenamefont {Rokhinson}\ \emph {et~al.}(2012)\citenamefont
  {Rokhinson}, \citenamefont {Liu},\ and\ \citenamefont
  {Furdyna}}]{Rokhinson:12}%
  \BibitemOpen
  \bibfield  {author} {\bibinfo {author} {\bibfnamefont {L.~P.}\ \bibnamefont
  {Rokhinson}}, \bibinfo {author} {\bibfnamefont {X.}~\bibnamefont {Liu}}, \
  and\ \bibinfo {author} {\bibfnamefont {J.~K.}\ \bibnamefont {Furdyna}},\
  }\href {\doibase 10.1038/nphys2429} {\bibfield  {journal} {\bibinfo
  {journal} {Nat. Phys.}\ }\textbf {\bibinfo {volume} {8}},\ \bibinfo {pages}
  {795} (\bibinfo {year} {2012})}\BibitemShut {NoStop}%
\bibitem [{\citenamefont {Deng}\ \emph
  {et~al.}(2012{\natexlab{a}})\citenamefont {Deng}, \citenamefont {Yu},
  \citenamefont {Huang}, \citenamefont {Larsson}, \citenamefont {Caroff},\ and\
  \citenamefont {Xu}}]{MDeng:12}%
  \BibitemOpen
  \bibfield  {author} {\bibinfo {author} {\bibfnamefont {M.~T.}\ \bibnamefont
  {Deng}}, \bibinfo {author} {\bibfnamefont {C.~L.}\ \bibnamefont {Yu}},
  \bibinfo {author} {\bibfnamefont {G.~Y.}\ \bibnamefont {Huang}}, \bibinfo
  {author} {\bibfnamefont {M.}~\bibnamefont {Larsson}}, \bibinfo {author}
  {\bibfnamefont {P.}~\bibnamefont {Caroff}}, \ and\ \bibinfo {author}
  {\bibfnamefont {H.~Q.}\ \bibnamefont {Xu}},\ }\href {\doibase
  10.1021/nl303758w} {\bibfield  {journal} {\bibinfo  {journal} {Nano Lett.}\
  }\textbf {\bibinfo {volume} {12}},\ \bibinfo {pages} {6414} (\bibinfo {year}
  {2012}{\natexlab{a}})}\BibitemShut {NoStop}%
\bibitem [{\citenamefont {Finck}\ \emph {et~al.}(2013)\citenamefont {Finck},
  \citenamefont {Van~Harlingen}, \citenamefont {Mohseni}, \citenamefont
  {Jung},\ and\ \citenamefont {Li}}]{Finck:13}%
  \BibitemOpen
  \bibfield  {author} {\bibinfo {author} {\bibfnamefont {A.~D.~K.}\
  \bibnamefont {Finck}}, \bibinfo {author} {\bibfnamefont {D.~J.}\ \bibnamefont
  {Van~Harlingen}}, \bibinfo {author} {\bibfnamefont {P.~K.}\ \bibnamefont
  {Mohseni}}, \bibinfo {author} {\bibfnamefont {K.}~\bibnamefont {Jung}}, \
  and\ \bibinfo {author} {\bibfnamefont {X.}~\bibnamefont {Li}},\ }\href
  {\doibase 10.1103/PhysRevLett.110.126406} {\bibfield  {journal} {\bibinfo
  {journal} {Phys. Rev. Lett.}\ }\textbf {\bibinfo {volume} {110}},\ \bibinfo
  {pages} {126406} (\bibinfo {year} {2013})}\BibitemShut {NoStop}%
\bibitem [{\citenamefont {Churchill}\ \emph {et~al.}(2013)\citenamefont
  {Churchill}, \citenamefont {Fatemi}, \citenamefont {Grove-Rasmussen},
  \citenamefont {Deng}, \citenamefont {Caroff}, \citenamefont {Xu},\ and\
  \citenamefont {Marcus}}]{Churchill:13}%
  \BibitemOpen
  \bibfield  {author} {\bibinfo {author} {\bibfnamefont {H.~O.~H.}\
  \bibnamefont {Churchill}}, \bibinfo {author} {\bibfnamefont {V.}~\bibnamefont
  {Fatemi}}, \bibinfo {author} {\bibfnamefont {K.}~\bibnamefont
  {Grove-Rasmussen}}, \bibinfo {author} {\bibfnamefont {M.~T.}\ \bibnamefont
  {Deng}}, \bibinfo {author} {\bibfnamefont {P.}~\bibnamefont {Caroff}},
  \bibinfo {author} {\bibfnamefont {H.~Q.}\ \bibnamefont {Xu}}, \ and\ \bibinfo
  {author} {\bibfnamefont {C.~M.}\ \bibnamefont {Marcus}},\ }\href {\doibase
  10.1103/PhysRevB.87.241401} {\bibfield  {journal} {\bibinfo  {journal} {Phys.
  Rev. B}\ }\textbf {\bibinfo {volume} {87}},\ \bibinfo {pages} {241401}
  (\bibinfo {year} {2013})}\BibitemShut {NoStop}%
\bibitem [{\citenamefont {Freedman}\ \emph {et~al.}(2002)\citenamefont
  {Freedman}, \citenamefont {Larsen},\ and\ \citenamefont
  {Wang}}]{FreedmanLarsenWang:2002a}%
  \BibitemOpen
  \bibfield  {author} {\bibinfo {author} {\bibfnamefont {M.~H.}\ \bibnamefont
  {Freedman}}, \bibinfo {author} {\bibfnamefont {M.~J.}\ \bibnamefont
  {Larsen}}, \ and\ \bibinfo {author} {\bibfnamefont {Z.}~\bibnamefont
  {Wang}},\ }\href {\doibase 10.1007/s002200200645} {\bibfield  {journal}
  {\bibinfo  {journal} {Commun. Math. Phys.}\ }\textbf {\bibinfo {volume}
  {227}},\ \bibinfo {pages} {605} (\bibinfo {year} {2002})}\BibitemShut
  {NoStop}%
\bibitem [{\citenamefont {Fidkowski}\ \emph {et~al.}(2009)\citenamefont
  {Fidkowski}, \citenamefont {Freedman}, \citenamefont {Nayak}, \citenamefont
  {Walker},\ and\ \citenamefont {Wang}}]{Fidkowski:DFib:09}%
  \BibitemOpen
  \bibfield  {author} {\bibinfo {author} {\bibfnamefont {L.}~\bibnamefont
  {Fidkowski}}, \bibinfo {author} {\bibfnamefont {M.}~\bibnamefont {Freedman}},
  \bibinfo {author} {\bibfnamefont {C.}~\bibnamefont {Nayak}}, \bibinfo
  {author} {\bibfnamefont {K.}~\bibnamefont {Walker}}, \ and\ \bibinfo {author}
  {\bibfnamefont {Z.}~\bibnamefont {Wang}},\ }\href {\doibase
  10.1007/s00220-009-0757-9} {\bibfield  {journal} {\bibinfo  {journal}
  {Commun. Math. Phys.}\ }\textbf {\bibinfo {volume} {287}},\ \bibinfo {pages}
  {805} (\bibinfo {year} {2009})}\BibitemShut {NoStop}%
\bibitem [{\citenamefont {Liu}\ \emph {et~al.}(2013)\citenamefont {Liu},
  \citenamefont {Bergholtz},\ and\ \citenamefont {Kapit}}]{PhysRevB.88.205101}%
  \BibitemOpen
  \bibfield  {author} {\bibinfo {author} {\bibfnamefont {Z.}~\bibnamefont
  {Liu}}, \bibinfo {author} {\bibfnamefont {E.~J.}\ \bibnamefont {Bergholtz}},
  \ and\ \bibinfo {author} {\bibfnamefont {E.}~\bibnamefont {Kapit}},\ }\href
  {\doibase 10.1103/PhysRevB.88.205101} {\bibfield  {journal} {\bibinfo
  {journal} {Phys. Rev. B}\ }\textbf {\bibinfo {volume} {88}},\ \bibinfo
  {pages} {205101} (\bibinfo {year} {2013})}\BibitemShut {NoStop}%
\bibitem [{\citenamefont {Wang}\ \emph {et~al.}(2015)\citenamefont {Wang},
  \citenamefont {Liu}, \citenamefont {Liu}, \citenamefont {Cao},\ and\
  \citenamefont {Fan}}]{PhysRevB.91.125138}%
  \BibitemOpen
  \bibfield  {author} {\bibinfo {author} {\bibfnamefont {D.}~\bibnamefont
  {Wang}}, \bibinfo {author} {\bibfnamefont {Z.}~\bibnamefont {Liu}}, \bibinfo
  {author} {\bibfnamefont {W.-M.}\ \bibnamefont {Liu}}, \bibinfo {author}
  {\bibfnamefont {J.}~\bibnamefont {Cao}}, \ and\ \bibinfo {author}
  {\bibfnamefont {H.}~\bibnamefont {Fan}},\ }\href {\doibase
  10.1103/PhysRevB.91.125138} {\bibfield  {journal} {\bibinfo  {journal} {Phys.
  Rev. B}\ }\textbf {\bibinfo {volume} {91}},\ \bibinfo {pages} {125138}
  (\bibinfo {year} {2015})}\BibitemShut {NoStop}%
\bibitem [{\citenamefont {Barkeshli}\ \emph {et~al.}(2015)\citenamefont
  {Barkeshli}, \citenamefont {Jiang}, \citenamefont {Thomale},\ and\
  \citenamefont {Qi}}]{Barkeshli:GeneralizedKitaevModel:15}%
  \BibitemOpen
  \bibfield  {author} {\bibinfo {author} {\bibfnamefont {M.}~\bibnamefont
  {Barkeshli}}, \bibinfo {author} {\bibfnamefont {H.-C.}\ \bibnamefont
  {Jiang}}, \bibinfo {author} {\bibfnamefont {R.}~\bibnamefont {Thomale}}, \
  and\ \bibinfo {author} {\bibfnamefont {X.-L.}\ \bibnamefont {Qi}},\ }\href
  {\doibase 10.1103/PhysRevLett.114.026401} {\bibfield  {journal} {\bibinfo
  {journal} {Phys. Rev. Lett.}\ }\textbf {\bibinfo {volume} {114}},\ \bibinfo
  {pages} {026401} (\bibinfo {year} {2015})}\BibitemShut {NoStop}%
\bibitem [{\citenamefont {Stoudenmire}\ \emph {et~al.}(2015)\citenamefont
  {Stoudenmire}, \citenamefont {Clarke}, \citenamefont {Mong},\ and\
  \citenamefont {Alicea}}]{Stoudenmire:Fibonacci:15}%
  \BibitemOpen
  \bibfield  {author} {\bibinfo {author} {\bibfnamefont {E.~M.}\ \bibnamefont
  {Stoudenmire}}, \bibinfo {author} {\bibfnamefont {D.~J.}\ \bibnamefont
  {Clarke}}, \bibinfo {author} {\bibfnamefont {R.~S.~K.}\ \bibnamefont {Mong}},
  \ and\ \bibinfo {author} {\bibfnamefont {J.}~\bibnamefont {Alicea}},\ }\href
  {\doibase 10.1103/PhysRevB.91.235112} {\bibfield  {journal} {\bibinfo
  {journal} {Phys. Rev. B}\ }\textbf {\bibinfo {volume} {91}},\ \bibinfo
  {pages} {235112} (\bibinfo {year} {2015})}\BibitemShut {NoStop}%
\bibitem [{\citenamefont {Xia}\ \emph {et~al.}(2004)\citenamefont {Xia},
  \citenamefont {Pan}, \citenamefont {Vicente}, \citenamefont {Adams},
  \citenamefont {Sullivan}, \citenamefont {Stormer}, \citenamefont {Tsui},
  \citenamefont {Pfeiffer}, \citenamefont {Baldwin},\ and\ \citenamefont
  {West}}]{Xia_12_5}%
  \BibitemOpen
  \bibfield  {author} {\bibinfo {author} {\bibfnamefont {J.~S.}\ \bibnamefont
  {Xia}}, \bibinfo {author} {\bibfnamefont {W.}~\bibnamefont {Pan}}, \bibinfo
  {author} {\bibfnamefont {C.~L.}\ \bibnamefont {Vicente}}, \bibinfo {author}
  {\bibfnamefont {E.~D.}\ \bibnamefont {Adams}}, \bibinfo {author}
  {\bibfnamefont {N.~S.}\ \bibnamefont {Sullivan}}, \bibinfo {author}
  {\bibfnamefont {H.~L.}\ \bibnamefont {Stormer}}, \bibinfo {author}
  {\bibfnamefont {D.~C.}\ \bibnamefont {Tsui}}, \bibinfo {author}
  {\bibfnamefont {L.~N.}\ \bibnamefont {Pfeiffer}}, \bibinfo {author}
  {\bibfnamefont {K.~W.}\ \bibnamefont {Baldwin}}, \ and\ \bibinfo {author}
  {\bibfnamefont {K.~W.}\ \bibnamefont {West}},\ }\href {\doibase
  10.1103/PhysRevLett.93.176809} {\bibfield  {journal} {\bibinfo  {journal}
  {Phys. Rev. Lett.}\ }\textbf {\bibinfo {volume} {93}},\ \bibinfo {pages}
  {176809} (\bibinfo {year} {2004})}\BibitemShut {NoStop}%
\bibitem [{\citenamefont {Kumar}\ \emph {et~al.}(2010)\citenamefont {Kumar},
  \citenamefont {Cs\'athy}, \citenamefont {Manfra}, \citenamefont {Pfeiffer},\
  and\ \citenamefont {West}}]{Kumar:12fifths}%
  \BibitemOpen
  \bibfield  {author} {\bibinfo {author} {\bibfnamefont {A.}~\bibnamefont
  {Kumar}}, \bibinfo {author} {\bibfnamefont {G.~A.}\ \bibnamefont {Cs\'athy}},
  \bibinfo {author} {\bibfnamefont {M.~J.}\ \bibnamefont {Manfra}}, \bibinfo
  {author} {\bibfnamefont {L.~N.}\ \bibnamefont {Pfeiffer}}, \ and\ \bibinfo
  {author} {\bibfnamefont {K.~W.}\ \bibnamefont {West}},\ }\href {\doibase
  10.1103/PhysRevLett.105.246808} {\bibfield  {journal} {\bibinfo  {journal}
  {Phys. Rev. Lett.}\ }\textbf {\bibinfo {volume} {105}},\ \bibinfo {pages}
  {246808} (\bibinfo {year} {2010})}\BibitemShut {NoStop}%
\bibitem [{\citenamefont {Choi}\ \emph {et~al.}(2008)\citenamefont {Choi},
  \citenamefont {Kang}, \citenamefont {Das~Sarma}, \citenamefont {Pfeiffer},\
  and\ \citenamefont {West}}]{ChoiPhysRevB.77.081301}%
  \BibitemOpen
  \bibfield  {author} {\bibinfo {author} {\bibfnamefont {H.~C.}\ \bibnamefont
  {Choi}}, \bibinfo {author} {\bibfnamefont {W.}~\bibnamefont {Kang}}, \bibinfo
  {author} {\bibfnamefont {S.}~\bibnamefont {Das~Sarma}}, \bibinfo {author}
  {\bibfnamefont {L.~N.}\ \bibnamefont {Pfeiffer}}, \ and\ \bibinfo {author}
  {\bibfnamefont {K.~W.}\ \bibnamefont {West}},\ }\href {\doibase
  10.1103/PhysRevB.77.081301} {\bibfield  {journal} {\bibinfo  {journal} {Phys.
  Rev. B}\ }\textbf {\bibinfo {volume} {77}},\ \bibinfo {pages} {081301}
  (\bibinfo {year} {2008})}\BibitemShut {NoStop}%
\bibitem [{\citenamefont {Pan}\ \emph {et~al.}(2008)\citenamefont {Pan},
  \citenamefont {Xia}, \citenamefont {Stormer}, \citenamefont {Tsui},
  \citenamefont {Vicente}, \citenamefont {Adams}, \citenamefont {Sullivan},
  \citenamefont {Pfeiffer}, \citenamefont {Baldwin},\ and\ \citenamefont
  {West}}]{PanPhysRevB.77.075307}%
  \BibitemOpen
  \bibfield  {author} {\bibinfo {author} {\bibfnamefont {W.}~\bibnamefont
  {Pan}}, \bibinfo {author} {\bibfnamefont {J.~S.}\ \bibnamefont {Xia}},
  \bibinfo {author} {\bibfnamefont {H.~L.}\ \bibnamefont {Stormer}}, \bibinfo
  {author} {\bibfnamefont {D.~C.}\ \bibnamefont {Tsui}}, \bibinfo {author}
  {\bibfnamefont {C.}~\bibnamefont {Vicente}}, \bibinfo {author} {\bibfnamefont
  {E.~D.}\ \bibnamefont {Adams}}, \bibinfo {author} {\bibfnamefont {N.~S.}\
  \bibnamefont {Sullivan}}, \bibinfo {author} {\bibfnamefont {L.~N.}\
  \bibnamefont {Pfeiffer}}, \bibinfo {author} {\bibfnamefont {K.~W.}\
  \bibnamefont {Baldwin}}, \ and\ \bibinfo {author} {\bibfnamefont {K.~W.}\
  \bibnamefont {West}},\ }\href {\doibase 10.1103/PhysRevB.77.075307}
  {\bibfield  {journal} {\bibinfo  {journal} {Phys. Rev. B}\ }\textbf {\bibinfo
  {volume} {77}},\ \bibinfo {pages} {075307} (\bibinfo {year}
  {2008})}\BibitemShut {NoStop}%
\bibitem [{\citenamefont {Jain}(2007)}]{jainbook}%
  \BibitemOpen
  \bibfield  {author} {\bibinfo {author} {\bibfnamefont {J.~K.}\ \bibnamefont
  {Jain}},\ }\href@noop {} {\emph {\bibinfo {title} {Composite Fermions}}}\
  (\bibinfo  {publisher} {Cambridge University Press},\ \bibinfo {year}
  {2007})\BibitemShut {NoStop}%
\bibitem [{\citenamefont {Read}\ and\ \citenamefont
  {Rezayi}(1999)}]{ReadRezayi:99}%
  \BibitemOpen
  \bibfield  {author} {\bibinfo {author} {\bibfnamefont {N.}~\bibnamefont
  {Read}}\ and\ \bibinfo {author} {\bibfnamefont {E.}~\bibnamefont {Rezayi}},\
  }\href {\doibase 10.1103/PhysRevB.59.8084} {\bibfield  {journal} {\bibinfo
  {journal} {Phys. Rev. B}\ }\textbf {\bibinfo {volume} {59}},\ \bibinfo
  {pages} {8084} (\bibinfo {year} {1999})}\BibitemShut {NoStop}%
\bibitem [{\citenamefont {Halperin}(1983)}]{Halperin83}%
  \BibitemOpen
  \bibfield  {author} {\bibinfo {author} {\bibfnamefont {B.~I.}\ \bibnamefont
  {Halperin}},\ }\href {\doibase 10.5169/seals-115362} {\bibfield  {journal}
  {\bibinfo  {journal} {Helv. Phys. Acta.}\ }\textbf {\bibinfo {volume} {56}},\
  \bibinfo {pages} {75} (\bibinfo {year} {1983})}\BibitemShut {NoStop}%
\bibitem [{\citenamefont {Haldane}(1983)}]{Haldane1983}%
  \BibitemOpen
  \bibfield  {author} {\bibinfo {author} {\bibfnamefont {F.~D.~M.}\
  \bibnamefont {Haldane}},\ }\href {\doibase 10.1103/PhysRevLett.51.605}
  {\bibfield  {journal} {\bibinfo  {journal} {Phys. Rev. Lett.}\ }\textbf
  {\bibinfo {volume} {51}},\ \bibinfo {pages} {605} (\bibinfo {year}
  {1983})}\BibitemShut {NoStop}%
\bibitem [{\citenamefont {Bonderson}\ and\ \citenamefont
  {Slingerland}(2008)}]{BondersonSlingerland:08}%
  \BibitemOpen
  \bibfield  {author} {\bibinfo {author} {\bibfnamefont {P.}~\bibnamefont
  {Bonderson}}\ and\ \bibinfo {author} {\bibfnamefont {J.~K.}\ \bibnamefont
  {Slingerland}},\ }\href {\doibase 10.1103/PhysRevB.78.125323} {\bibfield
  {journal} {\bibinfo  {journal} {Phys. Rev. B}\ }\textbf {\bibinfo {volume}
  {78}},\ \bibinfo {pages} {125323} (\bibinfo {year} {2008})}\BibitemShut
  {NoStop}%
\bibitem [{\citenamefont {Willett}\ \emph {et~al.}(2009)\citenamefont
  {Willett}, \citenamefont {Pfeiffer},\ and\ \citenamefont
  {West}}]{Willett:5HalvesInterf:2009}%
  \BibitemOpen
  \bibfield  {author} {\bibinfo {author} {\bibfnamefont {R.~L.}\ \bibnamefont
  {Willett}}, \bibinfo {author} {\bibfnamefont {L.~N.}\ \bibnamefont
  {Pfeiffer}}, \ and\ \bibinfo {author} {\bibfnamefont {K.~W.}\ \bibnamefont
  {West}},\ }\href {\doibase 10.1073/pnas.0812599106} {\bibfield  {journal}
  {\bibinfo  {journal} {Proc. Natl. Acad. Sci.}\ }\textbf {\bibinfo {volume}
  {106}},\ \bibinfo {pages} {8853} (\bibinfo {year} {2009})}\BibitemShut
  {NoStop}%
\bibitem [{\citenamefont {An}\ \emph {et~al.}(2011)\citenamefont {An},
  \citenamefont {Jiang}, \citenamefont {Choi}, \citenamefont {Kang},
  \citenamefont {Simon}, \citenamefont {Pfeiffer}, \citenamefont {West},\ and\
  \citenamefont {Baldwin}}]{Kang:5HalvesInterf:2011}%
  \BibitemOpen
  \bibfield  {author} {\bibinfo {author} {\bibfnamefont {S.}~\bibnamefont
  {An}}, \bibinfo {author} {\bibfnamefont {P.}~\bibnamefont {Jiang}}, \bibinfo
  {author} {\bibfnamefont {H.}~\bibnamefont {Choi}}, \bibinfo {author}
  {\bibfnamefont {W.}~\bibnamefont {Kang}}, \bibinfo {author} {\bibfnamefont
  {S.~H.}\ \bibnamefont {Simon}}, \bibinfo {author} {\bibfnamefont {L.~N.}\
  \bibnamefont {Pfeiffer}}, \bibinfo {author} {\bibfnamefont {K.~W.}\
  \bibnamefont {West}}, \ and\ \bibinfo {author} {\bibfnamefont {K.~W.}\
  \bibnamefont {Baldwin}},\ }\href@noop {} {\enquote {\bibinfo {title}
  {{Braiding of Abelian and Non-Abelian Anyons in the Fractional Quantum Hall
  Effect}},}\ } (\bibinfo {year} {2011}),\ \bibinfo {note} {unpublished},\
  \Eprint {http://arxiv.org/abs/1112.3400} {arXiv:1112.3400
  [cond-mat.mes-hall]} \BibitemShut {NoStop}%
\bibitem [{\citenamefont {Willett}\ \emph {et~al.}(2013)\citenamefont
  {Willett}, \citenamefont {Nayak}, \citenamefont {Shtengel}, \citenamefont
  {Pfeiffer},\ and\ \citenamefont {West}}]{WillettNayak:5HalvesInterf:2013}%
  \BibitemOpen
  \bibfield  {author} {\bibinfo {author} {\bibfnamefont {R.~L.}\ \bibnamefont
  {Willett}}, \bibinfo {author} {\bibfnamefont {C.}~\bibnamefont {Nayak}},
  \bibinfo {author} {\bibfnamefont {K.}~\bibnamefont {Shtengel}}, \bibinfo
  {author} {\bibfnamefont {L.~N.}\ \bibnamefont {Pfeiffer}}, \ and\ \bibinfo
  {author} {\bibfnamefont {K.~W.}\ \bibnamefont {West}},\ }\href {\doibase
  10.1103/PhysRevLett.111.186401} {\bibfield  {journal} {\bibinfo  {journal}
  {Phys. Rev. Lett.}\ }\textbf {\bibinfo {volume} {111}},\ \bibinfo {pages}
  {186401} (\bibinfo {year} {2013})}\BibitemShut {NoStop}%
\bibitem [{\citenamefont {Rezayi}\ and\ \citenamefont
  {Read}(2009)}]{RezayiRead:twelvefifths:09}%
  \BibitemOpen
  \bibfield  {author} {\bibinfo {author} {\bibfnamefont {E.~H.}\ \bibnamefont
  {Rezayi}}\ and\ \bibinfo {author} {\bibfnamefont {N.}~\bibnamefont {Read}},\
  }\href {\doibase 10.1103/PhysRevB.79.075306} {\bibfield  {journal} {\bibinfo
  {journal} {Phys. Rev. B}\ }\textbf {\bibinfo {volume} {79}},\ \bibinfo
  {pages} {075306} (\bibinfo {year} {2009})}\BibitemShut {NoStop}%
\bibitem [{\citenamefont {Eisenstein}\ \emph {et~al.}(2002)\citenamefont
  {Eisenstein}, \citenamefont {Cooper}, \citenamefont {Pfeiffer},\ and\
  \citenamefont {West}}]{Eisenstein2002}%
  \BibitemOpen
  \bibfield  {author} {\bibinfo {author} {\bibfnamefont {J.~P.}\ \bibnamefont
  {Eisenstein}}, \bibinfo {author} {\bibfnamefont {K.~B.}\ \bibnamefont
  {Cooper}}, \bibinfo {author} {\bibfnamefont {L.~N.}\ \bibnamefont
  {Pfeiffer}}, \ and\ \bibinfo {author} {\bibfnamefont {K.~W.}\ \bibnamefont
  {West}},\ }\href {\doibase 10.1103/PhysRevLett.88.076801} {\bibfield
  {journal} {\bibinfo  {journal} {Phys. Rev. Lett.}\ }\textbf {\bibinfo
  {volume} {88}},\ \bibinfo {pages} {076801} (\bibinfo {year}
  {2002})}\BibitemShut {NoStop}%
\bibitem [{\citenamefont {Deng}\ \emph
  {et~al.}(2012{\natexlab{b}})\citenamefont {Deng}, \citenamefont {Kumar},
  \citenamefont {Manfra}, \citenamefont {Pfeiffer}, \citenamefont {West},\ and\
  \citenamefont {Cs\'athy}}]{Deng2012_1}%
  \BibitemOpen
  \bibfield  {author} {\bibinfo {author} {\bibfnamefont {N.}~\bibnamefont
  {Deng}}, \bibinfo {author} {\bibfnamefont {A.}~\bibnamefont {Kumar}},
  \bibinfo {author} {\bibfnamefont {M.~J.}\ \bibnamefont {Manfra}}, \bibinfo
  {author} {\bibfnamefont {L.~N.}\ \bibnamefont {Pfeiffer}}, \bibinfo {author}
  {\bibfnamefont {K.~W.}\ \bibnamefont {West}}, \ and\ \bibinfo {author}
  {\bibfnamefont {G.~A.}\ \bibnamefont {Cs\'athy}},\ }\href {\doibase
  10.1103/PhysRevLett.108.086803} {\bibfield  {journal} {\bibinfo  {journal}
  {Phys. Rev. Lett.}\ }\textbf {\bibinfo {volume} {108}},\ \bibinfo {pages}
  {086803} (\bibinfo {year} {2012}{\natexlab{b}})}\BibitemShut {NoStop}%
\bibitem [{\citenamefont {Deng}\ \emph
  {et~al.}(2012{\natexlab{c}})\citenamefont {Deng}, \citenamefont {Watson},
  \citenamefont {Rokhinson}, \citenamefont {Manfra},\ and\ \citenamefont
  {Cs\'athy}}]{Deng2012_2}%
  \BibitemOpen
  \bibfield  {author} {\bibinfo {author} {\bibfnamefont {N.}~\bibnamefont
  {Deng}}, \bibinfo {author} {\bibfnamefont {J.~D.}\ \bibnamefont {Watson}},
  \bibinfo {author} {\bibfnamefont {L.~P.}\ \bibnamefont {Rokhinson}}, \bibinfo
  {author} {\bibfnamefont {M.~J.}\ \bibnamefont {Manfra}}, \ and\ \bibinfo
  {author} {\bibfnamefont {G.~A.}\ \bibnamefont {Cs\'athy}},\ }\href {\doibase
  10.1103/PhysRevB.86.201301} {\bibfield  {journal} {\bibinfo  {journal} {Phys.
  Rev. B}\ }\textbf {\bibinfo {volume} {86}},\ \bibinfo {pages} {201301}
  (\bibinfo {year} {2012}{\natexlab{c}})}\BibitemShut {NoStop}%
\bibitem [{\citenamefont {Prange}\ and\ \citenamefont
  {Girvin}(1990)}]{prangegirvin}%
  \BibitemOpen
  \bibfield  {author} {\bibinfo {author} {\bibfnamefont {R.~E.}\ \bibnamefont
  {Prange}}\ and\ \bibinfo {author} {\bibfnamefont {S.~M.}\ \bibnamefont
  {Girvin}},\ }\href@noop {} {\emph {\bibinfo {title} {The Quantum Hall
  Effect}}}\ (\bibinfo  {publisher} {Springer-Verlag, New York},\ \bibinfo
  {year} {1990})\BibitemShut {NoStop}%
\bibitem [{\citenamefont {Peterson}\ \emph {et~al.}(2008)\citenamefont
  {Peterson}, \citenamefont {Jolicoeur},\ and\ \citenamefont
  {Das~Sarma}}]{PhysRevB.78.155308}%
  \BibitemOpen
  \bibfield  {author} {\bibinfo {author} {\bibfnamefont {M.~R.}\ \bibnamefont
  {Peterson}}, \bibinfo {author} {\bibfnamefont {T.}~\bibnamefont {Jolicoeur}},
  \ and\ \bibinfo {author} {\bibfnamefont {S.}~\bibnamefont {Das~Sarma}},\
  }\href {\doibase 10.1103/PhysRevB.78.155308} {\bibfield  {journal} {\bibinfo
  {journal} {Phys. Rev. B}\ }\textbf {\bibinfo {volume} {78}},\ \bibinfo
  {pages} {155308} (\bibinfo {year} {2008})}\BibitemShut {NoStop}%
\bibitem [{\citenamefont {McCulloch}(2008)}]{McCulloch:2008}%
  \BibitemOpen
  \bibfield  {author} {\bibinfo {author} {\bibfnamefont {I.~P.}\ \bibnamefont
  {McCulloch}},\ }\href@noop {} {\enquote {\bibinfo {title} {Infinite size
  density matrix renormalization group, revisited},}\ } (\bibinfo {year}
  {2008}),\ \Eprint {http://arxiv.org/abs/0804.2509} {arXiv:0804.2509}
  \BibitemShut {NoStop}%
\bibitem [{\citenamefont {Zaletel}\ \emph {et~al.}(2015)\citenamefont
  {Zaletel}, \citenamefont {Mong}, \citenamefont {Pollmann},\ and\
  \citenamefont {Rezayi}}]{ZaletelMixing}%
  \BibitemOpen
  \bibfield  {author} {\bibinfo {author} {\bibfnamefont {M.~P.}\ \bibnamefont
  {Zaletel}}, \bibinfo {author} {\bibfnamefont {R.~S.~K.}\ \bibnamefont
  {Mong}}, \bibinfo {author} {\bibfnamefont {F.}~\bibnamefont {Pollmann}}, \
  and\ \bibinfo {author} {\bibfnamefont {E.~H.}\ \bibnamefont {Rezayi}},\
  }\href {\doibase 10.1103/PhysRevB.91.045115} {\bibfield  {journal} {\bibinfo
  {journal} {Phys. Rev. B}\ }\textbf {\bibinfo {volume} {91}},\ \bibinfo
  {pages} {045115} (\bibinfo {year} {2015})}\BibitemShut {NoStop}%
\bibitem [{Note1()}]{Note1}%
  \BibitemOpen
  \bibinfo {note} {We choose $\mathinner {|{\Omega _\protect \mathds
  {1}}\delimiter "526930B }$ to have lower entanglement entropy than that of
  $\mathinner {|{\Omega _\tau }\delimiter "526930B }$.}\BibitemShut {Stop}%
\bibitem [{\citenamefont {Li}\ and\ \citenamefont {Haldane}(2008)}]{LiHaldane}%
  \BibitemOpen
  \bibfield  {author} {\bibinfo {author} {\bibfnamefont {H.}~\bibnamefont
  {Li}}\ and\ \bibinfo {author} {\bibfnamefont {F.}~\bibnamefont {Haldane}},\
  }\href {\doibase 10.1103/PhysRevLett.101.010504} {\bibfield  {journal}
  {\bibinfo  {journal} {Phys. Rev. Lett.}\ }\textbf {\bibinfo {volume} {101}},\
  \bibinfo {pages} {010504} (\bibinfo {year} {2008})}\BibitemShut {NoStop}%
\bibitem [{\citenamefont {Kitaev}\ and\ \citenamefont
  {Preskill}(2006)}]{KitaevPreskill}%
  \BibitemOpen
  \bibfield  {author} {\bibinfo {author} {\bibfnamefont {A.}~\bibnamefont
  {Kitaev}}\ and\ \bibinfo {author} {\bibfnamefont {J.}~\bibnamefont
  {Preskill}},\ }\href {\doibase 10.1103/PhysRevLett.96.110404} {\bibfield
  {journal} {\bibinfo  {journal} {Phys. Rev. Lett.}\ }\textbf {\bibinfo
  {volume} {96}},\ \bibinfo {pages} {110404} (\bibinfo {year}
  {2006})}\BibitemShut {NoStop}%
\bibitem [{\citenamefont {Qi}\ \emph {et~al.}(2012)\citenamefont {Qi},
  \citenamefont {Katsura},\ and\ \citenamefont {Ludwig}}]{QiKatsuraLudwig}%
  \BibitemOpen
  \bibfield  {author} {\bibinfo {author} {\bibfnamefont {X.-L.}\ \bibnamefont
  {Qi}}, \bibinfo {author} {\bibfnamefont {H.}~\bibnamefont {Katsura}}, \ and\
  \bibinfo {author} {\bibfnamefont {A.~W.~W.}\ \bibnamefont {Ludwig}},\ }\href
  {\doibase 10.1103/PhysRevLett.108.196402} {\bibfield  {journal} {\bibinfo
  {journal} {Phys. Rev. Lett.}\ }\textbf {\bibinfo {volume} {108}},\ \bibinfo
  {pages} {196402} (\bibinfo {year} {2012})}\BibitemShut {NoStop}%
\bibitem [{\citenamefont {Zamolodchikov}\ and\ \citenamefont
  {Fateev}(1985)}]{ZamolodchikovParafermion}%
  \BibitemOpen
  \bibfield  {author} {\bibinfo {author} {\bibfnamefont {A.~B.}\ \bibnamefont
  {Zamolodchikov}}\ and\ \bibinfo {author} {\bibfnamefont {V.}~\bibnamefont
  {Fateev}},\ }\href {http://www.jetp.ac.ru/cgi-bin/e/index/e/62/2/p215?a=list}
  {\bibfield  {journal} {\bibinfo  {journal} {JETP}\ }\textbf {\bibinfo
  {volume} {62}},\ \bibinfo {pages} {215} (\bibinfo {year} {1985})},\ \bibinfo
  {note} {[Zh. Eksp. Teor. Fiz., {\bf 89}, 380--399]}\BibitemShut {NoStop}%
\bibitem [{Note2()}]{Note2}%
  \BibitemOpen
  \bibinfo {note} {For data in the \protect \ensuremath {\protect \overline
  {\protect \mathrm {RR}}_3}\ phase, we impose translational invariance of the
  ground state wavefunction to bias against CDO states.}\BibitemShut {Stop}%
\bibitem [{\citenamefont {Zaletel}\ \emph {et~al.}(2013)\citenamefont
  {Zaletel}, \citenamefont {Mong},\ and\ \citenamefont
  {Pollmann}}]{ZaletelQHdmrg13}%
  \BibitemOpen
  \bibfield  {author} {\bibinfo {author} {\bibfnamefont {M.~P.}\ \bibnamefont
  {Zaletel}}, \bibinfo {author} {\bibfnamefont {R.~S.~K.}\ \bibnamefont
  {Mong}}, \ and\ \bibinfo {author} {\bibfnamefont {F.}~\bibnamefont
  {Pollmann}},\ }\href {\doibase 10.1103/PhysRevLett.110.236801} {\bibfield
  {journal} {\bibinfo  {journal} {Phys. Rev. Lett.}\ }\textbf {\bibinfo
  {volume} {110}},\ \bibinfo {pages} {236801} (\bibinfo {year}
  {2013})}\BibitemShut {NoStop}%
\bibitem [{\citenamefont {Tu}\ \emph {et~al.}(2013)\citenamefont {Tu},
  \citenamefont {Zhang},\ and\ \citenamefont {Qi}}]{HHTuMomPol:13}%
  \BibitemOpen
  \bibfield  {author} {\bibinfo {author} {\bibfnamefont {H.-H.}\ \bibnamefont
  {Tu}}, \bibinfo {author} {\bibfnamefont {Y.}~\bibnamefont {Zhang}}, \ and\
  \bibinfo {author} {\bibfnamefont {X.-L.}\ \bibnamefont {Qi}},\ }\href
  {\doibase 10.1103/PhysRevB.88.195412} {\bibfield  {journal} {\bibinfo
  {journal} {Phys. Rev. B}\ }\textbf {\bibinfo {volume} {88}},\ \bibinfo
  {pages} {195412} (\bibinfo {year} {2013})}\BibitemShut {NoStop}%
\bibitem [{\citenamefont {Park}\ and\ \citenamefont
  {Haldane}(2014)}]{YeJeHaldane2014}%
  \BibitemOpen
  \bibfield  {author} {\bibinfo {author} {\bibfnamefont {Y.}~\bibnamefont
  {Park}}\ and\ \bibinfo {author} {\bibfnamefont {F.~D.~M.}\ \bibnamefont
  {Haldane}},\ }\href {\doibase 10.1103/PhysRevB.90.045123} {\bibfield
  {journal} {\bibinfo  {journal} {Phys. Rev. B}\ }\textbf {\bibinfo {volume}
  {90}},\ \bibinfo {pages} {045123} (\bibinfo {year} {2014})}\BibitemShut
  {NoStop}%
\bibitem [{\citenamefont {Wen}\ and\ \citenamefont {Zee}(1992)}]{wenzee}%
  \BibitemOpen
  \bibfield  {author} {\bibinfo {author} {\bibfnamefont {X.~G.}\ \bibnamefont
  {Wen}}\ and\ \bibinfo {author} {\bibfnamefont {A.}~\bibnamefont {Zee}},\
  }\href {\doibase 10.1103/PhysRevLett.69.953} {\bibfield  {journal} {\bibinfo
  {journal} {Phys. Rev. Lett.}\ }\textbf {\bibinfo {volume} {69}},\ \bibinfo
  {pages} {953} (\bibinfo {year} {1992})}\BibitemShut {NoStop}%
\bibitem [{\citenamefont {Bonderson}\ \emph {et~al.}(2012)\citenamefont
  {Bonderson}, \citenamefont {Feiguin}, \citenamefont {M\"oller},\ and\
  \citenamefont {Slingerland}}]{BSTwelveFifths:12}%
  \BibitemOpen
  \bibfield  {author} {\bibinfo {author} {\bibfnamefont {P.}~\bibnamefont
  {Bonderson}}, \bibinfo {author} {\bibfnamefont {A.~E.}\ \bibnamefont
  {Feiguin}}, \bibinfo {author} {\bibfnamefont {G.}~\bibnamefont {M\"oller}}, \
  and\ \bibinfo {author} {\bibfnamefont {J.~K.}\ \bibnamefont {Slingerland}},\
  }\href {\doibase 10.1103/PhysRevLett.108.036806} {\bibfield  {journal}
  {\bibinfo  {journal} {Phys. Rev. Lett.}\ }\textbf {\bibinfo {volume} {108}},\
  \bibinfo {pages} {036806} (\bibinfo {year} {2012})}\BibitemShut {NoStop}%
\bibitem [{\citenamefont {Morf}\ \emph {et~al.}(2002)\citenamefont {Morf},
  \citenamefont {d'Ambrumenil},\ and\ \citenamefont {Das~Sarma}}]{Morf2002}%
  \BibitemOpen
  \bibfield  {author} {\bibinfo {author} {\bibfnamefont {R.~H.}\ \bibnamefont
  {Morf}}, \bibinfo {author} {\bibfnamefont {N.}~\bibnamefont {d'Ambrumenil}},
  \ and\ \bibinfo {author} {\bibfnamefont {S.}~\bibnamefont {Das~Sarma}},\
  }\href {\doibase 10.1103/PhysRevB.66.075408} {\bibfield  {journal} {\bibinfo
  {journal} {Phys. Rev. B}\ }\textbf {\bibinfo {volume} {66}},\ \bibinfo
  {pages} {075408} (\bibinfo {year} {2002})}\BibitemShut {NoStop}%
\bibitem [{\citenamefont {Dean}\ \emph {et~al.}(2008)\citenamefont {Dean},
  \citenamefont {Piot}, \citenamefont {Hayden}, \citenamefont {Das~Sarma},
  \citenamefont {Gervais}, \citenamefont {Pfeiffer},\ and\ \citenamefont
  {West}}]{Dean2008}%
  \BibitemOpen
  \bibfield  {author} {\bibinfo {author} {\bibfnamefont {C.~R.}\ \bibnamefont
  {Dean}}, \bibinfo {author} {\bibfnamefont {B.~A.}\ \bibnamefont {Piot}},
  \bibinfo {author} {\bibfnamefont {P.}~\bibnamefont {Hayden}}, \bibinfo
  {author} {\bibfnamefont {S.}~\bibnamefont {Das~Sarma}}, \bibinfo {author}
  {\bibfnamefont {G.}~\bibnamefont {Gervais}}, \bibinfo {author} {\bibfnamefont
  {L.~N.}\ \bibnamefont {Pfeiffer}}, \ and\ \bibinfo {author} {\bibfnamefont
  {K.~W.}\ \bibnamefont {West}},\ }\href {\doibase
  10.1103/PhysRevLett.100.146803} {\bibfield  {journal} {\bibinfo  {journal}
  {Phys. Rev. Lett.}\ }\textbf {\bibinfo {volume} {100}},\ \bibinfo {pages}
  {146803} (\bibinfo {year} {2008})}\BibitemShut {NoStop}%
\bibitem [{\citenamefont {Bonderson}\ \emph {et~al.}(2008)\citenamefont
  {Bonderson}, \citenamefont {Freedman},\ and\ \citenamefont
  {Nayak}}]{Bonderson:MeasurementOnlyQC:08}%
  \BibitemOpen
  \bibfield  {author} {\bibinfo {author} {\bibfnamefont {P.}~\bibnamefont
  {Bonderson}}, \bibinfo {author} {\bibfnamefont {M.}~\bibnamefont {Freedman}},
  \ and\ \bibinfo {author} {\bibfnamefont {C.}~\bibnamefont {Nayak}},\ }\href
  {\doibase 10.1103/PhysRevLett.101.010501} {\bibfield  {journal} {\bibinfo
  {journal} {Phys. Rev. Lett.}\ }\textbf {\bibinfo {volume} {101}},\ \bibinfo
  {pages} {010501} (\bibinfo {year} {2008})}\BibitemShut {NoStop}%
\bibitem [{\citenamefont {Zhang}\ \emph {et~al.}(2012)\citenamefont {Zhang},
  \citenamefont {Huan}, \citenamefont {Xia}, \citenamefont {Sullivan},
  \citenamefont {Pan}, \citenamefont {Baldwin}, \citenamefont {West},
  \citenamefont {Pfeiffer},\ and\ \citenamefont {Tsui}}]{PhysRevB.85.241302}%
  \BibitemOpen
  \bibfield  {author} {\bibinfo {author} {\bibfnamefont {C.}~\bibnamefont
  {Zhang}}, \bibinfo {author} {\bibfnamefont {C.}~\bibnamefont {Huan}},
  \bibinfo {author} {\bibfnamefont {J.~S.}\ \bibnamefont {Xia}}, \bibinfo
  {author} {\bibfnamefont {N.~S.}\ \bibnamefont {Sullivan}}, \bibinfo {author}
  {\bibfnamefont {W.}~\bibnamefont {Pan}}, \bibinfo {author} {\bibfnamefont
  {K.~W.}\ \bibnamefont {Baldwin}}, \bibinfo {author} {\bibfnamefont {K.~W.}\
  \bibnamefont {West}}, \bibinfo {author} {\bibfnamefont {L.~N.}\ \bibnamefont
  {Pfeiffer}}, \ and\ \bibinfo {author} {\bibfnamefont {D.~C.}\ \bibnamefont
  {Tsui}},\ }\href {\doibase 10.1103/PhysRevB.85.241302} {\bibfield  {journal}
  {\bibinfo  {journal} {Phys. Rev. B}\ }\textbf {\bibinfo {volume} {85}},\
  \bibinfo {pages} {241302} (\bibinfo {year} {2012})}\BibitemShut {NoStop}%
\bibitem [{\citenamefont {Eisenstein}\ \emph {et~al.}(1990)\citenamefont
  {Eisenstein}, \citenamefont {Stormer}, \citenamefont {Pfeiffer},\ and\
  \citenamefont {West}}]{Eisenstein90}%
  \BibitemOpen
  \bibfield  {author} {\bibinfo {author} {\bibfnamefont {J.~P.}\ \bibnamefont
  {Eisenstein}}, \bibinfo {author} {\bibfnamefont {H.~L.}\ \bibnamefont
  {Stormer}}, \bibinfo {author} {\bibfnamefont {L.~N.}\ \bibnamefont
  {Pfeiffer}}, \ and\ \bibinfo {author} {\bibfnamefont {K.~W.}\ \bibnamefont
  {West}},\ }\href {\doibase 10.1103/PhysRevB.41.7910} {\bibfield  {journal}
  {\bibinfo  {journal} {Phys. Rev. B}\ }\textbf {\bibinfo {volume} {41}},\
  \bibinfo {pages} {7910} (\bibinfo {year} {1990})}\BibitemShut {NoStop}%
\bibitem [{\citenamefont {Kumada}\ \emph {et~al.}(2004)\citenamefont {Kumada},
  \citenamefont {Terasawa}, \citenamefont {Morino}, \citenamefont {Tagashira},
  \citenamefont {Sawada}, \citenamefont {Ezawa}, \citenamefont {Muraki},
  \citenamefont {Hirayama},\ and\ \citenamefont {Saku}}]{Hirayama2004}%
  \BibitemOpen
  \bibfield  {author} {\bibinfo {author} {\bibfnamefont {N.}~\bibnamefont
  {Kumada}}, \bibinfo {author} {\bibfnamefont {D.}~\bibnamefont {Terasawa}},
  \bibinfo {author} {\bibfnamefont {M.}~\bibnamefont {Morino}}, \bibinfo
  {author} {\bibfnamefont {K.}~\bibnamefont {Tagashira}}, \bibinfo {author}
  {\bibfnamefont {A.}~\bibnamefont {Sawada}}, \bibinfo {author} {\bibfnamefont
  {Z.~F.}\ \bibnamefont {Ezawa}}, \bibinfo {author} {\bibfnamefont
  {K.}~\bibnamefont {Muraki}}, \bibinfo {author} {\bibfnamefont
  {Y.}~\bibnamefont {Hirayama}}, \ and\ \bibinfo {author} {\bibfnamefont
  {T.}~\bibnamefont {Saku}},\ }\href {\doibase 10.1103/PhysRevB.69.155319}
  {\bibfield  {journal} {\bibinfo  {journal} {Phys. Rev. B}\ }\textbf {\bibinfo
  {volume} {69}},\ \bibinfo {pages} {155319} (\bibinfo {year}
  {2004})}\BibitemShut {NoStop}%
\bibitem [{\citenamefont {Papi\ifmmode~\acute{c}\else
  \'{c}\fi{}}(2013)}]{PhysRevB.87.245315}%
  \BibitemOpen
  \bibfield  {author} {\bibinfo {author} {\bibfnamefont {Z.}~\bibnamefont
  {Papi\ifmmode~\acute{c}\else \'{c}\fi{}}},\ }\href {\doibase
  10.1103/PhysRevB.87.245315} {\bibfield  {journal} {\bibinfo  {journal} {Phys.
  Rev. B}\ }\textbf {\bibinfo {volume} {87}},\ \bibinfo {pages} {245315}
  (\bibinfo {year} {2013})}\BibitemShut {NoStop}%
\bibitem [{\citenamefont {Stern}\ \emph {et~al.}(2012)\citenamefont {Stern},
  \citenamefont {Piot}, \citenamefont {Vardi}, \citenamefont {Umansky},
  \citenamefont {Plochocka}, \citenamefont {Maude},\ and\ \citenamefont
  {Bar-Joseph}}]{KnightShift5halves}%
  \BibitemOpen
  \bibfield  {author} {\bibinfo {author} {\bibfnamefont {M.}~\bibnamefont
  {Stern}}, \bibinfo {author} {\bibfnamefont {B.~A.}\ \bibnamefont {Piot}},
  \bibinfo {author} {\bibfnamefont {Y.}~\bibnamefont {Vardi}}, \bibinfo
  {author} {\bibfnamefont {V.}~\bibnamefont {Umansky}}, \bibinfo {author}
  {\bibfnamefont {P.}~\bibnamefont {Plochocka}}, \bibinfo {author}
  {\bibfnamefont {D.~K.}\ \bibnamefont {Maude}}, \ and\ \bibinfo {author}
  {\bibfnamefont {I.}~\bibnamefont {Bar-Joseph}},\ }\href {\doibase
  10.1103/PhysRevLett.108.066810} {\bibfield  {journal} {\bibinfo  {journal}
  {Phys. Rev. Lett.}\ }\textbf {\bibinfo {volume} {108}},\ \bibinfo {pages}
  {066810} (\bibinfo {year} {2012})}\BibitemShut {NoStop}%
\bibitem [{\citenamefont {Tiemann}\ \emph {et~al.}(2012)\citenamefont
  {Tiemann}, \citenamefont {Gamez}, \citenamefont {Kumada},\ and\ \citenamefont
  {Muraki}}]{Tiemann5halvesSpin}%
  \BibitemOpen
  \bibfield  {author} {\bibinfo {author} {\bibfnamefont {L.}~\bibnamefont
  {Tiemann}}, \bibinfo {author} {\bibfnamefont {G.}~\bibnamefont {Gamez}},
  \bibinfo {author} {\bibfnamefont {N.}~\bibnamefont {Kumada}}, \ and\ \bibinfo
  {author} {\bibfnamefont {K.}~\bibnamefont {Muraki}},\ }\href {\doibase
  10.1126/science.1216697} {\bibfield  {journal} {\bibinfo  {journal}
  {Science}\ }\textbf {\bibinfo {volume} {335}},\ \bibinfo {pages} {828}
  (\bibinfo {year} {2012})}\BibitemShut {NoStop}%
\bibitem [{\citenamefont {Haldane}\ \emph {et~al.}(2000)\citenamefont
  {Haldane}, \citenamefont {Rezayi},\ and\ \citenamefont
  {Yang}}]{HaldaneRezayiYang2000}%
  \BibitemOpen
  \bibfield  {author} {\bibinfo {author} {\bibfnamefont {F.~D.~M.}\
  \bibnamefont {Haldane}}, \bibinfo {author} {\bibfnamefont {E.~H.}\
  \bibnamefont {Rezayi}}, \ and\ \bibinfo {author} {\bibfnamefont
  {K.}~\bibnamefont {Yang}},\ }\href {\doibase 10.1103/PhysRevLett.85.5396}
  {\bibfield  {journal} {\bibinfo  {journal} {Phys. Rev. Lett.}\ }\textbf
  {\bibinfo {volume} {85}},\ \bibinfo {pages} {5396} (\bibinfo {year}
  {2000})}\BibitemShut {NoStop}%
\bibitem [{\citenamefont {{Rossokhaty}}\ \emph {et~al.}(2014)\citenamefont
  {{Rossokhaty}}, \citenamefont {{L{\"u}scher}}, \citenamefont {{Folk}},
  \citenamefont {{Watson}}, \citenamefont {{Gardner}},\ and\ \citenamefont
  {{Manfra}}}]{Rossokhaty2014}%
  \BibitemOpen
  \bibfield  {author} {\bibinfo {author} {\bibfnamefont {A.~V.}\ \bibnamefont
  {{Rossokhaty}}}, \bibinfo {author} {\bibfnamefont {S.}~\bibnamefont
  {{L{\"u}scher}}}, \bibinfo {author} {\bibfnamefont {J.~A.}\ \bibnamefont
  {{Folk}}}, \bibinfo {author} {\bibfnamefont {J.~D.}\ \bibnamefont
  {{Watson}}}, \bibinfo {author} {\bibfnamefont {G.~C.}\ \bibnamefont
  {{Gardner}}}, \ and\ \bibinfo {author} {\bibfnamefont {M.~J.}\ \bibnamefont
  {{Manfra}}},\ }\href@noop {} {\enquote {\bibinfo {title} {{Bias-induced
  breakdown of electron solids in the second Landau level}},}\ } (\bibinfo
  {year} {2014}),\ \Eprint {http://arxiv.org/abs/1412.1921} {arXiv:1412.1921}
  \BibitemShut {NoStop}%
\bibitem [{\citenamefont {Fogler}\ \emph {et~al.}(1996)\citenamefont {Fogler},
  \citenamefont {Koulakov},\ and\ \citenamefont {Shklovskii}}]{Fogler1996}%
  \BibitemOpen
  \bibfield  {author} {\bibinfo {author} {\bibfnamefont {M.~M.}\ \bibnamefont
  {Fogler}}, \bibinfo {author} {\bibfnamefont {A.~A.}\ \bibnamefont
  {Koulakov}}, \ and\ \bibinfo {author} {\bibfnamefont {B.~I.}\ \bibnamefont
  {Shklovskii}},\ }\href {\doibase 10.1103/PhysRevB.54.1853} {\bibfield
  {journal} {\bibinfo  {journal} {Phys. Rev. B}\ }\textbf {\bibinfo {volume}
  {54}},\ \bibinfo {pages} {1853} (\bibinfo {year} {1996})}\BibitemShut
  {NoStop}%
\bibitem [{\citenamefont {Moessner}\ and\ \citenamefont
  {Chalker}(1996)}]{Moessner1996}%
  \BibitemOpen
  \bibfield  {author} {\bibinfo {author} {\bibfnamefont {R.}~\bibnamefont
  {Moessner}}\ and\ \bibinfo {author} {\bibfnamefont {J.~T.}\ \bibnamefont
  {Chalker}},\ }\href {\doibase 10.1103/PhysRevB.54.5006} {\bibfield  {journal}
  {\bibinfo  {journal} {Phys. Rev. B}\ }\textbf {\bibinfo {volume} {54}},\
  \bibinfo {pages} {5006} (\bibinfo {year} {1996})}\BibitemShut {NoStop}%
\bibitem [{\citenamefont {Goerbig}\ \emph {et~al.}(2003)\citenamefont
  {Goerbig}, \citenamefont {Lederer},\ and\ \citenamefont
  {Morais~Smith}}]{Goerbig2003}%
  \BibitemOpen
  \bibfield  {author} {\bibinfo {author} {\bibfnamefont {M.~O.}\ \bibnamefont
  {Goerbig}}, \bibinfo {author} {\bibfnamefont {P.}~\bibnamefont {Lederer}}, \
  and\ \bibinfo {author} {\bibfnamefont {C.}~\bibnamefont {Morais~Smith}},\
  }\href {\doibase 10.1103/PhysRevB.68.241302} {\bibfield  {journal} {\bibinfo
  {journal} {Phys. Rev. B}\ }\textbf {\bibinfo {volume} {68}},\ \bibinfo
  {pages} {241302} (\bibinfo {year} {2003})}\BibitemShut {NoStop}%
\bibitem [{\citenamefont {MacDonald}\ and\ \citenamefont
  {Fisher}(2000)}]{MacDonald2000}%
  \BibitemOpen
  \bibfield  {author} {\bibinfo {author} {\bibfnamefont {A.~H.}\ \bibnamefont
  {MacDonald}}\ and\ \bibinfo {author} {\bibfnamefont {M.~P.~A.}\ \bibnamefont
  {Fisher}},\ }\href {\doibase 10.1103/PhysRevB.61.5724} {\bibfield  {journal}
  {\bibinfo  {journal} {Phys. Rev. B}\ }\textbf {\bibinfo {volume} {61}},\
  \bibinfo {pages} {5724} (\bibinfo {year} {2000})}\BibitemShut {NoStop}%
\bibitem [{\citenamefont {Barci}\ \emph {et~al.}(2002)\citenamefont {Barci},
  \citenamefont {Fradkin}, \citenamefont {Kivelson},\ and\ \citenamefont
  {Oganesyan}}]{Barci2002}%
  \BibitemOpen
  \bibfield  {author} {\bibinfo {author} {\bibfnamefont {D.~G.}\ \bibnamefont
  {Barci}}, \bibinfo {author} {\bibfnamefont {E.}~\bibnamefont {Fradkin}},
  \bibinfo {author} {\bibfnamefont {S.~A.}\ \bibnamefont {Kivelson}}, \ and\
  \bibinfo {author} {\bibfnamefont {V.}~\bibnamefont {Oganesyan}},\ }\href
  {\doibase 10.1103/PhysRevB.65.245319} {\bibfield  {journal} {\bibinfo
  {journal} {Phys. Rev. B}\ }\textbf {\bibinfo {volume} {65}},\ \bibinfo
  {pages} {245319} (\bibinfo {year} {2002})}\BibitemShut {NoStop}%
\bibitem [{\citenamefont {Nuebler}\ \emph {et~al.}(2010)\citenamefont
  {Nuebler}, \citenamefont {Umansky}, \citenamefont {Morf}, \citenamefont
  {Heiblum}, \citenamefont {von Klitzing},\ and\ \citenamefont
  {Smet}}]{Nuebler2010}%
  \BibitemOpen
  \bibfield  {author} {\bibinfo {author} {\bibfnamefont {J.}~\bibnamefont
  {Nuebler}}, \bibinfo {author} {\bibfnamefont {V.}~\bibnamefont {Umansky}},
  \bibinfo {author} {\bibfnamefont {R.}~\bibnamefont {Morf}}, \bibinfo {author}
  {\bibfnamefont {M.}~\bibnamefont {Heiblum}}, \bibinfo {author} {\bibfnamefont
  {K.}~\bibnamefont {von Klitzing}}, \ and\ \bibinfo {author} {\bibfnamefont
  {J.}~\bibnamefont {Smet}},\ }\href {\doibase 10.1103/PhysRevB.81.035316}
  {\bibfield  {journal} {\bibinfo  {journal} {Phys. Rev. B}\ }\textbf {\bibinfo
  {volume} {81}},\ \bibinfo {pages} {035316} (\bibinfo {year}
  {2010})}\BibitemShut {NoStop}%
\bibitem [{\citenamefont {Zhu}\ \emph {et~al.}(2015)\citenamefont {Zhu},
  \citenamefont {Gong}, \citenamefont {Haldane},\ and\ \citenamefont
  {Sheng}}]{WZhuDSheng:ReadRezayi}%
  \BibitemOpen
  \bibfield  {author} {\bibinfo {author} {\bibfnamefont {W.}~\bibnamefont
  {Zhu}}, \bibinfo {author} {\bibfnamefont {S.~S.}\ \bibnamefont {Gong}},
  \bibinfo {author} {\bibfnamefont {F.~D.~M.}\ \bibnamefont {Haldane}}, \ and\
  \bibinfo {author} {\bibfnamefont {D.~N.}\ \bibnamefont {Sheng}},\ }\href
  {\doibase 10.1103/PhysRevLett.115.126805} {\bibfield  {journal} {\bibinfo
  {journal} {Phys. Rev. Lett.}\ }\textbf {\bibinfo {volume} {115}},\ \bibinfo
  {pages} {126805} (\bibinfo {year} {2015})}\BibitemShut {NoStop}%
\bibitem [{\citenamefont {{Pakrouski}}\ \emph {et~al.}(2016)\citenamefont
  {{Pakrouski}}, \citenamefont {{Troyer}}, \citenamefont {{Wu}}, \citenamefont
  {{Das Sarma}},\ and\ \citenamefont {{Peterson}}}]{Pakrouski12fifths}%
  \BibitemOpen
  \bibfield  {author} {\bibinfo {author} {\bibfnamefont {K.}~\bibnamefont
  {{Pakrouski}}}, \bibinfo {author} {\bibfnamefont {M.}~\bibnamefont
  {{Troyer}}}, \bibinfo {author} {\bibfnamefont {Y.-L.}\ \bibnamefont {{Wu}}},
  \bibinfo {author} {\bibfnamefont {S.}~\bibnamefont {{Das Sarma}}}, \ and\
  \bibinfo {author} {\bibfnamefont {M.~R.}\ \bibnamefont {{Peterson}}},\
  }\href@noop {} {\enquote {\bibinfo {title} {{The enigmatic 12/5 fractional
  quantum Hall effect}},}\ } (\bibinfo {year} {2016}),\ \bibinfo {note}
  {unpublished},\ \Eprint {http://arxiv.org/abs/1604.04610} {arXiv:1604.04610}
  \BibitemShut {NoStop}%
\bibitem [{\citenamefont {Avron}\ \emph {et~al.}(1995)\citenamefont {Avron},
  \citenamefont {Seiler},\ and\ \citenamefont
  {Zograf}}]{AvronSeilerZograf:HallViscosity:1995}%
  \BibitemOpen
  \bibfield  {author} {\bibinfo {author} {\bibfnamefont {J.~E.}\ \bibnamefont
  {Avron}}, \bibinfo {author} {\bibfnamefont {R.}~\bibnamefont {Seiler}}, \
  and\ \bibinfo {author} {\bibfnamefont {P.~G.}\ \bibnamefont {Zograf}},\
  }\href {\doibase 10.1103/PhysRevLett.75.697} {\bibfield  {journal} {\bibinfo
  {journal} {Phys. Rev. Lett.}\ }\textbf {\bibinfo {volume} {75}},\ \bibinfo
  {pages} {697} (\bibinfo {year} {1995})}\BibitemShut {NoStop}%
\bibitem [{\citenamefont {Levin}\ and\ \citenamefont
  {Wen}(2006)}]{LevinWen:06}%
  \BibitemOpen
  \bibfield  {author} {\bibinfo {author} {\bibfnamefont {M.}~\bibnamefont
  {Levin}}\ and\ \bibinfo {author} {\bibfnamefont {X.-G.}\ \bibnamefont
  {Wen}},\ }\href {\doibase 10.1103/PhysRevLett.96.110405} {\bibfield
  {journal} {\bibinfo  {journal} {Phys. Rev. Lett.}\ }\textbf {\bibinfo
  {volume} {96}},\ \bibinfo {pages} {110405} (\bibinfo {year}
  {2006})}\BibitemShut {NoStop}%
\bibitem [{\citenamefont {Haldane}(1985)}]{Haldane-PhysRevLett.55.2095}%
  \BibitemOpen
  \bibfield  {author} {\bibinfo {author} {\bibfnamefont {F.~D.~M.}\
  \bibnamefont {Haldane}},\ }\href {\doibase 10.1103/PhysRevLett.55.2095}
  {\bibfield  {journal} {\bibinfo  {journal} {Phys. Rev. Lett.}\ }\textbf
  {\bibinfo {volume} {55}},\ \bibinfo {pages} {2095} (\bibinfo {year}
  {1985})}\BibitemShut {NoStop}%
\bibitem [{\citenamefont {{Metlitski}}\ and\ \citenamefont
  {{Grover}}(2011)}]{Metlitski2011}%
  \BibitemOpen
  \bibfield  {author} {\bibinfo {author} {\bibfnamefont {M.~A.}\ \bibnamefont
  {{Metlitski}}}\ and\ \bibinfo {author} {\bibfnamefont {T.}~\bibnamefont
  {{Grover}}},\ }\href@noop {} {\  (\bibinfo {year} {2011})},\ \Eprint
  {http://arxiv.org/abs/1112.5166} {arXiv:1112.5166} \BibitemShut {NoStop}%
\end{thebibliography}%

\end{document}